\documentclass[amssymb,11pt]{article}
\usepackage{graphicx}
\begin{document}
\title{A Theory of Computation Based on Quantum Logic (I)}
\author{Mingsheng Ying\thanks{This work was partly
supported by the National Key Project for Fundamental Research of
China (Grant No: 1998030905) and the National
Foundation of Natural Sciences of China (Grant No: 60273003)}\\
\\
\small \em State Key Laboratory of Intelligent Technology and Systems,\\
\small \em Department of Computer Science and Technology,\\
\small \em Tsinghua University, Beijing 100084, China,\\
\small \em Email: yingmsh@tsinghua.edu.cn}
\date{}
\maketitle

\begin{abstract}
The (meta)logic underlying classical theory of computation is
Boolean (two-valued) logic. Quantum logic was proposed by Birkhoff
and von Neumann as a logic of quantum mechanics more than sixty
years ago. It is currently understood as a logic whose truth
values are taken from an orthomodular lattice. The major
difference between Boolean logic and quantum logic is that the
latter does not enjoy distributivity in general. The rapid
development of quantum computation in recent years stimulates us
to establish a theory of computation based on quantum logic. The
present paper is the first step toward such a new theory and it
focuses on the simplest models of computation, namely finite
automata. We introduce the notion of orthomodular lattice-valued
(quantum) automaton. Various properties of automata are carefully
reexamined in the framework of quantum logic by employing an
approach of semantic analysis. We define the class of regular
languages accepted by orthomodular lattice-valued automata. The
acceptance abilities of orthomodular lattice-valued
nondeterministic automata and their various modifications (such as
deterministic automata and automata with $\varepsilon-$moves) are
compared. The closure properties of orthomodular lattice-valued
regular languages are derived. The Kleene theorem about
equivalence of regular expressions and finite automata is
generalized into quantum logic. We also present a pumping lemma
for orthomodular lattice-valued regular languages. It is found
that the universal validity of many properties (for example, the
Kleene theorem, the equivalence of deterministic and
nondeterministic automata) of automata depend heavily upon the
distributivity of the underlying logic. This indicates that these
properties does not universally hold in the realm of quantum
logic. On the other hand, we show that a local validity of them
can be recovered by imposing a certain commutativity to the
(atomic) statements about the automata under consideration. This
reveals an essential difference between the classical theory of
computation and the computation theory based on quantum logic.
\end{abstract}\par

\smallskip\


finite automata, regular languages, quantum computation

\vspace{2em}

\textbf{Contents}

\smallskip\

1. Introduction (page 2)

2. Quantum logic (page 11)

\ \ \ \ 2.1. Orthomodular lattices (page 11)

\ \ \ \ 2.2. The language of quantum logic (page 20)

\ \ \ \ 2.3. The algebraic semantics of quantum logic (page 20)

\ \ \ \ 2.4. The operations of quantum sets (page 22)

3. Orthomodular lattice-valued automata (page 25)

4. Orthomodular lattice-valued deterministic automata (page 33)

5. Orthomodular lattice-valued automata with $\epsilon-$moves
(page 40)

6. Closure properties of orthomodular lattice-valued regularity
(page 44)

7. Orthomodular lattice-valued regular expressions (page 57)

8. Pumping lemma for orthomodular lattice-valued regular languages
(page 69)

9. Conclusion (page 71)

\ \ \ \ References (page 73)

\bigskip\

\smallskip\

\textbf{1. Introduction}

\smallskip\

It is well-known that an axiomatization of a mathematical theory
consists of a system of fundamental notions as well as a set of
axioms about these notions. The mathematical theory is then the
set of theorems which can be derived from the axioms. Obviously,
one needs a certain logic to provide tools for reasoning in the
derivation of these theorems from the axioms. As pointed out by A.
Heyting [He63, page 5], in elementary axiomatics logic was used in
an unanalyzed form. Afterwards, in the studies for foundations of
mathematics beginning in the early of twentieth century, it had
been realized that a major part of mathematics has to exploit the
full power of classical (Boolean) logic [Ha82], the strongest one
in the family of existing logics. For example, group theory is
based on first-order logic, and point-set topology is built on a
fragment of second-order logic. However, a few mathematicians,
including the big names L. E. J. Brouwer, H. Poincare, L.
Kronecker and H. Weyl, took some kind of constructive position
which is in more or less explicit opposition to certain forms of
mathematical reasoning used by the majority of the mathematical
community. Some of them even endeavored to establish so-called
constructive mathematics, the part of mathematics that could be
rebuilt on constructivist principles. The logic employed in the
development of constructive mathematics is intuitionistic logic
[TD88] which is truly weaker than classical logic.

Since many logics different from classical logic and
intuitionistic logic have been invented in the last century, one
may naturally ask the question whether we are able to establish
some mathematical theories based on other nonclassical logics
besides intuitionistic logic. Indeed, as early as the first
nonclassical logics appeared, the possibility of building
mathematics upon them was conceived. As mentioned by A. Mostowski
[M65], J. Lukasiewicz hoped that there would be some nonclassical
logics which can be properly used in mathematics as non-Euclidean
geometry does. In 1952, J. B. Rosser and A. R. Turquette [RT52,
page 109] proposed a similar and even more explicit idea:
$$$$
\textit{"The fact that it is thus possible to generalize the
ordinary two-valued logic so as not only to cover the case of
many-valued statement calculi, but of many-valued quantification
theory as well, naturally suggests the possibility of further
extending our treatment of many-valued logic to cover the case of
many-valued sets, equality, numbers, etc. Since we now have a
general theory of many-valued predicate calculi, there is little
doubt about the possibility of successfully developing such
extended many-valued theories. ... we shall consider their careful
study one of the major unsolved problems of many-valued logic."}
$$$$
Unfortunately, the above idea has not attracted much attention in
logical community. For such a situation, A. Mostowski [M65]
pointed out that most of nonclassical logics invented so far have
not been really used in mathematics, and intuitionistic logic
seems the unique one of nonclassical logics which still has an
opportunity to carry out the Lukasiewicz's project. A similar
opinion was also expressed by J. Dieudonne [Di78], and he said
that mathematical logicians have been developing a variety of
nonclassical logics such as second-order logic, modal logic and
many-valued logic, but these logics are completely useless for
mathematicians working in other research areas.

One reason for this situation might be that there is no suitable
method to develop mathematics within the framework of nonclassical
logics. As was pointed out above, classical logic is applied as
the deduction tool in almost all mathematical theories. It should
be noted that what is used in these theories is the deductive
(proof-theoretical) aspect of classical logic. However, the proof
theory of nonclassical logics is much more complicated than that
of classical logic, and it is not an easy task to conduct
reasoning in the realm of the proof theory of nonclassical logics.
It is the case even for the simplest nonclassical logics,
three-valued logics. This is explicitly indicated by the following
excerpt from H. Hodes [Ho89]:
$$$$
\textit{"Of course three-valued logics will be somewhat more
complicated than classical two-valued logic. In fact,
proof-theoretically they are at least twice as complicated: ....
But model-theoretically they are only 50 percent more
complicated,....}
$$$$
And much worse, some nonclassical logics were introduced only in a
semantic way, and the axiomatizations of some among them are still
to be found, and some of them may be not (finitely) axiomatizable.
Thus, our experience in studying classical mathematics may be not
suited, or at least cannot directly apply, to develop mathematics
based on nonclassical logics. In the early 1990's an attempt had
been made by the author [Yi91-93; Yi93] to give a partial and
elementary answer in the case of point-set topology to the J. B.
Rosser and A. R. Turquette's question raised above. We employed a
semantical analysis approach to establish topology based on
residuated lattice-valued logic, especially the Lukasiewicz system
of continuous-valued logic. Roughly speaking, the semantical
analysis approach transforms our intended conclusions in
mathematics, which are usually expressed as implication formulas
in our logical language, into certain inequalities in the
truth-value lattice by truth valuation rules, and then we
demonstrate these inequalities in an algebraic way and conclude
that the original conclusions are semantically valid. We believe
that semantical analysis approach is an effective method to
develop mathematics based on nonclassical logics.

A much more essential reason for the situation that few
nonclassical logics have been applied in mathematics is absence of
appealing from other subjects or applications in the real world.
One major exception may be the case of quantum logic. Quantum
logic was introduced by G. Birkhoff and J. von Neumann [BN36] in
the thirties of the twentieth century as the logic of quantum
mechanics. They realized that quantum mechanical systems are not
governed by classical logical laws. Their proposed logic stems
from von Neumann's Hilbert space formalism of quantum mechanics.
The starting point was explained very well by the following
excerpt from G. Birkhoff and J. von Neumann [BN36]:
$$$$
\textit{"what logical structure one may hope to find in physical
theories which, like quantum mechanics, do not conform to
classical logic. Our main conclusion, based on admittedly
heuristic arguments, is that one can reasonably expect to find a
calculus of propositions which is formally indistinguishable from
the calculus of linear subspaces [of Hilbert space] with respect
to set products, linear sums, and orthogonal complements - and
resembles the usual calculus of propositions with respect to
'and', 'or', and 'not'."}
$$$$
Thus linear (closed) subspaces of Hilbert space are identified
with propositions concerning a quantum mechanical system, and the
operations of set product, linear sum and orthogonal complement
are treated as connectives. By observing that the set of linear
subspaces of a finite-dimensional Hilbert space together with
these operations enjoys Dedekind's modular law, G. Birkhoff and J.
von Neumann [BN36] suggested to use modular lattices as the
algebraic version of the logic of quantum mechanics, just like
that Boolean algebras act as an algebraic counterpart of classical
logic. However, the modular law does not hold in an
infinite-dimensional Hilbert space. In 1937, K. Husimi [Hu37]
found a new law, called now the orthomodular law, which is valid
for the set of linear subspaces of any Hilbert space. Nowadays,
what is usually called quantum logic in the mathematical physics
literatures refers to the theory of orthomodular lattices.
Obviously, this kind of quantum logic is not very logical. Indeed,
there is also another much more 'logical' point of view on quantum
logic in which quantum logic is seen as a logic whose truth values
range over an orthomodular lattice (for an excellent exposition
for the latter approach of quantum logic, see M. L. Dalla Chiara
[DC86], or J. P. Rawling and S. A. Selesnick [RS00]). After the
invention of quantum logic, quite a few mathematicians have tried
to establish mathematics based on quantum logic. Indeed, J. von
Neumann [N62] himself proposed the idea of considering a quantum
set theory, corresponding to quantum logic, as does classical set
theory to classical logic. One important contribution in this
direction was made by G. Takeuti [T81]. His main idea was
explained, and the nature of mathematics based on quantum logic
was analyzed very well by the following citation from the
introduction of [T81]:
$$$$
\textit{"Since quantum logic is an intrinsic logic, i.e. the logic
of the quantum world, it is an important problem to develop
mathematics based on quantum logic, more specifically set theory
based on quantum logic. It is also a challenging problem for
logicians since quantum logic is drastically different from the
classical logic or the intuitionistic logic and consequently
mathematics based on quantum logic is extremely difficult. On the
other hand, mathematics based on quantum logic has a very rich
mathematical content. This is clearly shown by the fact that there
are many complete Boolean algebras inside quantum logic. For each
complete Boolean algebra $B$, mathematics based on $B$ has been
shown by our work on Boolean valued analysis to have rich
mathematical meaning. Since mathematics based on $B$ can be
considered as a sub-theory of mathematics based on quantum logic,
there is no doubt about the fact that mathematics based on quantum
logic is very rich. The situation seems to be the following.
Mathematics based on quantum logic is too gigantic to see through
clearly."}
$$$$
The main technical result of G. Takeuti [T81] is a construction of
orthomodular lattice-valued universe. He built up such an universe
in a way similar to Boolean-valued models of ZF + AC, and showed
that a reasonable set theory, including some axioms from ZF + AC
or their slight modifications, holds in this universe. Recently,
K. -G. Schlesinger [Sc99] developed a theory of quantum sets by
using a categorical approach in the spirit of topos theory. He
started with the category of complex (pre-)Hilbert spaces and
linear maps. This category was seen as the (basic) quantum set
universe. Then he was able to introduce the analog of number
systems and to deal with the analog of some algebraic structures
in quantum set theory. Indeed, K. -G. Schlesinger's terminal goal
is to build a quantum mathematics, i.e., a mathematical theory
where all the ingredients (like logic and set theory) adhere to
the rules of quantum mechanics. Quantum set theory is the
quantization of the mathematical theory of pure objects, and so it
is just the first step toward his goal. It is worth noting that
the role of quantum logic in such a quantum mathematics is
different from that in G. Takeuti's quantum set theory, and
quantum logic appears as an internal logic in the former.

After a careful examination on the development of mathematics
based on nonclassical logics, we now come to explore the
possibility of establishing a theory of computation based on
nonclassical logics. A formal formulation of the notion of
computation is one of the greatest scientific achievements in the
twentieth century. Since the middle of 1930's, various models of
computation have been introduced, such as Turing machines, Post
systems, $\lambda-$calculus and $\mu-$recursive functions. In
classical computing theory, these models of computations are
investigated in the framework of classical logic; more explicitly,
all properties of them are deduced by classical logic as a
(meta)logical tool. So, it is reasonable to say that classical
computing theory is a part of classical mathematics. Knowing the
basic idea of mathematics based on nonclassical logics, we may
naturally ask the question: is it possible to build a theory of
computation based on nonclassical logics, and what are the same of
and difference between the properties of the models of
computations in classical logic and the corresponding ones in
non-classical logics? There has been a very big population of
non-classical logics. Of course, it is unnecessary to construct
models of computations in each nonclassical logic and to compare
them with the ones in classical logic because some nonclassical
logics are completely irrelative to behaviors of computations.
Nevertheless, as will be explained shortly, it is absolutely worth
studying deeply and systematically models of computations based on
quantum logic.

It seems that both points of views on quantum logic mentioned
above have no obvious links to computations; but appearance of the
idea of quantum computers changed dramatically the long-standing
situation. The idea of quantum computation came from the studies
of connections between physics and computation. The first step
toward it was the understanding of the thermodynamics of classical
computation. In 1973, C. H. Bennet [Ben73] noted that a logically
reversible operation need not dissipate any energy and found that
a logically reversible Turing machine is a theoretical
possibility. In 1980, further progress was made by P. A. Benioff
[Be80] who constructed a quantum mechanical model of Turing
machine. His construction is the first quantum mechanical
description of computer, but it is not a real quantum computer. It
should be noted that in P. A. Benioff's model between computation
steps the machine may exist in an intrinsically quantum state, but
at the end of each computation step the tape of the machine always
goes back to one of its classical states. Quantum computers were
first envisaged by R. P. Feynman [Fe82; Fe86]. In 1982, he [Fe82]
conceived that no classical Turing machine could simulate certain
quantum phenomena without an exponential slowdown, and so he
realized that quantum mechanical effects should offer something
genuinely new to computation. Although R. P. Feynman proposed the
idea of universal quantum simulator, he did not give a concrete
design of such a simulator. His ideas were elaborated and
formalized by D. Deutsch in a seminal paper [De85]. In 1985, D.
Deutsch described the first true quantum Turing machine. In his
machine, the tape is able to exist in quantum states too. This is
different from P. A. Benioff's machine. In particular, D. Deutsch
introduced the technique of quantum parallelism by which quantum
Turing machine can encode many inputs on the same tape and perform
a calculation on all the inputs simultaneously. Furthermore, he
proposed that quantum computers might be able to perform certain
types of computations that classical computers can only perform
very inefficiently. One of the most striking advances was made by
P. W. Shor [S94] in 1994. By exploring the power of quantum
parallelism, he discovered a polynomial-time algorithm on quantum
computers for prime factorization of which the best known
algorithm on classical computers is exponential. In 1996, L. K.
Grover [Gr96] offered another apt killer of quantum computation,
and he found a quantum algorithm for searching a single item in an
unsorted database in square root of the time it would take on a
classical computer. Since both prime factorization and database
search are central problems in computer science and the quantum
algorithms for them are highly faster than the classical ones, P.
W. Shor and L. K. Grover's works stimulated an intensive
investigation on quantum computation. After that, quantum
computation has been an extremely exciting and rapidly growing
field of research.

The studies of quantum computation may be roughly divided into
four strata, arranged according increasing order of abstraction
degree: (1) physical implementations; (2) physical models; (3)
mathematical models; and (4) logical foundations. Almost all
pioneer works such as [Be80, F82, D85] in this field were devoted
to build physical models of quantum computing. In 1990's, a great
attention was paid to the physical implementation of quantum
computation. For example, S. Lloyd [L93] considered the practical
implementation by using electromagnetic pulses and J. I. Cirac and
P. Zoller [CZ95] used laser manipulations of cold trapped ions to
implement quantum computing. The current theoretical concerns in
the area of quantum computation have mainly been given to quantum
algorithms. But also there have been a few attempts to develop
mathematical models of quantum computation and to clarify the
relationship between different models. For example, except quantum
Turing machines, D. Deutsch [De89] also proposed the quantum
circuit model of computation, and A. C. Yao [Ya93] showed that the
quantum circuit model is equivalent to the quantum Turing machine
in the sense that they can simulate each other in polynomial time.
As is well known, in classical computing theory, there are still
two important classes of models of computation rather than Turing
machines; namely, finite automata and pushdown automata. They are
equipped with finite memory or finite memory with stack,
respectively, and so have weaker computing power than Turing
machines. Recently, J. P. Crutchfield and C. Moore [CM00], A.
Kondacs and J. Watrous [KW97], and S. Gudder [Gu00] tried to
introduce some quantum devices corresponding to these weaker
models of computation. Roughly speaking, quantum automata may be
seen as quantum counterparts of probabilistic automata. In a
probabilistic automaton, each transition is equipped with a number
in the unit interval to indicate the probability of the occurrence
of the transition; by contrast in a quantum automaton we associate
with each transition a vector in a Hilbert space which is
interpreted as the probability amplitude of the transition. In a
sense, these mathematical models of quantum computation can be
seen as abstractions of its physical models.

It should be noted that the theoretical models of quantum
computation mentioned above, including quantum Turing machines and
quantum automata, are still developed in classical (Boolean)
logic. Thus, their logical basis is the same as that of classical
computation, and we may argue that sometimes these models might be
not suitable for quantum computers that obey some logical laws
different from that in Boolean logic. Indeed, V. Vedral and M. B.
Plenio [VP98] already advocated that quantum computers require
quantum logic, something fundamentally different to classical
Boolean logic. As stated above, quantum logic has been existing
for a long time. So, the point is how to apply quantum logic in
the analysis and design of quantum computers. The background
exposed above highly motivates us to explore the possibility of
establishing a theory of computation based on quantum logic. The
purpose of the present paper and its continuations is exactly to
develop such a new theory. In a sense, our approach may be thought
of as a logical foundation of quantum computation and a further
abstraction of its mathematical models. The relation between
Crutchfield et al's studies [CM00, KW97, Gu00] on quantum automata
and our automata theory based on quantum logic is quite similar to
that between J. von Neumann's Hilbert space formalism of quantum
mechanics and quantum logic.

Since finite automata are the simplest models of computation (with
finite memory), in this paper we focus our attention on developing
a theory of finite automata based on quantum logic. The present
paper is organized as follows. In Section 2, we recall some basic
notions and results of quantum logic and its algebraic semantics
needed in the subsequent sections from the previous literature.
Some new lemmas on implication operators in quantum logic are
presented too. They are crucial in the proofs of several main
results in this paper. In Section 3, we introduce the notion of
orthomodular lattice-valued (quantum) automaton. Then two
different orthomodular lattice-valued predicates of regularity on
languages are proposed. These two predicates stands indeed for the
(orthomodular lattice-valued) class of languages accepted by
orthomodular lattice-valued automata. This provides us with a
framework in which various properties of automata can be
reexamined within quantum logic. The technique employed in this
paper is mainly the approach of semantic analysis developed in
[Yi91-93; Yi93]. The acceptance ability of orthomodular
lattice-valued nondeterministic automata are then compared with
that of their two kinds of modifications, namely deterministic
automata and automata with $\varepsilon-$moves, respectively in
Sections 4 and 5. The closure properties of orthomodular
lattice-valued regular languages are derived in Section 6. In
Section 7, we introduce the notion of orthomodular lattice-valued
regular expression, and the Kleene theorem about equivalence of
regular expressions and finite automata is generalized into
quantum logic. Section 8 is devoted to present a pumping lemma for
orthomodular lattice-valued regular languages. Some basic ideas of
this paper were announced in [Yi00], and Definitions 3.1 and 3.2,
Examples 3.1-4, and Propositions 6.1 and 6.3 were also presented
there. For completeness, however, they are included in the present
paper.

The most interesting thing is, in the author's opinion, the
discovery that the universal validity of many properties (for
example, the Kleene theorem, the equivalence of deterministic and
nondeterministic automata) of automata depend heavily upon the
distributivity of the underlying logic. It is shown that the
universal validity of these properties is equivalent to the
requirement that the set of truth values of the meta-logic
underlying our theory of automata is a Boolean algebra. This
indicates that these properties does not universally hold in the
realm of quantum logic, and it is in fact a negative conclusion in
our theory of automata based on quantum logic. Furthermore, it
implies the fact that an essential difference exists between the
classical theory of computation and the computation theory based
on quantum logic.

Observing that some important properties of automata cannot be
built within quantum logic, one may naturally ask the question
whether they may be partially recast without appealing to
distributivity of the underlying logic. Fortunately, we are able
to show that a local validity of these properties of automata can
be recovered by imposing a certain commutativity to the truth
values of the (atomic) statements about the automata under
consideration. Very surprisingly, almost all results in classical
automata theory that are not valid in a non-distributive logic can
be revived by a certain commutativity in quantum logic. This
further leads us to a new question: why commutativity plays such a
key role for quantum automata, and is there any physical
interpretation for it? To answer this question, let us first note
that all truth values in quantum logic are taken from an
orthomodular lattice. The prototype of orthomodular lattice is the
set of linear (closed) subspaces of a Hilbert space with the set
inclusion as its ordering relation. Suppose that $X$ and $Y$ are
two subspaces of a Hilbert space $H$. Moreover, we use $P_X$ and
$P_Y$ to denote the projections on $X$ and $Y$ respectively. Then
$P_X$ and $P_Y$ are Hermitian operators on $H$, and they may be
seen as two (physical) observables $A$ and $B$ in a quantum system
whose state space is $H$, according to the Hilbert space formalism
of quantum mechanics. If we write $\Delta (A)$ and $\Delta (B)$
for the respective standard deviations of measurement on $A$ and
$B$, then the Heisenberg uncertainty principle gives the following
inequality:
$$\Delta (A)\cdot \Delta (B)\geq \frac{1}{2} |<\psi
|[A,B]|\psi>|$$ for all quantum state $|\psi>$ in $H$, where
$[A,b]=AB-BA$ is the commutator between $A$ and $B$. We now turn
back to the orthomodular lattice of the linear subspaces of $H$.
The commutativity of $A$ and $B$ is defined by the condition
$X=(X\wedge Y)\vee (X\wedge Y^{\bot})$, where $\wedge$, $\vee$ and
$\bot$ are respectively the meet, union and orthocomplement. It
may be seen that the commutativity between $X$ and $Y$ is
equivalent to exactly the fact that $A$ and $B$ commutate, i.e.,
$AB=BA$. In this case, $|<\psi|[A,B]|\psi>|=0$, and $\Delta
(A)\cdot \Delta (B)$ may vanish; or in other words, $\Delta (A)$
and $\Delta (B)$ can simultaneously become arbitrarily small.
Remember that in our theory of automata based on quantum logic the
commutativity is attached to the basic statements describing the
considered automata. On the other hand, the basic statements are
indeed corresponding to some actions in these automata. Therefore,
a potential physical interpretation for the need of commutativity
is that some nice properties of automata require the standard
deviations of the observables concerning the basic actions in
these automata being able to reach simultaneously very small
values.

The results gained in our approach may offer some new insights on
the theory of computation. As an example, let us consider the
Church-Turing thesis. The realization that the intuitive notion of
"effective computation" can be identified with the mathematical
concept of "computation by the Turing machine" is based on the
fact that the Turing machine is computationally equivalent to some
vastly dissimilar formalisms for the same purpose, such as Post
systems, $\mu-$recursive functions, $\lambda-$calculus and
combinatory logic. As pointed out by J. E. Hopcroft and J. D.
Ullman [HU79], another reason for the acceptance of the Turing
machine as a general model of a computation is that the Turing
machine is equivalent to its many modified versions that would
seem off-hand to have increased computing power. We should note
that the equivalence between the Turing machine and its various
generalizations as well as other formalisms of computation has
been reached in classical Boolean logic. In addition, quantum
logic is known to be strictly weaker than Boolean logic. Thus, it
is reasonable to doubt that the same equivalence can be achieved
when our underlying meta-logic is replaced by quantum logic, and
the Church-Turing thesis needs to be reexamined in the realm of
quantum logic. Indeed, in a continuation of this paper we are
going to establish a theory of Turing machines based on quantum
logic. The details of such a theory is still to be exploited, but
the conclusion concerning the equivalence between deterministic
and nondeterministic automata obtained in this paper suggests us
to believe that the equivalence between deterministic and
nondeterministic Turing machines also depends upon the
distributivity of the underlying logic, and a certainty
commutativity for the basic actions in Turing machines will
guarantee such an equivalence. Keeping this belief in mind, we may
assert that a certain commutativity of the observables for some
basic actions in the Turing machine is a physical support of the
Church-Turing thesis in the framework of quantum logic.
Furthermore, with the above physical interpretation for
commutativity, this hints that there might be a deep connection
between the Heisenberg uncertainty principle and the Church-Turing
thesis, two of the greatest scientific discoveries in the
twentieth century. It is notable that such a connection could be
observed via an argument in a nonclassical logic (and it is
impossible to be found if we always work within the classical
logic). As early as in 1985, it was argued by D. Deutsch [De84]
that underlying the Church-Turing thesis there is an implicit
physical assertion. There is certainly no doubt about the
existence of such a physical assertion. The true problem here is:
what is it? The answer given by D. Deutsch himself is the
following physical principle: "every finitely realizable physical
system can be perfectly simulated by a universal model computing
machine operating by finite means". Our above analysis on the role
of commutativity in computation theory based on quantum logic
perhaps indicates that in order to be simulated by a universal
computing machine some observables of the physical system are
required to possess a certain commutativity. So, it is fair to say
that the observation on commutativity presented above provides a
complement to D. Deutsch's argument from a logical point of view.

\bigskip\

\textbf{2. Quantum Logic}

\smallskip\

The aim of this section is to recall some basic notions and
results about quantum logic needed in the subsequent sections and
to fix notations. In this paper, quantum logic is understood as a
complete orthomodular lattice-valued logic. This section is mainly
concerned with the semantic aspect of such a logic, and it will be
divided into four subsections. The first subsection will briefly
review some fundamental results on orthomodular lattices; for more
details, we refer to [Ka83] and [BH00]. In the second one we will
introduce the language of first-order quantum logic. The third
will discuss the algebraic semantics of first-order quantum logic.
Some useful properties of orthomodular lattice-valued sets are
given in the fourth subsection.

\smallskip\

\textbf{2.1. Orthomodular Lattices}

\smallskip\

The set of truth values of a quantum logic will be taken to be an
orthomodular lattice. So we first introduce the notion of
orthomodular lattice. An ortholattice is a 7-tuple
$$\ell =<L,\leq ,\wedge ,\vee ,\bot ,0,1>$$ where:

(1) $<L,\leq ,\wedge ,\vee ,0,1>$ is a bounded lattice, $0,1$ are
the least and greatest elements of $L,$ respectively, $\leq $ is
the partial ordering in $L,$ and for any $a,b\in L,$ $a\wedge b,$
and $a\vee b$ stand for the greatest lower bound and the least
upper bound of $a$ and $b$, respectively;

(2) $\bot $ is a unary operation on $L,$ called orthocomplement,
and required to satisfy the following conditions: for any $a,b\in
L$,

\ \ \ (i) $a\wedge a^{\bot }=0,\ a\vee a^{\bot }=1;$

\ \ \ (ii) $a^{\bot \bot }=a;$ and

\ \ \ (iii) $a\leq b$ implies $b^{\bot }\leq a^{\bot }.$

It is easy to see that the condition (iii) is equivalent to one of
the De Morgan laws: for any $a,b\in L$,

\ \ \ (iii') $(a\wedge b)^{\bot} = a^{\bot}\vee b^{\bot},\ (a\vee
b)^{\bot} = a^{\bot}\wedge b^{\bot}.$

Let $\ell =<L,\leq ,\wedge ,\vee ,\bot ,0,1>$ be an ortholattice,
and let $a,b\in L$. We say that $a$ commutes with $b$, in symbols
$aCb$, if $$a=(a\wedge b)\vee (a\wedge b^{\bot}).$$

An orthomodular lattice is an ortholattice $\ell =<L,\leq ,\wedge
,\vee ,\bot ,0,1>$ satisfying the orthomodular law: for all
$a,b\in L$,

\ \ \ (iv) $a\leq b\ {\rm implies}\ a\vee (a^{\bot}\wedge b)=b.$

The orthomodular law can be replaced by the following equation:

\ \ \ (iv') $a\vee (a^{\bot }\wedge (a\vee b))=a\vee b\ {\rm for\
any}\ a,b\in L.$

A Boolean algebra is an ortholattice $\ell =<L,\leq ,\wedge ,\vee
,\bot ,0,1>$ fulfilling the distributive law of join over meet:
for all $a,b,c\in L$,

\ \ \ (v) $a\vee (b\wedge c)=(a\vee b)\wedge (a\vee c).$

With the De Morgan law it is easy to know that the condition (v)
is equivalent to the distributive law of meet over join: for any
$a,b,c\in L$,

\ \ \ (v') $a\wedge (b\vee c)=(a\wedge b)\vee (a\wedge c).$

Obviously, the distributive law implies the orthomodular law, and
so a Boolean algebra is an orthomodular lattice.

The following lemma gives a characterization of orthomodular
lattices and it distinguishes orthomodular lattices from
ortholattices.

\smallskip\

\textbf{Lemma 2.1.} ([BH00], Propositions 2.1 and 2.2) Let $\ell
=<L,\leq ,\wedge ,\vee ,\bot ,0,1>$ be an ortholattice. Then the
following seven statements are equivalent:

(1) $\ell$ is an orthomodular lattice;

(2) For any $a,b\in L$, if $a\leq b$ and $a^{\bot}\wedge b=0$ then
$a=b$;

(3) For any $a,b\in L$, if $aCb$ then $bCa$;

(4) For any $a,b\in L$, if $aCb$ then $a^{\bot}Cb$;

(5) For any $a,b\in L$, if $aCb$ then $a\vee (a^{\bot}\wedge
b)=a\vee b$;

(6) The benzene ring $O_6$ (see Figure 1) is not a subalgebra of
$\ell$.
\begin{figure}\centering
\includegraphics{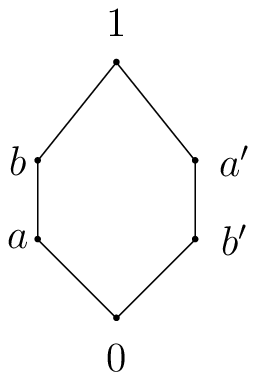}
\caption{Benzene ring} \label{fig 1 }
\end{figure}

(7) For any $a,b\in L$, if $a\leq b$ then the subalgebra $[a,b]$
of $\ell$ generated by $a$ and $b$ is a Boolean
algebra.$\heartsuit$

\smallskip\

The set of truth values of classical logic is a Boolean algebra;
whereas quantum logic is an orthomodular lattice-valued logic. It
is well-known that a Boolean algebra must be an orthomodular
lattice, but the inverse is not true. Thus, quantum logic is
weaker than classical logic. The major difference between a
Boolean algebra and an orthomodular lattice is that distributivity
is not valid in the latter. However, many cases still appeal an
application of the distributivity even when we manipulate elements
in an orthomodular lattice. This requires us to regain a certain
(weaker) version of distributivity in the realm of orthomodular
lattices. The key technique for this purpose is commutativity
which is able to provide a localization of distributivity. The
following lemma together with Lemma 2.1(4) indicates that
commutativity is preserved by all operations of orthomodular
lattice.

\smallskip\

\textbf{Lemma 2.2}. ([BH00], Proposition 2.4) Let $\ell =<L,\leq
,\wedge ,\vee ,\bot ,0,1>$ be an orthomodular lattice, and let
$a\in L$ and $b_i\in L$ $(i\in I)$. If $aCb_i$ for any $i\in I$,
then $$aC\wedge_{i\in I}b_i\ {\rm and}\ aC\vee_{i\in I}b_i$$
provided $\wedge_{i\in I}b_i$ and $\vee_{i\in I}b_i$
exist.$\heartsuit$

\smallskip\

The local distributivity implied by commutativity is then given by
the following

\smallskip\

\textbf{Lemma 2.3}. ([BH00], Proposition 2.3) Let $\ell =<L,\leq
,\wedge ,\vee ,\bot ,0,1>$ be an orthomodular lattice and let
$A\subseteq L$. For any $a\in A$ and $b_i\in A$ $(i\in I)$, if
$aCb_i$ for all $i\in I$, then
$$a\wedge \vee_{i\in I}b_i=\vee_{i\in I}(a\wedge b_i),$$
$$a\vee \wedge_{i\in I}b_i=\wedge_{i\in I}(a\vee b_i)$$
provided $\wedge_{i\in I} b_i$ and $\vee_{i\in I} b_i$
exist.$\heartsuit$

\smallskip\

The above lemma is very useful, and it often enables us to recover
distributivity in an orthomodular lattice. However, its condition
that all elements involved commute each other is quite strong, and
not easy to meet. This suggests us to find a way to weaken this
condition. One solution was found by G. Takeuti [T81], and he
introduced the notion of commutator which can be seen as an index
measuring the degree to which the commutativity is valid.

\smallskip\

\textbf{Definition 2.1}. ([T81], pages 305 and 307) Let $\ell
=<L,\leq ,\wedge ,\vee ,\bot ,0,1>$ be an orthomodular lattice and
let $A\subseteq L$.

(1) If $A$ is finite, then the commutator $\gamma (A)$ of $A$ is
defined by
$$\gamma (A)=\vee\{\wedge_{a\in A}a^{f(a)}:f\ {\rm is\ a\ mapping\ from}\ A\ {\rm into}\
\{1,-1\}\},$$ where $a^{1}$ denotes $a$ itself and $a^{-1}$
denotes $a^{\bot}$.

(2) The strong commutator $\Gamma (A)$ of $A$ is defined by
$$\Gamma (A)=\vee\{b: C(a,b)\ {\rm for\ all}\ a\in A\ {\rm and}\
C(a_1\wedge b, a_2\wedge b)\ {\rm for\ all}\ a_1, a_2\in A\}.$$

\smallskip\

The relation  between commutator and strong commutator is
clarified by the following lemma. In addition, the third item of
the following lemma shows that commutator is a relativization of
the notion of commutativity.

\smallskip\

\textbf{Lemma 2.4}. ([T81], Proposition 4 and its corollary) Let
$\ell =<L,\leq ,\wedge ,\vee ,\bot ,0,1>$ be an orthomodular
lattice and let $A\subseteq L$. Then

(1) $\Gamma(A)\leq \gamma(A)$.

(2) If $A$ is finite, then $\Gamma(A)= \gamma(A)$.

(3) $\gamma(A)=1$ if and only if all the members of $A$ are
mutually commutable.$\heartsuit$

\smallskip\

We now can present a generalization of Lemma 2.3 by using the tool
of commutator. It is easy to see from Lemmas 2.4(2) and (3) that
the following lemma degenerates to Lemma 2.3 when $aCb_i$ for all
$i\in I$.

\smallskip\

\textbf{Lemma 2.5}. ([T81], Propositions 5 and 6) Let $\ell
=<L,\leq ,\wedge ,\vee ,\bot ,0,1>$ be an orthomodular lattice and
let $A\subseteq L$. Then for any $a\in A$ and $b_i\in A$ $(i\in
I)$,
$$\Gamma(A)\wedge (a\wedge \vee_{i\in I}b_i)\leq \vee_{i\in I}(a\wedge b_i),$$
$$\Gamma(A)\wedge \wedge_{i\in I}(a\vee b_i)\leq a\vee \wedge_{i\in I}b_i.\heartsuit$$

\smallskip\

Suppose that we want to use the above lemma on a formula of the
form $a\wedge \vee_{i\in I}b_i$ or $a\vee \wedge_{i\in I}b_i$ in
order to get a local distributivity. In many situations, the
elements $a$ and $b_i$ $(i\in I)$ may be very complicated, and the
operations $\bot$, $\wedge$ and $\vee$ are involved in them. Then
the above lemma cannot be applied directly, and it needs the help
of the following

\smallskip\

\textbf{Lemma 2.6}. Let $\ell =<L,\leq ,\wedge ,\vee ,\bot ,0,1>$
be an orthomodular lattice and let $A\subseteq L$. Then for any
$B\subseteq [A]$ we have $$\Gamma (A)\leq \Gamma (B),$$ where
$[A]$ stands for the subalgebra of $\ell$ generated by $A$.

\smallskip\

\textbf{Proof}. For any $X\subseteq L$, we write
$$K(X)=\{b\in L:aCb\ {\rm and}\ (a_1\wedge b)C(a_2\wedge b)\
{\rm for\ all}\ a, a_1, a_2 \in X \}$$ Furthermore, we set $A_0=A$
and
$$A_{i+1} = A_i \cup \{a^{\bot}: a\in A_i\}\cup\{a_1\wedge a_2:
a_1, a_2\in A_i\}\ (i=0,1,2,...)$$

First, we prove that $K(A_i)=K(A)$ for all $i\geq 0$ by induction
on $i$. It is obvious that $K(A_{i+1})\subseteq K(A)$. Conversely,
suppose that $b\in K(A)$ and we want to show that $b\in
K(A_{i+1})$. It is easy to see that $aCb$ for any $a\in A_{i+1}$.
Thus, we only need to demonstrate the following

\textit{Claim:} $(a_1\wedge b)C(a_2\wedge b)$ for any $a_1, a_2\in
A_{i+1}$.

The essential part of the proof of the above claim is the
following two cases, and the other cases are clear, or can be
treated as iterations of them:

Case 1. $a_1\in A_i$, $a_2=c_1\wedge c_2$ and $c_1, c_2\in A_i$.
From the induction hypothesis we have $$(a_1\wedge b)C(c_1\wedge
b)\ {\rm and}\ (a_1\wedge b)C(c_2\wedge b).$$ This yields
$$(a_1\wedge b)C(c_1\wedge b)\wedge (c_2\wedge b)=(c_1\wedge
c_2)\wedge b=a_2\wedge b.$$

Case 2. $a_1\in A_i$, $a_2=c^{\bot}$ and $c\in A_i$. Then from the
induction hypothesis we obtain $(a_1\wedge b)C(c\wedge b)$, and
further $(a_1\wedge b)C(c\wedge b)^{\bot}$ by using Lemma 2.1(4).
In addition, $(a_1\wedge b)Cb$. This together with Lemma 2.2
yields $(a_1\wedge b)Cb\wedge(c\wedge b)^{\bot}$. Note that $cCb$
and so $b^{\bot}Cc^{\bot}$. Then by Lemma 2.3 we assert that
$$b^{\bot}\vee(c\wedge b)=b^{\bot}\vee c\ {\rm and}\ b\wedge (c\wedge
b)^{\bot}=b\wedge c^{\bot}$$. Hence, it follows that $(a_1\wedge
b)Cb\wedge c^{\bot}=a_2\wedge b.$

We now write $$A_\infty = \cup_{i=0}^{\infty}A_i.$$ Then
$$K(A_\infty)=\cap_{i=0}^{\infty}K(A_i)=K(A).$$ It is easy to see
that $A\subseteq A_\infty$ is a subalgebra of $\ell$. So,
$[A]\subseteq A_\infty$, $$K(A)=K(A_\infty)\subseteq
K([A])\subseteq K(B),$$ and $$\Gamma (A)=\vee K(A)\leq \vee
K(B)=\Gamma (B).\heartsuit$$

\smallskip\

As stated in the introduction, the aim of this paper is to develop
a theory of computation based on quantum logic. The logical
language for a theory of computation has to contain the universal
and existential quantifiers, and the two quantifiers are usually
interpreted as (infinite) meet and join, respectively. Hence, we
should assume that the lattice of the truth values of our quantum
logic is complete. A complete orthomodular lattice is an
orthomodular lattice $\ell =<L,\leq ,\wedge ,\vee ,\bot ,0,1>$ in
which for any $M\subseteq L,$ both the greatest lower bound
$\wedge M$ and the least upper bound $\vee M$ exist.

The function of a logic is provide us with a certain reasoning
ability, and the implication connective is an intrinsic
representative of inference within the logic. Thus each logic
should reasonably contain a connective of implication. To make a
complete orthomodular lattice available as the set of truth values
of quantum logic, we need to define a binary operation, called
implication operator, on it such that this operation may serve as
the interpretation of implication in this logic. Unfortunately, it
is a very vexed problem to define a reasonable implication
operator for quantum logic. All implication operators that one can
reasonably introduce in an orthomodular lattice are more or less
anomalous in the sense that they do not share most of the
fundamental properties of the implication in classical logic. This
is different from the cases of most weak logics. (For a thorough
discussion on the implication problem in quantum logic, see
[DC86], Section 3.)

A minimal condition for an implication operator $\rightarrow$ is
the requirement proposed by G. Birkhoff and J. von Neumann [BN36]:
$$a\rightarrow b=1\ {\rm if\ and\ only\ if}\ a\leq b$$ for any $a,b\in L$.
Usually in a logic, there are two ways in which implication is
introduced. The first one is to treat implication as a derived
connective; that is, implication is explicitly defined in terms of
other connectives such as negation, conjunction and disjunction.
All implications of this kind were found by G. Kalmbach [Ka74],
and they are presented by the following:

\smallskip\

\textbf{Lemma 2.7}. ([Ka74]; see also [Ka83], Theorem 15.3) The
orthomodular lattice freely generated by two elements is
isomorphic to $2^{4}\times MO2$, where $2$ stands for the Boolean
algebra of two elements. The elements of $2^{4}\times MO2$
satisfying the Birkhoff-von Neumann requirement are exactly the
following five polynomials of two variables:
$$a\rightarrow_1 b=(a^{\bot}\wedge b)\vee (a^{\bot}\wedge b^{\bot})\vee (a\wedge (a^{\bot}\vee b)),$$
$$a\rightarrow_2 b=(a^{\bot}\wedge b)\vee (a\wedge b)\vee ((a^{\bot}\vee b)\wedge b^{\bot}),$$
$$a\rightarrow_3 b=a^{\bot}\vee (a\wedge b),$$
$$a\rightarrow_4 b=b\vee (a^{\bot}\wedge b^{\bot}),$$
$$a\rightarrow_5 b=(a^{\bot}\wedge b)\vee (a\wedge b)\vee (a^{\bot}\wedge b^{\bot}).\heartsuit$$

\smallskip\

Obviously, this lemma implies that the above five polynomials are
all implication operators definable in orthomodular lattices. It
was shown by G. Kalmbach [Ka74, Ka83] that the orthomodular
lattice-valued (propositional) logic can be (finitely)
axiomatizable by using the modus ponens with
implication$\rightarrow_1$ as the only one rule of inference, but
the same conclusion does not hold for the other implications
$\rightarrow_i$ $(2\leq i\leq 5).$

We may also define the material conditional $\rightarrow_0$ in an
orthomodular $\ell=<L,\leq ,\wedge ,\vee ,\bot ,0,1>$ by
$$a\rightarrow _0 b=a^{\bot}\vee b$$
for all $a,b\in L.$ It is easy to see that $\rightarrow_0$ does
not fulfil the Birkhoff-von Neumann requirement. On the other
hand, the following lemma shows that the five implication
operators given in Lemma 2.7 degenerate to the material
conditional whenever the two operands are compatible.

\smallskip\

\textbf{Lemma 2.8}. ([DC86], Theorem 3.2) Let $\ell =<L,\leq
,\wedge ,\vee ,\bot ,0,1>$ be an orthomodular lattice. Then for
any $a,b\in L$, $$a\rightarrow_i b=a\rightarrow_0 b$$ if and only
if $aCb$, where $1\leq i\leq 5.\heartsuit$

\smallskip\

The second way of defining an implication is to take its truth
function as the adjunctor (i.e., residuation) of the truth
function of conjunction. Note that in this case the implication is
usually not definable from negation, conjunction and disjunction,
and it has been treated as a primitive connective. Indeed, L.
Herman, E. Marsden and R. Piziak [HMP75] introduced an implication
in the style of residuation. Furthermore, the following lemma
shows that the five polynomial implication operators
$\rightarrow_i$ $(1\leq i\leq 5)$ cannot be defined as the
residuation of the conjunction unless $\ell$ is a Boolean algebra.

\smallskip\

\textbf{Lemma 2.9}. ([DC86], the revised version, page 25) Let
$\ell =<L,\leq ,\wedge ,\vee ,\bot ,0,1>$ be an orthomodular
lattice, and let $1\leq i\leq 5$. Then the following two
statements are equivalent:

(i) $\ell$ is a Boolean algebra.

(ii) the import-export law: for all $a,b\in L$,
$$a\wedge b\leq c\ {\rm if\ and\ only\ if}\ a\leq b\rightarrow_i c.\heartsuit$$

\smallskip\

Among the five orthomodular polynomial implications,
$\rightarrow_3$, named the Sasaki-hook, has often been preferred
since it enjoys some properties resembling those in intuitionistic
logic. The Sasaki-hook was originally introduced by P. D. Finch
[Fi70]. For a detailed discussion of the Sasaki-hook, see L.
Rom\'{a}n and B. Rumbos [RR91] and L. Rom\'{a}n and R. E. Zuazua
[RZ99]. Here we first point out that the Sasaki-hook possesses a
modification of residual characterization although it is defined
as a polynomial in orthomodular lattice. A weakening of the
import-export law is the resulting condition, called compatible
import-export law, by restricting the import-export law for any
$a,b\in L$ with $aCb$; that is, if $aCb$, then $a\wedge b\leq c$
if and only if $a\leq b\rightarrow c$.

\smallskip\

\textbf{Lemma 2.10}. ([T81], Proposition 1 and its corollary;
[DC86], the revised version, page 25) Let $\ell =<L,\leq ,\wedge
,\vee ,\bot ,0,1>$ be an orthomodular lattice, and let $a,$ $b$,
$c\in L$. Then
$$a\rightarrow b=\vee \{x:xCa\ {\rm and}\ x\wedge a\leq b\}.$$
Moreover, among the five implications $\rightarrow_i$ $(1\leq
i\leq 5)$, the Sasaki-hook $\rightarrow_3$ is the only one
satisfying the compatible import-export law.$\heartsuit$

\smallskip\

Our mathematical reasoning frequently require that implication
relation is preserved by conjunction and disjunction. Also, the
negation is needed to be compatible with implication in the sense
that the negation can reverse the direction of implication. And,
to warrant the validity of a chain of inferences, the transitivity
of implication is required. However, this is not the case in
general if we are working in an orthomodular lattice. Fortunately,
if we adopt the Sasaki-hook, then these properties of implication
can be recovered by attaching a certain commutator.

\smallskip\

\textbf{Lemma 2.11.} Let $\ell =<L,\leq ,\wedge ,\vee ,\perp
,0,1>$ be an orthomodular lattice. Then

(1) for any $a_{i},b_{i}\in L$ $(i=1,...,n),$ let
$X=\{a_{1},...,a_{n}\}\cup \{b_{1},...,b_{n}\},$
$$\Gamma
(X)\wedge \wedge _{i=1}^{n}(a_{i}\rightarrow _{3}b_{i})\leq \wedge
_{i=1}^{n}a_{i}\rightarrow _{3}\wedge _{i=1}^{n}b_{i},$$
$$\Gamma (X)\wedge \wedge _{i=1}^{n}(a_{i}\rightarrow
_{3}b_{i})\leq \vee _{i=1}^{n}a_{i}\rightarrow _{3}\vee
_{i=1}^{n}b_{i}.$$

(2) for any $a,b\in L$, $$\Gamma (a,b)\wedge (a\rightarrow
_{3}b)\leq b^{\perp }\rightarrow _{3}a^{\perp }.$$

(3) for any $a,b,c\in L,$
$$\Gamma (a,b,c)\wedge (a\rightarrow _{3}b)\wedge (b\rightarrow
_{3}c)\leq a\rightarrow _{3}c.$$

\smallskip\

\textbf{Proof.} (1) We only prove the first inequality, and the
proof of the second is similar. With Lemmas 2.5 and 2.6 we obtain
$$\wedge
_{i=1}^{n}a_{i}\rightarrow _{3}\wedge _{i=1}^{n}b_{i}=(\wedge
_{i=1}^{n}a_{i})^{\perp }\vee (\wedge _{i=1}^{n}a_{i}\wedge \wedge
_{i=1}^{n}b_{i})$$
$$=\vee _{i=1}^{n}a_{i}{}^{\perp }\vee \wedge
_{i=1}^{n}(a_{i}\wedge b_{i})$$
$$\geq \Gamma (X)\wedge \wedge
_{i=1}^{n}(\vee _{j=1}^{n}a_{j}{}^{\perp }\vee (a_{i}\wedge
b_{i}))$$
$$\geq \Gamma (X)\wedge \wedge _{i=1}^{n}(a_{i}{}^{\perp }\vee
(a_{i}\wedge b_{i}))$$
$$=\Gamma (X)\wedge \wedge
_{i=1}^{n}(a_{i}\rightarrow b_{i}).$$

(2) First, we note that $a\wedge b,$ $a^{\perp }\wedge b,$ $
a^{\perp }\wedge b^{\perp }\leq b\vee (a^{\perp }\wedge b^{\perp
})=b^{\perp }\rightarrow _{3}a^{\perp }.$ Thus,
$$\Gamma (a,b)=(a\wedge b)\vee (a\wedge b^{\perp })\vee (
a^{\perp }\wedge b)\vee (a^{\perp }\wedge b^{\perp })$$
$$\leq (b^{\perp }\rightarrow _{3}a^{\perp })\vee (a\wedge b^{\perp
}),$$ and furthermore with Lemmas 2.5 and 2.6 we have
$$\Gamma (a,b)\wedge (a\rightarrow _{3}b)=\Gamma (a,b)\wedge
(a^{\perp }\vee (a\wedge b))$$
$$\leq \Gamma (a,b)\wedge (a^{\perp }\vee b)$$
$$=\Gamma (a,b)\wedge \Gamma (a,b)\wedge (a^{\perp }\vee b)$$
$$\leq \Gamma (a,b)\wedge \lbrack (b^{\perp }\rightarrow
_{3}a^{\perp })\vee (a\wedge b^{\perp })]\wedge (a^{\perp }\vee
b)$$
$$\leq \ [(b^{\perp }\rightarrow _{3}a^{\perp })\wedge (a^{\perp
}\vee b)]\vee \lbrack (a\wedge b^{\perp })\wedge (a^{\perp }\vee
b)]$$
$$\leq (b^{\perp }\rightarrow _{3}a^{\perp })\vee \lbrack (a\wedge
b^{\perp })\wedge (a^{\perp }\vee b)].$$ Note that $(a\wedge
b^{\perp })^{\perp }=a^{\perp }\vee b$ and $(a\wedge b^{\perp
})\wedge (a^{\perp }\vee b)=0.$ Then
$$\Gamma (a,b)\wedge (a\rightarrow _{3}b)\leq b^{\perp }\rightarrow
_{3}a^{\perp }.$$

(3) Again, we use Lemmas 2.5 and 2.6. This enables us to assert
that
$$\Gamma
(a,b,c)\wedge (a\rightarrow _{3}b)\wedge (b\rightarrow
_{3}c)=\Gamma (a,b,c)\wedge (a^{\perp }\vee (a\wedge b))\wedge
(b^{\perp }\vee (b\wedge c)) $$
$$\leq \Gamma (a,b,c)\wedge
([a^{\perp }\wedge (b^{\perp }\vee (b\wedge c))]\vee \lbrack
(a\wedge b)\wedge (b^{\perp }\vee (b\wedge c))])$$
$$\leq \Gamma
(a,b,c)\wedge (a^{\perp }\vee \lbrack (a\wedge b)\wedge (b^{\perp
}\vee (b\wedge c))]).$$

We note that $\Gamma (a,b,c)Ca^{\perp }$ and $$\Gamma
(a,b,c)C[(a\wedge b)\wedge (b^{\perp }\vee (b\wedge c))]).$$ Then
$$\Gamma (a,b,c)\wedge (a\rightarrow _{3}b)\wedge (b\rightarrow
_{3}c)\leq (\Gamma (a,b,c)\wedge a^{\perp })\vee (\Gamma
(a,b,c)\wedge \lbrack (a\wedge b)\wedge (b^{\perp }\vee (b\wedge
c))])$$
$$\leq a^{\perp }\vee (\Gamma (a,b,c)\wedge \lbrack (a\wedge
b)\wedge (b^{\perp }\vee (b\wedge c))])$$
$$\leq a^{\perp }\vee
\lbrack (a\wedge b)\wedge b^{\perp }]\vee \lbrack (a\wedge
b)\wedge (b\wedge c)]$$
$$=a^{\perp }\vee \lbrack (a\wedge b)\wedge
(b\wedge c)]$$
$$\leq a^{\perp }\vee (a\wedge c)$$
$$=a\rightarrow _{3}c.\heartsuit$$

\smallskip\

For simplicity of presentation, we finally introduce an
abbreviation. For each implication operator $\rightarrow$, the
bi-implication operator on $\ell $ is defined as follows:
$$a\leftrightarrow b\stackrel{def}{=}(a\rightarrow b)\wedge
(b\rightarrow a)$$ for any $a,b\in L.$

\smallskip\

\textbf{2.2. The Language of Quantum Logic}

\smallskip\

In this subsection we present the syntax of quantum logic. Given a
complete orthomodular lattice $\ell =<L,\leq ,\wedge ,\vee ,\bot
,0,1>.$ We require that the language of an $\ell -$valued
(quantum) logic possesses a nullary connective $ \mathbf{a}$ for
each $a\in L$ as well as three other primitive connectives: an
unary one $\lnot $ (negation) and two binary ones $\wedge $
(conjunction), $\rightarrow$ (implication). The language also has
a primitive quantifier $\forall $ (universal quantifier).

It deserves an explanation for our design decision of choosing
implication as a primitive connective. In the sequel, many results
only need to suppose that the implication operator satisfies the
Birkhoff-von Neumann requirement. It is known that there are five
polynomials fulfilling the Birkhoff-von Neumann requirement. If we
treated implication as a derived connective defined in terms of
negation, conjunction and disjunction, then it would be necessary
to assume five different connectives of implication in our logical
language. This would often complicate our presentation very much.
On the other hand, in some cases, the Birkhoff-von Neumann
condition is not enough and it requires the implication operator
to be the Sasaki-hook. So, we decide to use implication as a
primitive connective, and specify it when needed.

The syntax of $\ell-$valued logic is defined in a familiar way; we
omit its details. To simplify the notations in what follows, it is
necessary to introduce several derived formulas:

(i) $\varphi \vee \psi \stackrel{def}{=}\lnot (\lnot \varphi
\wedge \lnot \psi );$

(ii) $\varphi \leftrightarrow \psi \stackrel{def}{=}(\varphi
\rightarrow \psi )\wedge (\psi \rightarrow \varphi );$

(iii) $(\exists x)\varphi \stackrel{def}{=}\lnot (\forall x)\lnot
\varphi ;$

(iv) $A\subseteq B\stackrel{def}{=}(\forall x)(x\in A\rightarrow
x\in B);$ and

(v) $A\equiv B\stackrel{def}{=}(A\subseteq B)\wedge (B\subseteq
A).$

Suppose that $\Delta$ is a finite set of formulas. The commutator
of $\Delta$ is defined to be
$$\gamma(\Delta)\stackrel{def}{=}\vee\{\wedge_{\varphi\in\Delta}
\varphi^{f(\varphi)}:f\in \{1,-1\}^{\Delta}\},$$ where
$\varphi^{1},\ \varphi^{-1}$ express $\varphi$ and $\lnot\varphi,$
respectively. It is obvious that the above formula is the
counterpart of Definition 2.1(1) in the language of our quantum
logic.

\smallskip\

\textbf{2.3. The Algebraic Semantics of Quantum Logic}

\smallskip\

We now turn to give the semantics of quantum logic. There are
several different versions of semantics for quantum logic; for
example, quantum logic enjoys a semantics in the Kripke style
[DC86; RS00]. What concerns us here is its algebraic semantics.
Assume that $\ell =<L,\leq ,\wedge ,\vee ,\bot ,0,1>$ be an
orthomodular lattice equipped with additionally a binary operation
$\rightarrow$ over it. The operation $\rightarrow$ is required to
be suited to serve as the truth function of implication
connective. According to our explanation of the connective of
implication in the last subsection, we leave the operation
$\rightarrow$ unspecified but suppose that it satisfies the
Birkhoff-von Neumann requirement. An $\ell -$valued interpretation
is an interpretation in which every predicate symbol is associated
with an $ \ell -$valued relation, i.e., a mapping from the product
of some copies of the discourse universe into $L,$ where the
number of copies is exactly the arity of the predicate symbol. The
other items in $\ell-$valued logical language are interpreted as
usual. For every (well-formed) formula $\varphi ,$ its truth value
$\lceil \varphi \rceil$ is assumed in $L,$ and the truth valuation
rules for logical and set-theoretical formulas are given as
follows:

(i) $\lceil \mathbf{a}\rceil =a;$

(ii) $\lceil \lnot \varphi \rceil =\lceil \varphi \rceil ^{\bot
};$

(iii) $\lceil \varphi \wedge \psi \rceil =\lceil \varphi \rceil
\wedge \lceil \psi \rceil ;$

(iv) $\lceil \varphi \rightarrow \psi \rceil =\lceil \varphi
\rceil \rightarrow \lceil \psi \rceil ;$

(v) if $U$ is the universe of discourse, then
$$\lceil (\forall x)\varphi (x)\rceil =\wedge _{u\in U}\lceil
\varphi (u)\rceil ;$$ and

(vi) $\lceil x\in A\rceil =A(x),$ where $A$ is a set constant
(unary predicate symbol) and it is interpreted as a mapping, also
denoted as $A,$ from the universe into $L,$ i.e., an $ \ell
-$valued set (more exactly, an $\ell -$valued subset of the
universe).

Note that in the above truth valuation rules $\wedge $ and $\vee $
in the left-hand side are connectives in quantum logic whereas
$\wedge $ and $\vee $ in the right-hand side stand for operations
in the orthomodular lattice $ \ell $ of truth values. Also, the
symbol $\rightarrow$ in the left-hand side of (iv) is a connective
in the language of quantum logic, but the symbol $\rightarrow$ in
the right-hand side of (iv) is the binary operation attached to
$\ell$ that is explained at the beginning of this subsection.

As we claimed in the introduction, quantum logic will act as our
meta-logic in the theory of computation developed in this paper.
Then we still have to introduce several meta-logical notions for
quantum logic. For every orthomodular lattice $\ell =<L,\leq
,\wedge ,\vee ,\bot ,0,1>,$ if $\Gamma $ is a set of formulas and
$\varphi $ a formula, then $\varphi $ is a semantic consequence of
$\Gamma $ in $\ell -$valued logic, written $\Gamma \stackrel{\ell
}{\models }\varphi ,$ whenever $$\wedge _{\psi \in \Gamma }\lceil
\psi \rceil \leq \lceil \varphi \rceil $$ for all $\ell -$valued
interpretations. In particular, $\stackrel{\ell }{\models }\varphi
$ means that $\phi \stackrel{\ell }{\models }\varphi ,$ i.e.,
$\lceil \varphi \rceil =1$ always holds for every $\ell -$valued
interpretation; in other words, $1$ is the unique designated truth
value in $\ell .$ Furthermore, if $\Gamma \stackrel{ \ell
}{\models }\varphi $ (resp. $\stackrel{\ell }{\models }\varphi $)
for all orthomodular lattice $\ell $ then we say that $\varphi $
is a semantic consequence of $\Gamma $ (resp. $\varphi $ is valid)
in quantum logic and write $\Gamma \models \varphi $ (resp.
$\models \varphi $).

We here are not going to give a detailed exposition on quantum
logic, but would like to point out that quantum logic gives rise
to many counterexamples to some meta-logical properties which hold
for classical logic and for a large class of weaker logics; for
example, M. L. Dalla Chiara [DC81] showed that a minimal version
of quantum logic fails to enjoy the Lindenbaum property, and J.
Malinowski [Ma90] found that the deduction theorem fails in
quantum logic and some of its variants.

\smallskip\

\textbf{2.4. The Operations of Quantum Sets}

\smallskip\

Beside the language of quantum logic introduced in Section 2.2, we
will also need some notations such as $\in $ (membership) from
set-theoretical language in our study of computing theory based on
quantum logic. As mentioned in the introduction, a theory of
quantum sets has already been developed by G. Takeuti [T81]. A
careful review of quantum set theory is out of the scope of the
present paper. What mainly concerned G. Takeuti [T81] is how some
axioms of classical set theory could be modified so that they will
holds in the framework of quantum logic. In other words, he tried
to clarify the relation of quantum set theory with the classical
mathematics. Here, we instead propose some operations of
$\ell-$valued sets and also introduce several notations for
$\ell-$valued sets. These are needed in the subsequent sections.
We write $L^{X}$ for the set of all $\ell-$valued subsets of $X$,
i.e., all mappings from $X$ into $L$. For any non-empty set $X,$
if $x\in X$ and $\lambda \in L-\{0\},$ then $x_\lambda $ is
defined to be a mapping from $X$ into $L$ such that
$$x_\lambda (x^{\prime })=\left\{
\begin{array}{c}
\lambda \ {\rm if }\ x^{\prime }=x, \\
0\ {\rm otherwise,}
\end{array}
\right.$$ and it is often called an $\ell-$valued point in $X.$ We
write $p_\ell(X)$ for the set of all $\ell-$valued points in $X;$
that is, $$p_\ell(X)=\{x_\lambda :x\in X\ {\rm and}\ \lambda \in
L-\{0\}\}.$$ For each $e=x_\lambda \in p_\ell(X),$ $x$ is called
the support of $e$ and denoted $s(e),$ and $\lambda $ is called
the height of $e$ and written $h(e).$ In particular, an
$\ell-$valued point of height $1$ is always identified with its
support. The predicate $\in$ can be extended to a predicate
between $\ell-$valued points and $\ell-$valued sets in a natural
way: $$x_{\lambda}\in A\stackrel{def}{=}x_{\lambda}\subseteq A.$$
Then it is easy to see that $$\lceil x_{\lambda}\in A\rceil =
\lambda\rightarrow A(x)$$ for any $x\in X$, $\lambda\in L$ and
$A\in L^{X}$, where $\rightarrow$ is the implication operator
under consideration. For any $A\subseteq X$, its characteristic
function is a mapping from $X$ into the Boolean algebra
$\mathbf{2}=\{0,1\}$ of two elements, and so it can also be seen
as a mapping from $X$ into $L$, namely, an $\ell-$valued subset of
$X$. We will identify $A$ with its characteristic function. For
any $a\in L$ and $A,B\in L^{X}$, we define all of the scalar
product $aA,$ complement $A^{c}$, intersection $A\cap B$ and union
$A\cup B$ to be $\ell-$valued subsets of $X$ and for all $x\in X$,

\smallskip\

(i) $x\in aA\stackrel{def}{=}\mathbf{a}\wedge (x\in A);$

(ii) $x\in A^{c}\stackrel{def}{=}\lnot (x \in A);$

(iii) $x\in A\cap B\stackrel{def}{=}(x\in A)\wedge (x\in B);$

(iv) $x\in A\cup B\stackrel{def}{=}(x\in A)\vee (x\in B).$

\smallskip\

From the truth valuation rules and the definition of derived
formulas in the $\ell-$valued logical and set-theoretical
language, we know that for all $x\in X$,

\smallskip\

(i') $(aA)(x)=a \wedge A(x);$

(ii') $(A^{c})(x)=A(x)^{\bot};$

(iii') $(A\cap B)(s)=A(s)\wedge B(s);$ and

(iv') $(A\cup B)(s)=A(s)\vee B(s).$

\smallskip\

It is easy to see that in the domain of $\ell-$valued sets the
intersection and union operations are idempotent, commutative and
associative, and they have $X$ and $\phi$, respectively as their
unit elements. The intersection and union together with the
complement satisfy the De Morgan law, but the distributivity of
intersection over union or union over intersection is no longer
valid. Clearly, the laws for operations of $\ell-$valued sets are
essentially determined by the algebraic properties of the lattice
$\ell$ of truth values.

Assume that $X$ and $Y$ are two non-empty sets, and
$h:X\longrightarrow Y$ is a mapping. For any $A\in L^{X}$, its
image $h(A)$ under $h$ is defined by
$$y\in h(A)\stackrel{def}{=}(\exists x\in X)(y=f(x)\wedge x\in
A,$$ and for any $B\in L^{Y}$, its pre-image $h^{-1}(B)$ under $h$
is defined by $$x\in h^{-1}(B)\stackrel{def}{=}h(x)\in B.$$ The
defining equations of $h(A)$ and $h^{-1}(B)$ may be rewritten,
respectively, as follows: $$h(A)(y)=\vee \{A(X):x\in X\ {\rm and}\
f(x)=y\},\ {\rm and}$$ $$h^{-1}(B)(x)=B(h(x)).$$

\smallskip\

\textbf{Lemma 2.12.} Let $\ell =<L,\leq ,\wedge ,\vee ,\perp
,0,1>$ be an orthomodular lattice, let $\rightarrow $ enjoy the
Birkhoff-von Neumann requirement, and let $h:X\rightarrow Y$ be a
mapping. Then for any $A,B\in L^{Y},$
$$\stackrel{\ell }{\models} A\equiv B\rightarrow h^{-1}(A)\equiv h^{-1}(B).$$

\smallskip\

\textbf{Proof.} $$\lceil h^{-1}(A)\equiv h^{-1}(B)\rceil =\wedge
_{x\in X}(h^{-1}(A)(x)\longleftrightarrow h^{-1}(B)(x))$$
$$=\wedge _{x\in X}(A(h(x))\longleftrightarrow B(h(x)))$$
$$\geq \wedge _{y\in Y}(A(y)\longleftrightarrow B(y))$$
$$=\lceil A\equiv B\rceil .\heartsuit$$

\smallskip\

To conclude this section, we introduce the notion of $\ell-$valued
language as well as some operations of $\ell-$valued languages.
Suppose that $\Sigma$ is an alphabet; that is, a finite nonempty
set (of input symbols). We write $\Sigma^{*}$ for the set of
strings over $\Sigma$:
$$\Sigma^{*}=\cup_{n=0}^{\infty} \Sigma^{n}.$$
An $\ell-$valued language over $\Sigma$ is defined to be an
$\ell-$valued subset of $\Sigma^{\ast}$. Thus, the set of
$\ell-$valued languages over $\Sigma$ is exactly
$L^{\Sigma^{\ast}}$. Let $A,B\in L^{\Sigma ^{\ast }}$ be two
$\ell-$valued subsets of $\Sigma^{*}$. Then we define the
concatenation $A\cdot B$ of $A$ and $B$ and the Kleene closure
$A^{\ast }\in L^{\Sigma ^{\ast }}$ of $A$ as follows: for any
$s\in \Sigma ^{\ast },$

\smallskip\

(v) $s\in A\cdot B\stackrel{def}{=}(\exists u,v\in
\Sigma^{\ast})(s=uv\wedge u\in A\wedge v\in B);$

(vi) $s\in A^{\ast }\stackrel{def}{=}(\exists n\geq 0)(\exists
s_{1},...,s_{n}\in \Sigma ^{\ast }(s=s_{1}...s_{n}\wedge \wedge
_{i=1}^{n}(s_{i}\in A)).$

\smallskip\

The above defining equations can also be translated to the
following two formulas in the lattice of truth values by employing
the truth valuation rules: for every $s\in \Sigma^{*}$,

\smallskip\

(v') $(A\cdot B)(s)=\vee \{A(u)\wedge B(v):u,v\in \Sigma ^{\ast }\
{\rm and}\ s=uv\};$

(vi') $A^{\ast }(s)=\vee \{\wedge _{i=1}^{n}A(s_{i}):n\geq
0,s_{1},...,s_{n}\in \Sigma ^{\ast }\ {\rm and}\
s=s_{1}...s_{n}\}$.

\smallskip\

It is easy to demonstrate that if the meet $\wedge$ is
distributive over the join $\vee$ in $\ell$ (in other words,
$\ell$ is a Boolean algebra), then we have
$$A^{*}=\cup_{n=0}^{\infty}A^{n}$$ where
$$\left\{
\begin{array}{c}
A^{0}=\{\varepsilon \}, \\
A^{n+1}=A^{n}\cdot A\ {\rm for\ all }\ n\geq 0.
\end{array}
\right. $$

\smallskip\

\textbf{3. Orthomodular Lattice-Valued Automata}

\smallskip\

For convenience we first recall some basic notions in classical
automata theory. Let $\Sigma $ be a finite input alphabet whose
elements are called input symbols or labels. Then a
nondeterministic finite automaton (NFA for short) over $\Sigma $
is a quadruple
$$\Re =<Q,I,T,E>$$
in which:

(i) $Q$ is a finite set whose elements are called states;

(ii) $I\subseteq Q$ and states in $I$ are said to be initial;

(iii) $T\subseteq Q$ and states in $T$ are said to be terminal;
and

(iv) $E\subseteq Q\times \Sigma \times E,$ and each $(p,\sigma
,q)\in E$ is called a transition in (or an edge of) $\Re $ and it
means that input $\sigma $ makes state $p$ evolves to $q.$

An NFA is said to be deterministic if $I$ is a singleton, and for
any $p$ in $Q$ and $\sigma$ in $\Sigma$, there is exactly one $q$
in  $Q$ such that $(p,\sigma,q)\in E$. Thus, the transition
relation $E$ in a deterministic finite automaton (DFA, for short)
may be seen as a mapping from $Q\times \Sigma $ into $Q,$ and it
is called the transition function.

A path in $\Re $ is a finite sequence of the form
$$c=q_0\sigma_1q_1...q_{k-1}\sigma _kq_k$$
such that $(q_i,\sigma _{i+1},q_{i+1})\in E$ for each $i<k.$ In
this case, the sequence $\sigma _1...\sigma _k$ is called the
label of $c.$ A path $c=q_0\sigma _1q_1...q_{k-1}\sigma _kq_k$ is
said to be successful if $q_0\in I$ and $q_k\in T.$ The language
accepted by an automaton $\Re $ is the set of labels of all
successful paths in $\Re .$ Let
$$A\subseteq \Sigma ^{*}=\cup _{n=0}^\infty \Sigma ^n.$$
Then $A$ is said to be regular if there is an automaton $\Re $
over $\Sigma $ such that $A$ is the language accepted by $\Re .$

The notion of orthomodular lattice-valued automata is a natural
generalization of NFAs. Let $\ell =<L,\leq ,\wedge ,\vee ,\bot
,0,1>$ be an orthomodular lattice, and let $\Sigma $ be a finite
alphabet. Then an $\ell -$valued (quantum) automaton over $\Sigma
$ is a quadruple $$\Re =<Q,I,T,\delta >$$ where:

(i) $Q$ is the same as in an NFA;

(ii) $I$ is an $\ell-$valued subset of $Q$; that is, a mapping
from $Q$ into $L$. For each $q\in Q$, $I(q)$ indicates the truth
value (in the underlying quantum logic) of the proposition that
$q$ is an initial state;

(iii) $T$ is also an $\ell-$valued subset of $Q$, and for every
$q\in Q$, $T(q)$ expresses the truth value (in our quantum logic)
of the proposition that $q$ is terminal; and

(iv) $\delta $ is an $\ell -$valued subset of $Q\times \Sigma
\times Q$; that is, a mapping from $Q\times \Sigma \times Q$ into
$L,$ and it is called the $\ell -$valued (quantum) transition
relation of $\Re .$ Intuitively, $\delta$ is an $\ell -$valued
(ternary) predicate over $ Q,\Sigma $ and $Q$, and for any $p,q\in
Q$ and $\sigma\in \Sigma$, $\delta (p,\sigma ,q)$ stands for the
truth value (in quantum logic) of the proposition that input
$\sigma $ causes state $p$ to become $q.$

The propositions of the form
$$"q\ {\rm is\ an\ initial\ state}",\ {\rm written}\ "q\in I",$$
$$"q\ {\rm is\ a\ terminal\ state}",\ {\rm written}\ "q\in T",$$
and
$$"{\rm input}\ \sigma\ {\rm causes\ state}\ p\ {\rm to\ become}\ q,\
{\rm according\ to\ the\ specification}$$ $${\rm given\ by}\
\delta,"\ {\rm written}\ "p\stackrel{\delta,\sigma
}{\longrightarrow }q"$$ are assumed to be atomic propositions in
our logical language designated for describing $\ell-$valued
automata $\Re$. The truth values of the above three propositions
are respectively $I(q)$, $T(q)$ and $\delta(p,\sigma,q)$. The set
of these atomic propositions is denoted $atom(\Re)$. Formally, we
have
$$atom(\Re)=\{"q\in I":q\in Q\}\cup \{"q\in T":q\in Q\}\cup \{"p\stackrel{\delta,\sigma }
{\longrightarrow }q":p,q\in Q\ {\rm and}\ \sigma\in \Sigma\}.$$

We write $\mathbf{A}(\Sigma ,\ell )$ for the (proper) class of all
$\ell -$valued automata over $\Sigma.$

Before defining the concept of recognizability for $\ell -$valued
automata, we need to introduce some auxiliary notions and
notations. We set
$$T(Q,\Sigma )=(Q\Sigma )^{*}Q=\cup _{n=0}^\infty [(Q\Sigma )^n\ Q];$$
that is, the set of all alternative sequences of states and labels
beginning at a state and also ending at a state. For any
$c=q_0\sigma _1q_1...q_{k-1}\sigma _kq_k\in T(Q,\Sigma ),$ the
length of $c$ is defined to be $k$ and denoted by $|c|,$ $q_0$ is
the beginning of $c$ and denoted by $b(c),$ $q_k$ is the end of
$c$ and denoted by $e(c),$ and sequence $s=\sigma _1...\sigma _k$
is called the label of $c$ and denoted by $lb(c).$

Let $\Re \in \mathbf{A}(\Sigma ,\ell )$ be an $\ell-$valued
automaton over $\Sigma$. Then the $\ell -$valued (unary) predicate
$path_\Re $ on $T(Q,\Sigma )$ is defined as $path_\Re \in
L^{T(Q,\Sigma )}$ (the set of all mappings from $ T(Q,\Sigma )$
into $L$):
$$path_\Re (c)\stackrel{def}{=}\wedge _{i=0}^{k-1}[(q_i,\sigma
_{i+1},q_{i+1})\in \delta ]$$ for every $c=q_0\sigma
_1q_1...q_{k-1}\sigma _kq_k\in T(Q,\Sigma ).$

In intuition, the truth value of the proposition that $c=q_0\sigma
_1q_1...q_{k-1}\sigma _kq_k$ is a path in $\Re $ is
$$\lceil path_\Re (c)\rceil =\wedge _{i=0}^{k-1}\delta (q_i,\sigma _{i+1},q_{i+1}).$$
Note the difference between the symbols $\wedge$ in the above two
equations: the former is a logical connective, whereas the latter
is an operation on the lattice of truth values.

Now, we are ready to define one of the key notions in this paper,
namely, recognizability for $\ell -$valued automata. It will be
seen that the defining equation of $\ell-$valued recognizability
is the same as that in the classical theory of automata. The
essential difference between the quantum recognizability and the
corresponding classical notion implicitly resides in their truth
values.

\smallskip\

\textbf{Definition 3.1}. Let $\Re \in \mathbf{A}(\Sigma ,\ell )$.
Then the $\ell -$valued (unary) recognizability predicate $rec_\Re
$ on $\Sigma ^{*}$ is defined as $rec_\Re \in L^{\Sigma ^{*}}:$
for every $s\in \Sigma ^{*},$
$$rec_\Re (s)\stackrel{def}{=}(\exists c\in
T(Q,\Sigma ))(b(c)\in I\wedge e(c)\in T\wedge lb(c)=s\wedge
path_\Re (c)).$$ In other words, the truth value of the
proposition that $s$ is recognizable by $ \Re $ is given by
$$\lceil rec_\Re (s)\rceil
=\vee \{I(b(c))\wedge T(e(c))\wedge\lceil path_\Re (c)\rceil :c\in
T(Q,\Sigma )\ {\rm and}\ lb(c)=s\}.$$

\smallskip\

We note that $rec_\Re $ is defined above as an $\ell -$valued
unary predicate on $\Sigma ^{*},$ so it may also be seen as an
$\ell -$valued subset of $\Sigma ^{*};$ that is, a mapping
$rec_\Re :\Sigma ^{*}\rightarrow L$ with $rec_\Re (s)=\lceil
rec_\Re (s)\rceil $ for all $s\in \Sigma ^{*}.$

As a straightforward generalization of regular language, we can
also define regularity for $\ell-$valued languages.

\smallskip\

\textbf{Definition 3.2}. The $\ell -$valued (unary) regularity
predicate $Reg_\Sigma $ on $L^{\Sigma ^{*}}$ (the set of all $\ell
-$valued subsets of $\Sigma ^{*}$ ) is defined as $Reg_\Sigma \in
L^{(L^{\Sigma ^{*}})}:$ for each $A\in L^{\Sigma ^{*}},$
$$Reg_\Sigma (A)\stackrel{def}{=}(\exists \Re \in \mathbf{A}(\Sigma
,\ell ))(A\equiv rec_\Re ).$$ Thus, the truth value of the
proposition that $A$ is regular is
$$\lceil Reg_\Sigma (A)\rceil =\vee \{\lceil A\equiv rec_\Re \rceil
:\Re \in \mathbf{A}(\Sigma ,\ell )\}.$$

It should be noted that the (automaton) variable $\Re $ bounded by
the existential quantifier in the right-hand side of the defining
formula of $ Reg_\Sigma $ ranges over the proper class
$\mathbf{A}(\Sigma ,\ell ).$ Some readers who are familiar with
axiomatic set theory may worry about that this definition will
cause a certain set-theoretical difficulty, but we stay well away
from anything genuinely problematic. Indeed, for any $\ell
-$valued automaton $\Re =<Q,I,T,\delta >,$ there is a bijection
$\varsigma :Q\rightarrow |Q|$ (the cardinality of $Q$)
$=\{0,1,...,|Q|-1\}$ and we can construct a new $\ell -$valued
automaton $$\varsigma (\Re )=<|Q|,\varsigma (I),\varsigma
(T),\varsigma (\delta )>$$ where $$\varsigma (\delta )(m,\sigma
,n)=\delta (\varsigma ^{-1}(m),\sigma ,\varsigma ^{-1}(n))$$ for
any $m,n\in |Q|$ and $\sigma \in \Sigma .$ It is easy to see that
$rec_\Re =rec_{\varsigma (\Re )}.$ Then in Definition 3.2 we may
only require that the variable $\Re $ bounded by the existential
quantifier ranges over all $ \ell -$valued automata whose state
sets are subsets of $\omega $ (the set of all non-negative
integers) and the class of all $\ell -$valued automata with
subsets of $\omega $ as state sets is really a set, and in fact it
is a subset of $(2^\omega )^3\times \cup _{Q\subseteq \omega
}L^{Q\times \Sigma \times Q}.$ In most situations, however, the
original version of Definition 3.2 is much more convenient and
compatible with the corresponding definition in classical automata
theory.

Before investigating carefully various properties of regular
$\ell-$valued languages, we present some interesting examples. The
first one indicates that every $\ell-$valued language is regular.
It is well-known that a similar conclusion holds in classical
automata theory.

\smallskip\

\textbf{Example 3.1.} For any $A\in L^{\Sigma ^{*}},$ if $A$ is
finite, i..e., $suppA=\{s\in \Sigma^{\ast}:A(s)>0\}$ is finite,
then
$$\stackrel{\ell }{\models }Reg_\Sigma (A).$$
Indeed, suppose that $suppA=\{\sigma _{i1}...\sigma
_{im_i}:i=1,...,k\}.$ Then we construct an $\ell -$valued
automaton $\Re _A=(Q_A,I_A,T_A,\delta _A)$ in the following way:

(i) $Q_A=\cup _{i=1}^k\{q_{i0},q_{i1},...,q_{im_i}\};$

(ii) $I_A=\{q_{10},q_{20},...,q_{k0}\};$

(iii) $T_A=\{q_{1m_1},q_{2m_2},...,q_{km_k}\};$ and

(iv) We define $$\delta _A(q_{ij},\sigma
_{i(j+1)},q_{i(j+1)})=A(\sigma _{i1}...\sigma _{im_i})$$ for any
$1\leq i\leq k$ and $0\leq j<m_i,$ and we define $\delta
_A(p,\sigma ,q)=0$ for other $(p,\sigma ,q)\in Q_A\times \Sigma
\times Q_A.$ Then it is easy to see that $rec_{\Re _A}=A$ and
$$\lceil Reg_\Sigma (A)\rceil \geq \lceil A\equiv rec_{\Re
_A}\rceil =1.\heartsuit $$

\smallskip\

The following example may be seen as an extension of Example 3.1,
and it shows that the recognizability of a quantum language is not
less than the volume of its finite part.

\smallskip\

\textbf{Example 3.2.} For any $A\in L^{\Sigma ^{*}},$ we define
$$A\downarrow \lambda =\{s\in \Sigma ^{*}:A(s)\not\leq \lambda
\},$$ and
$$A\uparrow \lambda =\{s\in \Sigma ^{*}:A(s)\not\geq
\lambda \}.$$

Let $A\in L^{\Sigma ^{*}}.$ Then

(1) $\stackrel{\ell }{\models }\mathbf{\mu }\rightarrow Reg_\Sigma
(A),$ where $\mu =\vee \{\lambda ^{\bot }:A\downarrow \lambda $ is
finite$\};$ and

(2) $\stackrel{\ell }{\models }\mathbf{\theta }\rightarrow
Reg_\Sigma (A),$ where $\theta =\vee \{\lambda :A\uparrow \lambda
$ is finite$\}.$

Here, $\rightarrow$ may be interpreted as any implication operator
satisfying the Birkhoff-von Neumann requirement. We only prove (1)
and (2) may be proven likewise. For any $\lambda \in L,$ if
$A\downarrow \lambda $ is finite, then we define $A\Downarrow
\lambda \in L^{\Sigma ^{*}}$ as follows: for any $s\in \Sigma
^{*},$
$$(A\Downarrow \lambda )(s)=\left\{
\begin{array}{c}
A(s)\ {\rm if}\ A(s)\not\leq \lambda , \\
0\ {\rm if}\ A(s)\leq \lambda .
\end{array}
\right. $$ Clearly, $A\Downarrow \lambda $ is finite. Then from
Example 3.1 we know that there is an $\ell -$valued automata $\Re
[\lambda ]$ such that $rec_{\Re [\lambda ]}=A\Downarrow \lambda ,$
i.e., $ rec_{\Re [\lambda ]}=A(s)$ if $A(s)\not\leq \lambda $ and
$rec_{\Re [\lambda ]}=0$ if $A(s)\leq \lambda ,$ and
$$\lceil Rec_\Sigma (A)\rceil \geq \lceil A\equiv rec_{\Re [\lambda
]}\rceil =\wedge \{A(s)\leftrightarrow rec_{\Re [\lambda
]}:A(s)\not\leq \lambda \}\wedge \wedge \{A(s)\leftrightarrow
0:A(s)\leq \lambda \}$$
$$=\wedge \{A(s)\leftrightarrow 0:A(s)\leq
\lambda \}\geq \lambda ^{\bot }.\heartsuit $$

\smallskip\

The third example gives a simple connection between
recognizability in classical automata theory and the $\ell
-$valued predicate $Reg_\Sigma $ introduced above.

\smallskip\

\textbf{Example 3.3.} Let $A\subseteq \Sigma ^{*}$ be a regular
language (in classical automata theory), $B\in L^{\Sigma ^{*}}$
and
$$suppB=\{s\in \Sigma ^{*}:B(s)>0\}\subseteq A,$$ and let
$$\lambda =\vee \{\wedge _{s\in A}(a\leftrightarrow B(s)):a\in
L\}.$$ Then $$\stackrel{\ell }{\models }\mathbf{\lambda
}\rightarrow Reg_\Sigma (B).$$ In particular, if $A\subseteq
\Sigma ^{*}$ is regular then for every $a \in L,$
$$\stackrel{\ell }{\models }Reg_\Sigma (A[a]),$$ where $
A[a]\in L^{\Sigma ^{*}}$ is given as
$$A[a](s)=\left\{
\begin{array}{c}
a \ {\rm if}\ s\in A, \\
0\ {\rm otherwise.}
\end{array}
\right. $$

This conclusion is not difficult to prove. In fact, since $A$ is
regular, there must be an automaton $\Re =<Q,I,T,E>$ that accepts
the language $A.$ Now, for each $a\in L,$ we construct an $ \ell
-$valued automaton $\Re _a=<Q,I,T,\delta _a)$ such that
$$\delta _a(p,\sigma ,q)=\left\{
\begin{array}{c}
a\ {\rm if}\ (p,\sigma ,q)\in E, \\
0\ {\rm otherwise.}
\end{array}
\right.$$ Then it is easy to know that for all $s\in \Sigma ^{*},$
$$\lceil rec_{\Re _a}(s)\rceil =\left\{
\begin{array}{c}
a\ {\rm if}\ s\in A, \\
0\ {\rm otherwise,}
\end{array}
\right.$$ and
$$\lceil B\equiv rec_{\Re _a}\rceil =\wedge _{s\in
A}(a\leftrightarrow B(s)]).$$ Therefore, we have
$$\lceil Reg_\Sigma
(B)\rceil \geq \vee \{\lceil B\equiv rec_{\Re _a}\rceil :a\in
L\}=\lambda .\heartsuit $$

\smallskip\

The fourth example demonstrates that the $\ell -$valued predicate
$ Reg_\Sigma $ defined above is not trivial; that is, it does not
in general degenerate into a two-valued (Boolean) predicate.

\smallskip\

\textbf{Example 3.4}. We consider a canonical orthomodular
lattice. This lattice has a clear interpretation in quantum
physics. One pasts together observables of the spin one-half
system. Then he will obtain an orthomodular lattice $L(x)\oplus
L(\overline{x}),$ where
$$L(x)=\{0,p_{-},p_{+},1\}$$ corresponds to the outcomes of a
measurement of the spin states along the $x-$axis and
$$L(\overline{x})=\{
\overline{0}=1,\overline{p_{-}},\overline{p_{+}},\overline{1}=0\}$$
is obtained by measuring the spin states along a different spatial
direction; and $L(x)\oplus L(\overline{x})$ may be visualized as
the following ''Chinese lantern'' (see [Sv98] for a more detailed
description of $ L(x)\oplus L(\overline{x})$) (see Figure 2).

\begin{figure}\centering
\includegraphics{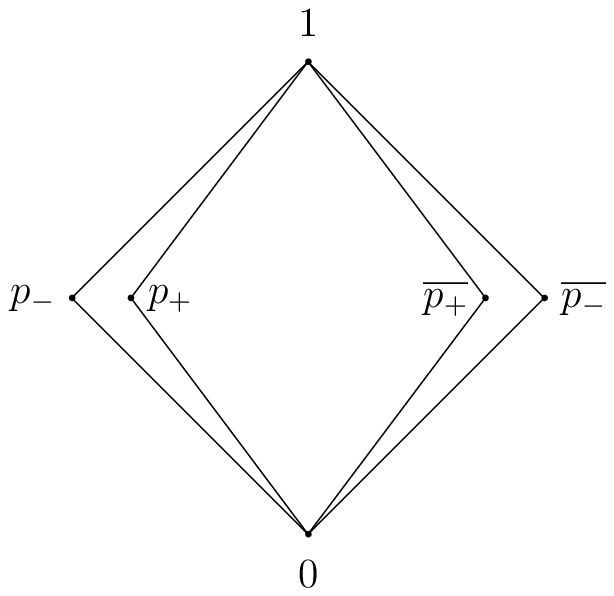}\caption{"Chinese lantern"} \label{fig 2}
\end{figure}

In this example, we set $\rightarrow =\rightarrow_3$ (the
Sasaki-hook). By a routine calculation we have
$$p_{-}\leftrightarrow p_{+}=p_{-}\leftrightarrow
\overline{p_{-}}=p_{-}\leftrightarrow \overline{ p_{+}}=0$$ and
$p_{-}\leftrightarrow 1=p_{-}.$ Thus, for each $\lambda \in
L(x)\oplus L(\overline{x}),$ $\lambda \not\leq p_{-}$ implies $
p_{-}\leftrightarrow \lambda \leq p_{-}.$

Furthermore, let $\Sigma =\{\sigma ,\tau \}$ and $A=\{\sigma
^n\tau ^n:n\in \omega \},$ and for any $t\in L(x)\oplus
L(\overline{x}),$ let $A_t\in L^{\Sigma ^{*}}$ be given as
follows:
$$A_t(s)=\left\{
\begin{array}{c}
1\ {\rm if}\ s\in A, \\
t\ {\rm otherwise.}
\end{array}
\right. $$ Then it holds that $$\stackrel{\ell }{\models
}\mathbf{p}_{-}\leftrightarrow Reg_\Sigma (A_{p_{-}});$$ that is,
$\lceil Reg_\Sigma (A_{p_{-}})\rceil =p_{-}.$ In fact, we know
that $\Sigma ^{*}$ is regular (see [E74], Example II.2.3), and
with Example 3.3 it is easy to see that $\lceil Reg_\Sigma
(A_{p_{-}})\rceil \geq p_{-}.$ Conversely, for any $\ell -$valued
automaton $ \Re =<Q,I,T,\delta >,$ if $|Q|=n$ then
$$\lceil A_{p_{-}}\equiv rec_\Re \rceil \leq [A_{p_{-}}(\sigma
^n\tau ^n)\leftrightarrow rec_\Re (\sigma ^n\tau ^n)]\wedge \wedge
_{k,l\in \omega \ {\rm s.t. }\ k\neq l}[A_{p_{-}}(\sigma ^k\tau
^l)\leftrightarrow rec_\Re (\sigma ^k\tau ^l)]$$
$$=rec_\Re (\sigma
^n\tau ^n)\wedge \wedge _{k,l\in \omega \ {\rm s.t. }\ k\neq
l}[p_{-}\leftrightarrow rec_\Re (\sigma ^k\tau ^l)].$$ If $rec_\Re
(\sigma ^n\tau ^n)\leq p_{-},$ then $\lceil A_{p_{-}}\equiv
rec_\Re \rceil \leq p_{-}.$ Now, we consider the case of $rec_\Re
(\sigma ^n\tau ^n)\not\leq p_{-}.$ For any $c\in T(Q,\Sigma ),$ if
$b(c)\in I,$ $ e(c)\in T$ and $lb(c)=\sigma ^n\tau ^n,$ then $c$
must be of the form $$ c=p_0\sigma p_1...p_{n-1}\sigma p_n\tau
q_1...q_{n-1}\tau q_n.$$ Since $|Q|=n, $ there are $i,j$ such that
$i<j\leq n$ and $p_i=p_j.$ We put
$$c^{+}=p_0\sigma p_1...p_{j-1}\sigma p_j(=p_i)\sigma
p_{i+1}...p_{j-1}\sigma p_j\sigma p_{j+1}...p_{n-1}\sigma p_n\tau
q_1...q_{n-1}\tau q_n.$$ Then $b(c^{+})\in I,$ $e(c^{+})\in T,$
$lb(c^{+})=\sigma ^{n+(j-i)}\tau ^n$ and $\lceil path_\Re
(c^{+})\rceil =\lceil path_\Re (c)\rceil .$ Therefore, it holds
that
$$rec_\Re (\sigma ^{n+(j-i)}\tau ^n)\geq \vee \{\lceil
path_\Re (c^{+})\rceil :b(c)\in I, e(c)\in T\ {\rm and}\
 lb(c)=\sigma ^n\tau ^n\}$$
$$=\vee \{\lceil path_\Re (c)\rceil :b(c)\in I, e(c)\in T\ {\rm
and}\ lb(c)=\sigma ^n\tau ^n\}$$
$$=rec_\Re (\sigma ^n\tau ^n),$$
and $$rec_\Re (\sigma ^{n+(j-i)}\tau ^n)\not\leq p_{-}.$$
Furthermore, we have $$\lceil A_{p_{-}}\equiv rec_\Re \rceil \leq
p_{-}\leftrightarrow rec_\Re (\sigma ^{n+(j-i)}\tau ^n)\leq
p_{-}.$$

So, for all $\ell -$valued automata $\Re $ we have $\lceil
A_{p_{-}}\equiv rec_\Re \rceil \leq p_{-},$ and it follows that
$$\lceil Rec_\Sigma (A_{p_{-}})\rceil =\vee \{\lceil A\equiv
rec_\Re \rceil :\Re \in \mathbf{A} (\Sigma ,\ell )\}\leq p_{-}.$$
This together with $\lceil Reg_\Sigma (A_{p_{-}})\rceil \geq
p_{-}$ obtained before leads to $\lceil Reg_\Sigma
(A_{p_{-}})\rceil =p_{-}.$

Similarly, we have $\lceil Reg_\Sigma (A_t)\rceil =t$ for
$t=p_{+},\overline{ p_{-}}$ and $\overline{p_{+}}.\heartsuit$

\smallskip\

Motivated by the above example, we propose the open problem: how
to describe orthomodular lattices $\ell =<L,\leq ,\wedge ,\vee
,\bot ,0,1>$ which satisfy that
$$\{\lceil Reg_\Sigma (A)\rceil :A\in L^{\Sigma ^{*}}\}=L,$$ i.e., the truth
values of recognizability traverse all over $L,$ or more
explicitly, for every $\lambda \in L,$ there is $A\in L^{\Sigma
^{*}}$ such that $\lceil Reg_\Sigma (A)\rceil =\lambda .\ $ It
seems that this is a difficult problem.

The $\ell-$valued regularity predicate $Reg_\Sigma$ in Definition
3.2 is a direct generalization of the notion of regular language
in classical automata theory. In what follows, we will see that
the predicate $Reg_\Sigma$ does not work well in many cases. Why
this happens? Note that $Reg_\Sigma$ is merely a simple mimic of
the classical concept of regular language, and an essential
feature of quantum logic is missing here. In the defining equation
of $Reg_\Sigma$, the language $A$ to be recognized and the
automaton $\Re$ for recognizing $A$ are left completely
irrelevant. In the case of classical logic, this does not causes
any difficulty in manipulating regular languages. Nevertheless,
the thing changes when we work in quantum logic. After an analysis
it was found that a suitable link between $A$ and $\Re$ is a
commutativity of them. This motivates the following:

\smallskip\

\textbf{Definition 3.3.} The $\ell -$valued (unary and partial)
predicate $CReg_{\Sigma }$ on $L^{\Sigma ^{\ast }}$ is called
commutative regularity and it is defined as $CReg_{\Sigma }\in
L^{(L^{\Sigma ^{\ast }})}:$ for any $A\in L^{\Sigma ^{\ast }}$
with finite $ Range(A)=\{A(s):s\in \Sigma ^{\ast }\},$
$$CReg_{\Sigma }(A)\stackrel{def}{=}(\exists \Re
\in \mathbf{A}(\Sigma ,\ell ))(\gamma (atom(\Re )\cup r(A))\wedge
(A\equiv rec_{\Re })),$$ where $r(A)=\{\mathbf{a}:a\in
Range(A)\}.$

\smallskip\

The exposition concerning the automata variable $\Re$ in the
defining equation of $Reg_{\Sigma}$ in Definition 3.2 also applies
to $CReg_{\Sigma}$ in the above definition.

It is obvious that the notion of commutative regularity is
stronger than regularity. In other words, we have for any $A\in
L^{\Sigma^{\ast}}$,
$$\stackrel{\ell }{\models } CReg_{\Sigma}(A)\rightarrow
Reg_{\Sigma}(A).$$ On the other hand, if $\ell$ is a Boolean
algebra; that is, the underlying logic is the classical Boolean
logic, then these two notions are equivalent; or formally, for all
$A\in L^{\Sigma^{\ast}}$, it holds that
$$\stackrel{\ell }{\models } CReg_{\Sigma}(A)\leftrightarrow
Reg_{\Sigma}(A).$$ This is just why the predicate $Reg_\Sigma$
works very well in classical automata theory but not in the theory
of automata based on quantum logic.

\bigskip\

\textbf{4. Orthomodular Lattice-Valued Deterministic Automata}

\smallskip\

The notion of nondeterminism plays a central role in the theory of
computation. The nondeterministic mechanism enables a device to
change its states in a way that is only partially determined by
the current state and input symbol. Obviously, the concept of
$\ell-$valued automaton introduced in the last section is a
generalization of nondeterministic finite automaton. In classical
theory of automata, each nondeterministic finite automaton is
equivalent to a deterministic one; more exactly, there exists a
deterministic finite automaton which accepts the same language as
the originally given nondeterministic one does. The aim of this
section is just to see whether this result is still valid in the
framework of quantum logic. To this end, we first introduce the
concept of deterministic $\ell-$valued automaton.

Let $\Re=<Q,I,T,\delta>\in \mathbf{A}(\Sigma,\ell)$ be an
$\ell-$valued automata over $\Sigma$. If

(i) there is a unique $q_0$ in $Q$ with $I(q_0)>0$; and

(ii) for all $q$ in $Q$ and $\sigma$ in $\Sigma ,$ there is a
unique state $p$ in $Q$ such that $$ \delta (q,\sigma,p)>0,$$ then
$M$ is called an $\ell-$valued (quantum) deterministic finite
automaton ($\ell-$valued DFA for short). The $\ell-$valued
transition relation $\delta $ in an $\ell-$valued DFA may be
equivalently represented by a mapping from $Q\times \Sigma $ into
$Q\times(L-\{0\}).$ For any $q$ in $Q$ and $\sigma$ in $\Sigma,$
if $p$ is the unique element in $Q$ with $\delta (q,\sigma,p)>0,$
then $\delta (q,\sigma)=(p,\delta (q,\sigma,p))\in
Q\times(L-\{0\}).$

The class of $\ell-$valued DFAs over $\Sigma$ is denoted
$\mathbf{DA}(\Sigma,\ell)$.

Suppose that $\Re$ is an $\ell-$valued DFA, $\delta
(q_0,\sigma_1)=(q_1,\lambda _1)$ and $\delta
(q_i,\sigma_{i+1})=(q_{i+1},\lambda _{i+1})$ for all
$i=1,2,...,n-1.$ Then it is easy to see that $$\lceil rec_\Re
(\sigma_1 ...\sigma_n)\rceil =I(q_0)\wedge T(q_n)\wedge \wedge
_{i=1}^n\lambda _i.$$

Throughout this section, we always suppose that the lattice $\ell$
of truth values is finite.

The proof of the equivalence between classical deterministic
finite and nondeterministic finite automata is carried out by
building the power set construction of a nondeterministic finite
automaton that is deterministic and can simulate the given
nondeterministic one. The power set construction can be naturally
extended into the case of $\ell-$valued automata.

Let $\Re =<Q,I,T,\delta
>\in \mathbf{A}(\Sigma ,\ell )$ be an $\ell -$valued automaton
over $\Sigma .$ We define the $\ell -$valued power set
construction of $\Re $ to be $\ell -$valued automaton
$$\ell ^{\Re }=<L^{Q},I_{1},\overline{T},\overline{
\delta }>$$ over $\Sigma ,$ where:

(i) $L^{Q}$ is the set of all $\ell -$valued subsets of $Q;$ that
is, mappings from $Q$ into $L;$

(ii) $I_{1}$ is an $\ell -$valued point with height $1;$ that is,
$I_{1}\in L^{(L^{Q})}$ and
$$I_{1}(X)=\left\{
\begin{array}{c}
1\ {\rm if }X=I, \\
0\ {\rm otherwise}
\end{array}
\right.$$ for all $X\in L^{Q};$

(iii) $\overline{T}\in L^{(L^{Q})};$ that is, $\overline{T}$ is an
$\ell -$valued subset of $L^{Q},$ and
$$\overline{T}(X)=\vee _{q\in Q}[X(q)\wedge T(q)]$$
for any $X\in L^{Q};$ and

(iv) $\overline{\delta }$ is a mapping from $L^{Q}\times \Sigma $
into $ L^{Q},$ and for each $X\in L^{Q},$ $\overline{\delta
}(X,\sigma )\in L^{Q}$ and
$$\overline{\delta }(X,\sigma )(q)=\vee _{p\in Q}[X(p)\wedge \delta (p,\sigma
,q)]$$ for every $q\in Q.$

Since $L$ is assumed to be finite, $L^{Q}$ is finite too. Thus, it
is easy to see that $\ell ^{\Re }$ is an $\ell -$valued DFA.
Moreover, both the set of the initial states and the transition
relation are two-valued, namely, their truth values are either $0$
or $1,$ and only the set of terminal states carries $\ell -$valued
information.

The following theorem compares the abilities of an $\ell-$valued
automaton and its power set construction according to the
$\ell-$valued languages recognized by them.

\smallskip\

\textbf{Theorem 4.1.} Let $\ell =<L,\leq ,\wedge ,\vee ,\bot
,0,1>$ be a finite orthomodular lattice, and let $\longrightarrow$
be an implication operator satisfying the Birkhoff-von Neumann
requirement.

(1) For any $\Re \in \mathbf{A}(\Sigma ,\ell )$ and $s\in \Sigma
^{\ast },$
$$\stackrel{\ell }{ \models } rec_{\Re }(s)\longrightarrow rec_{(\ell
^{\Re })}(s).$$

(2) For any $\Re \in \mathbf{A}(\Sigma ,\ell )$ and $s\in \Sigma
^{\ast },$
$$\stackrel{\ell }{ \models } \gamma(atom(\Re))\wedge rec_{(\ell
^{\Re })}(s) \longrightarrow rec_{\Re }(s),$$ and in particular if
$\rightarrow =\rightarrow_3$, then
$$\stackrel{\ell }{ \models } \gamma(atom(\Re))\longrightarrow (rec_{(\ell
^{\Re })}(s) \longleftrightarrow rec_{\Re }(s)).$$

(3) The following two statements are equivalent to each other:

\ \ \ \ \ \ (i) $\ell$ is a Boolean algebra.

\ \ \ \ \ \ (ii) For any $\Re \in \mathbf{A}(\Sigma ,\ell )$ and
$s\in \Sigma ^{\ast },$
$$\stackrel{\ell }{ \models } rec_{\Re }(s)\longleftrightarrow rec_{(\ell
^{\Re })}(s).$$

\smallskip\

\textbf{Proof.} The proof of (1) is easy, and we omit it here.

(2) Suppose that $\Re =<Q,I,T,\delta
>\in \mathbf{A}(\Sigma ,\ell )$ and $\ell ^{\Re
}=<L^{Q},I_{1},\overline{T}, \overline{\delta }>$ is the $\ell
-$valued power set construction of $\Re .$ Our aim is to
demonstrate that
$$\lceil \gamma(atom(\Re))\rceil \wedge \lceil rec_{(\ell ^{\Re })}(s)\rceil\leq \lceil rec_{\Re }(s)\rceil
$$ for all $s\in \Sigma ^{\ast }.$ To this end, we first prove the
following
$$claim:\ \ \lceil \gamma(atom(\Re))\rceil \wedge \overline{\delta }(I,\sigma _{1}...\sigma
_{n})(q_{n}) \ \ \ \ \ \ \ \ \ \ \ \ \ \ \ \ \ \ \ \ \ \ \ \ \ \ \
\ \ \ \ \ \ \ \ \ \ \ \ \ \ \ \ \ \ \ \ \ \ \ $$
$$\ \ \ \ \ \ \ \ \ \ \ \ \ \ \ \ \ \ \ \ \leq\vee \{I(q_{0})\wedge \wedge
_{i=0}^{n-1}\delta (q_{i},\sigma
_{i+1},q_{i+1}):q_{0},q_{1},...,q_{n-1}\in Q\}$$ for any $\sigma
_{1},...,\sigma _{n}\in \Sigma $ and $q\in Q.$ We proceed by
induction on $n.$ For $n=0,$ it is clear. The definition of
$\overline{\delta}$ yields
$$\overline{\delta }(I,\sigma _{1}...\sigma
_{n})(q_{n})=\overline{\delta }( \overline{\delta }(I,\sigma
_{1}...\sigma _{n-1}),\sigma _{n})(q_{n})\ \ \ \ \ \ \ \ \ \ \ \ \
\ \ \ \ \ \ \ \ \ $$
$$=\vee _{q_{n-1}\in Q}[\overline{\delta }(I,\sigma _{1}...\sigma
_{n-1})(q_{n-1})\wedge \delta (q_{n-1},\sigma _{n},q_{n})].$$ We
write $$\lceil atom(\Re)\rceil =\{\lceil \varphi\rceil: \varphi\in
atom(\Re)\}.$$ Then it holds that
$$\lceil \gamma(atom(\Re))\rceil=\gamma(\lceil atom(\Re)\rceil).$$ Note
that the symbol $\gamma$ in the left-hand side applies to a set of
logical formulas, whereas the one in the right-hand side applies
to a subset of $L$. Furthermore, it is easy to see that
$\delta(q_{n-1},\sigma_n,q_n),\ \overline{\delta}(I,\sigma_1
...\sigma_{n-1})(q_{n-1})$ and $\lceil \gamma(atom(\Re))\rceil$
are all in $[\lceil atom(\Re)\rceil]$ (the subalgebra of $\ell$
generated by $\lceil atom(\Re)\rceil$). Thus, with Lemmas 2.5 and
2.6 and the induction hypothesis we obtain
$$\lceil \gamma(atom(\Re))\rceil\wedge \overline{\delta }(I,\sigma _{1}...\sigma
_{n})(q_{n}) =\lceil \gamma(atom(\Re))\rceil\wedge \lceil
\gamma(atom(\Re))\rceil $$ $$\ \ \ \ \ \ \ \ \ \ \wedge \vee
_{q_{n-1}\in Q}[\overline{\delta }(I,\sigma _{1}...\sigma
_{n-1})(q_{n-1})\wedge \delta (q_{n-1},\sigma _{n},q_{n})]$$
$$\leq \vee _{q_{n-1}\in Q}[\lceil \gamma(atom(\Re))\rceil\wedge
(\vee \{I(q_0)\wedge\wedge_{i=0}^{n-2}\delta(q_i,\sigma
_{i+1},q_{i+1}):$$ $$\ \ \ \ \ \ \ \ \ \ \ \ \ \ \ \ \ \ \ \ \ \ \
\ \ \ \ \ \ \ \ \ \ \ \ \ \ \ \ \ \ \ \ \ \ \ \ \ \ \ \ \ \ \ \ \
q_0,q_1,...,q_{n-2}\in Q\})\wedge \delta (q_{n-1},\sigma
_{n},q_{n})].$$ Using Lemmas 2.5 and 2.6 again, we complete the
proof of the above claim.

Now with this claim, we can use Lemmas 2.5 and 2.6 twice and
derive that
$$\lceil \gamma(atom(\Re))\rceil\wedge \lceil
rec_{(\ell^{Re})}(\sigma_1 ...\sigma_n)\rceil =\lceil
\gamma(atom(\Re))\rceil\wedge \overline{T}( \overline{\delta
}(I,\sigma _{1}...\sigma _{n}))$$
$$=\lceil \gamma(atom(\Re))\rceil\wedge \vee _{q_{n}\in Q}[\overline{\delta }(I,\sigma _{1}...\sigma
_{n})(q_{n})\wedge T(q_{n})]$$
$$\leq \vee _{q_{n}\in Q}[\lceil \gamma(atom(\Re))\rceil\wedge \overline{\delta }(I,\sigma _{1}...\sigma
_{n})(q_{n})\wedge T(q_{n})]$$
$$\leq \vee _{q_{n}\in Q}[\lceil \gamma(atom(\Re))\rceil\wedge (\vee \{I(q_{0})\wedge \wedge
_{i=0}^{n-1}\delta (q_{i},\sigma
_{i+1},q_{i+1}):q_{0},q_{1},...,q_{n-1}\in Q\})\wedge T(q_{n})]$$
$$\leq \vee \{I(q_{0})\wedge \wedge _{i=0}^{n-1}\delta (q_{i},\sigma
_{i+1},q_{i+1})\wedge T(q_{n}):q_{0},q_{1},...,q_{n-1}\in Q\}$$
$$=\lceil rec_{\Re }(\sigma _{1}...\sigma _{n})\rceil .$$

For the case of $\rightarrow =\rightarrow_3$, what we want to
prove is
$$\lceil \gamma(atom(\Re))\rceil \leq \lceil rec_{(\ell ^{\Re })}(s)\rceil\longrightarrow_3 \lceil rec_{\Re }(s)\rceil
.$$ With the above conclusion and Lemma 2.10, it suffices to show
that $\lceil \gamma(atom(\Re))\rceil C \lceil rec_{\Re
}(s)\rceil$. We observe that $$\lceil \gamma(atom(\Re))\rceil
=\vee \{\wedge_{\varphi\in \lceil
atom(\Re)\rceil}\varphi^{f(\varphi)}:f\in \{0,1\}^{\lceil
atom(\Re)\rceil}\}.$$ Then Lemma 2.2 tells us that we only need to
prove $\wedge_{\varphi\in \lceil
atom(\Re)\rceil}\varphi^{f(\varphi)}C \lceil rec_{\Re }(s)\rceil$
for all $f\in \{0,1\}^{\lceil atom(\Re)\rceil}$. For every
$\psi\in \lceil atom(\Re)\rceil$, note that
$$\wedge_{\varphi\in \lceil
atom(\Re)\rceil}\varphi^{f(\varphi)}\leq \psi^{f(\psi)}.$$ Then
$\wedge_{\varphi\in \lceil atom(\Re)\rceil}\varphi^{f(\varphi)} C
\psi^{f(\psi)},$ and furthermore it follows that
$\wedge_{\varphi\in \lceil atom(\Re)\rceil}\varphi^{f(\varphi)} C
\psi$ from Lemmas 2.1(3) and (4). Since $\lceil rec_\Re (s)\rceil$
is calculated from some elements in $\lceil atom(\Re)\rceil$ by
applying a finite number of meets or unions, we complete the proof
with Lemma 2.2.

(3) Note that $\lceil \gamma(atom(\Re))\rceil=1$ is always valid
when $\ell$ is a Boolean algebra. Thus, it is proved that (i)
implies (ii). We now turn to show that (ii) implies (i). It
suffices to show that the meet $\wedge$ is distributive over the
union $\vee$; that is, $$a\wedge (b\vee c)=(a\wedge b)\vee
(a\wedge c)$$ for all $a,b,c\in L$. Let $a,b,c\in L.$ We construct
an $\ell -$valued automaton
$$\Re =<\{u,v,w\},\{u,v\},\{w\},\delta >$$ over $ \Sigma $ which
has at least one element $\sigma ,$ where $\delta (u,\sigma
,u)=a,$ $\delta (u,\sigma ,w)=c,$ $\delta (v,\sigma ,u)=b,$ and
$\delta $ takes the value $0$ for other cases. It may be
visualized by Figure 3.

\begin{figure}\centering
\includegraphics[width=0.5\textwidth]{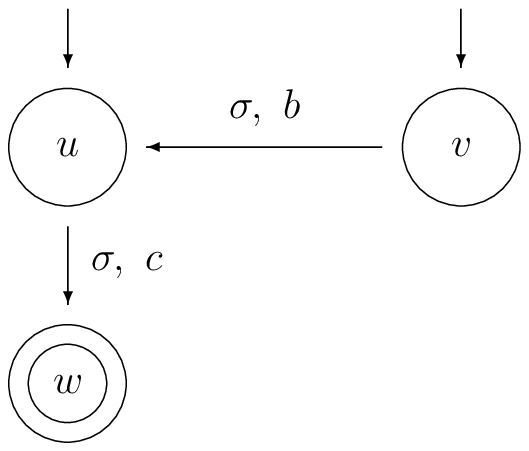}\caption{Automaton a} \label{fig 3}
\end{figure}

In the automaton $\Re $ we have
$$\lceil rec_{\Re }(\sigma \sigma )\rceil =\vee \{I(q_{0})\wedge
T(q_{2})\wedge \delta (q_{0},\sigma ,q_{1})\wedge \delta
(q_{1},\sigma ,q_{2}):q_{0},q_{1},q_{2}\in Q\}$$
$$=\vee \{\delta (u,\sigma ,q_{1})\wedge \delta (q_{1},\sigma
,w):q_{1}\in Q\}\vee \vee \{\delta (v,\sigma ,q_{1})\wedge \delta
(q_{1},\sigma ,v):q_{1}\in Q\}$$
$$=[\delta (u,\sigma ,u)\wedge \delta (u,\sigma ,w)]\vee \lbrack
\delta (v,\sigma ,u)\wedge \delta (u,\sigma ,w)]$$
$$=(a\wedge c)\vee (b\wedge c).$$

Consider the $\ell -$valued power set construction $\ell ^{\Re }$
of $\Re .$ Then
$$\overline{\delta }(I,\sigma )(u)=\vee _{q\in Q}[I(q)\wedge \delta
(q,\sigma ,u)]$$
$$=\delta (u,\sigma ,u)\vee \delta (v,\sigma ,u)]$$
$$=a\vee b.$$
Similarly, we obtain $\overline{\delta }(I,\sigma )(v)=0$ and
$\overline{ \delta }(I,\sigma )(w)=c.$ It follows that for any
$q\in Q,$
$$\overline{\delta }(I,\sigma \sigma )(q)=\overline{\delta
}(\overline{\delta }(I,\sigma ),\sigma )(q)$$
$$=\vee _{q^{\prime }\in Q}[\overline{\delta }(I,\sigma )(q^{\prime
})\wedge \delta (q^{\prime },\sigma ,q)]$$
$$=[\overline{\delta }(I,\sigma )(u)\wedge \delta (u,\sigma
,q)]\vee \lbrack \overline{\delta }(I,\sigma )(w)\wedge \delta
(w,\sigma ,q)]$$
$$=(a\vee b)\wedge \delta (u,\sigma ,q).$$
Thus, $$\overline{\delta }(I,\sigma \sigma )(u)=(a\vee b)\wedge
a=a,$$ $$ \overline{\delta }(I,\sigma \sigma )(v)=0$$ and
$$\overline{\delta }(I,\sigma \sigma )(w)=(a\vee b)\wedge c.$$
Therefore,
$$\lceil rec_{(\ell ^{\Re })}(\sigma \sigma )\rceil
=\overline{T}(\overline{ \delta }(I,\sigma \sigma ))$$
$$=\vee _{q\in Q}[\overline{\delta }(I,\sigma \sigma )(q)\wedge
T(q)]$$
$$=\overline{\delta }(I,\sigma \sigma )(w)$$
$$=(a\vee b)\wedge c.$$

Finally, from the assumption (ii) we assert that
$$(a\wedge c)\vee (b\wedge c)=\lceil rec_{\Re }(\sigma \sigma
)\rceil =\lceil rec_{(\ell ^{\Re })}(\sigma \sigma )\rceil =(a\vee
b)\wedge c.\heartsuit$$

\smallskip\

Many results in this paper appear in the same scheme as the above
theorem. So, we here give a detailed explanation of this theorem.
The above theorem points out that the ability of an $\ell-$valued
automaton for recognizing language is always weaker than that of
its power set construction. On the other hand, in order to warrant
that an $\ell-$valued automaton $\Re$ and its power set
construction have the same ability of accepting language, the
condition $\gamma(atom(\Re))$ has to be imposed. The intuitive
meaning of this condition is that (the truth values of) any two
atomic propositions describing $\Re$ should commute. (See also the
physical interpretation of commutativity presented in the
introductory section.) The third part of Theorem 4.1 indicates
that the equivalence between a nondeterministic finite automaton
and its power set construction is universally valid if and only if
the underlying logic degenerates to the classical Boolean logic.
In other words, if the meta-logic that we use in our reasoning
does not enjoy distributivity, then such a meta-logic is not
strong enough to guarantee the universal validity of any
nondeterministic finite automaton and its power set construction,
and we can always find a nondeterministic finite automaton such
that the equivalence between it and its power set construction is
not derivable with the mere inference power provided by such a
meta-logic.

In Section 3, we introduced the regularity and commutative
regularity predicates $Reg_{\Sigma}$ and $CReg_{\Sigma}$. They are
all given with respect to nondeterministic $\ell-$valued automata.
Now we propose a restricted version of them based on the smaller
class of deterministic $\ell-$valued automata.

\smallskip\

\textbf{Definition 4.1.} Let $\ell=<L,\leq ,\wedge ,\vee ,\bot
,0,1>$ be an orthomodular lattice. Then the $\ell -$valued (unary)
predicates $DReg_{\Sigma}$ and (unary and partial) predicate
$CDReg_{\Sigma }$ on $L^{\Sigma ^{\ast }}$ are called
deterministic regularity and commutative deterministic regularity,
respectively, and they are defined as $DReg_{\Sigma}$,
$CDReg_{\Sigma }\in L^{(L^{\Sigma ^{\ast }})}:$ for any $A\in
L^{\Sigma ^{\ast }}$,
$$DReg_{\Sigma }(A)\stackrel{def}{=}(\exists \Re
\in \mathbf{DA}(\Sigma ,\ell ))(A\equiv rec_{\Re }),$$ and for any
$A\in L^{\Sigma ^{\ast }}$ with finite $Range(A)=\{A(s):s\in
\Sigma^{\ast}\},$
$$CDReg_{\Sigma }(A)\stackrel{def}{=}(\exists \Re
\in \mathbf{DA}(\Sigma ,\ell ))(\gamma (atom(\Re )\cup r(A))\wedge
(A\equiv rec_{\Re })),$$ where $r(A)=\{\mathbf{a}:a\in
Range(A)\}.$

\smallskip\

It is similar to the relation between $Reg_\Sigma$ and
$CReg_\Sigma$ that $CDReg_\Sigma$ is stronger than $DReg_\Sigma$.
In other words, it holds that for any $A\in L^{\Sigma^{\ast}}$,
$$\stackrel{\ell }{\models } CDReg_{\Sigma}(A)\rightarrow
DReg_{\Sigma}(A).$$ The following corollary shows that a certain
commutativity condition guarantees that they are equivalent.
Furthermore, if $\ell$ is a Boolean algebra, then the four notions
$Reg_{\Sigma}$, $CReg_{\Sigma}$, $DReg_{\Sigma}$ and
$CDReg_{\Sigma}$ all coincide.

\smallskip\

\textbf{Corollary 4.2.} Let $\ell=<L,\leq ,\wedge ,\vee ,\bot
,0,1>$ be a finite orthomodular lattice, and let $\longrightarrow
= \longrightarrow_3$. Then for any $A\in L^{\Sigma^{\ast}}$,
$$\stackrel{\ell }{ \models }CReg_{\Sigma}(A)\longleftrightarrow CDReg_{\Sigma}(A).$$
In particular, if $\ell$ is a Boolean algebra, then for any $A\in
L^{\Sigma ^{\ast }},$
$$\stackrel{\ell }{ \models } Reg_{\Sigma}(A)\longleftrightarrow DReg_{\Sigma}(A) .$$

\smallskip\

\textbf{Proof}. It is clear that $$\stackrel{\ell }{ \models
}CDReg_{\Sigma}(A)\rightarrow CDReg_{\Sigma }(A).$$ Then we only
need to prove that $$\stackrel{\ell }{ \models }CReg_{\Sigma
}(A)\rightarrow CDReg_{\Sigma }(A);$$ that is, for any $\Re \in
\mathbf{A}(\Sigma ,\ell ),$
$$\lceil \gamma (atom(\Re )\cup r(A))\rceil \wedge \lceil A\equiv
rec_{\Re }\rceil \leq \vee \{\lceil \gamma (atom(\wp )\cup
r(A))\rceil \wedge \lceil A\equiv rec_{\wp }\rceil :\wp \in
\mathbf{DA}(\Sigma ,\ell )\}.$$

First, by using Lemmas 2.5, 2.6 and 2.11(2) we have $$\lceil
\gamma (atom(\Re )\cup r(A))\rceil \wedge \lceil A\equiv rec_{\Re
}\rceil \wedge \lceil rec_{\Re }\equiv rec_{(\ell ^{\Re })}\rceil
=\ \ \ \ \ \ \ \ \ \ \ \ \ \ \ \ \ \ \ \ \ \ \ \ \ \ \ \ \ \ \ \ \
\ \ \ $$
$$\lceil \gamma (atom(\Re )\cup r(A))\rceil \wedge \wedge _{s\in
\Sigma ^{\ast }}(A(s)\leftrightarrow rec_{\Re }(s))\wedge \wedge
_{s\in \Sigma ^{\ast }}(rec_{\Re }(s)\leftrightarrow rec_{(\ell
^{\Re })}(s))$$
$$=\wedge _{s\in \Sigma ^{\ast }}(\lceil \gamma (atom(\Re )\cup
r(A))\rceil \wedge (A(s)\rightarrow rec_{\Re }(s))\wedge (rec_{\Re
}(s)\rightarrow rec_{(\ell ^{\Re })}(s))\wedge$$
$$\wedge _{s\in \Sigma ^{\ast }}(\lceil \gamma (atom(\Re
)\cup r(A))\rceil \wedge (rec_{(\ell ^{\Re })}(s)\rightarrow
rec_{\Re }(s))\wedge (rec_{\Re }(s)\rightarrow A(s))$$
$$\leq \wedge _{s\in \Sigma ^{\ast }}(A(s)\rightarrow rec_{(\ell
^{\Re })}(s))\wedge \wedge _{s\in \Sigma ^{\ast }}(rec_{(\ell
^{\Re })}(s)\rightarrow A(s))$$
$$=\wedge _{s\in \Sigma ^{\ast }}(A(s)\leftrightarrow rec_{(\ell
^{\Re })}(s)) $$
$$=\lceil A\equiv rec_{(\ell ^{\Re })}\rceil .$$

Second, from Theorem 4.1(2) we obtain
$$\lceil \gamma (atom(\Re )\cup r(A))\rceil \leq \lceil \gamma
(atom(\Re ))\rceil \leq \lceil rec_{\Re }\equiv rec_{(\ell ^{\Re
})}\rceil ,$$ and
$$\lceil \gamma (atom(\Re )\cup r(A))\rceil \wedge \lceil A\equiv
rec_{\Re }\rceil \leq \lceil \gamma (atom(\Re )\cup r(A))\rceil
\wedge \lceil A\equiv rec_{\Re }\rceil \wedge \lceil rec_{\Re
}\equiv rec_{(\ell ^{\Re })}\rceil $$
$$\leq \lceil A\equiv rec_{(\ell ^{\Re })}\rceil .$$

In addition, it is easy to see that
$$\lceil \gamma (atom(\Re )\cup r(A))\rceil \leq \lceil \gamma
(atom(\ell ^{\Re })\cup r(A))\rceil $$ from Lemma 2.6. Therefore,
it follows that
$$\lceil \gamma (atom(\Re )\cup r(A))\rceil \wedge \lceil A\equiv
rec_{\Re }\rceil \leq \lceil \gamma (atom(\ell ^{\Re })\cup
r(A))\rceil \wedge \lceil A\equiv rec_{(\ell ^{\Re })}\rceil $$
$$\leq \vee \{\lceil \gamma (atom(\wp )\cup r(A))\rceil \wedge
\lceil A\equiv rec_{\wp }\rceil :\wp \in \mathbf{DA}(\Sigma ,\ell
)\},$$ and we complete the proof. $\heartsuit $

\smallskip\

It should be note that in the above corollary the second
conclusion is obtained from the first one by removing simply the
commutativity. The second conclusion is in general not correct.
The reason is that an essential application is needed in the
derivation of the implication $CReg_\Sigma\longrightarrow
CDReg_\Sigma.$

\bigskip\

\textbf{5. Orthomodular Lattice-Valued Automata with
$\varepsilon-$Moves}

\smallskip\

Automata with $\varepsilon-$moves are nondeterministic automata in
which transitions on the empty input $\varepsilon$ are included,
and they have the same power for accepting languages. In the
classical theory of automata, automata with $\varepsilon-$moves
are very convenient tools in building complex automata from simple
ones and in proving the closure properties of regular languages.
The aim of this section is to introduce an orthomodular
lattice-valued extension of automaton with $\varepsilon-$moves.
Let $\ell =<L,\leq ,\wedge ,\vee ,\perp ,0,1>$ be an orthomodular
lattice. Then an $\ell -$valued automaton with $\varepsilon
-$moves over $\Sigma$ is a quadruple $ \Re =<Q,I,T,\delta >$ in
which all components are the same as in an $\ell -$valued
automaton (without $\varepsilon -$moves), but the domain of the
quantum transition relation $\delta $ is changed to $Q\times
(\Sigma \cup \{\varepsilon \})\times Q;$ that is, $\delta $ is a
mapping from $Q\times (\Sigma \cup \{\varepsilon \})\times Q$ into
$L,$ where $\varepsilon $ stands for the empty string of input
symbols. Thus, in an $\ell -$valued automaton with $\varepsilon
-$moves, transitions of the form "$p\stackrel{\delta ,\varepsilon
}{\longrightarrow }q$" are allowed. So, $atom(\Re )$ contains the
atomic propositions "$p\stackrel{\delta ,\varepsilon
}{\longrightarrow }q$", and their truth values are given as
$\delta (p,\varepsilon ,q)$ for all $p,q\in Q.$

Now let $\Re =<Q,I,T,\delta >$ be an $\ell -$valued automaton with
$ \varepsilon -$moves. We put $$T_{\varepsilon }(Q,\Sigma
)=(Q(\Sigma \cup \{\varepsilon \}))^{\ast }Q=\cup _{n=0}^{\infty
\lbrack }(Q(\Sigma \cup \{\varepsilon \}))^{n}Q].$$ The difference
between $T(Q,\Sigma )$ and $T_{\varepsilon }(Q,\Sigma )$ is that
in the latter the empty string may be used as labels. For any $
c=q_{0}\sigma_{1} q_{1}...q_{k-1}\sigma _{k}q_{k}\in
T_{\varepsilon }(Q,\Sigma ), $ $lb_{\varepsilon }(c)$ is defined
to be the sequence $\sigma _{1}...\sigma _{k} $ with all
occurrences of $\varepsilon $ deleted. Note that it is possible
that the length of $lb_{\varepsilon }(c)$ is strictly smaller than
$k.$ Then the recognizability $ rec_{\Re }$ is also defined as an
$\ell -$valued unary predicate over $ \Sigma ^{\ast },$ and it is
given by
$$rec_{\Re }(s)\stackrel{def}{=}(\exists c\in T_{\varepsilon
}(Q,\Sigma ))(b(c)\in I\wedge e(c)\in T\wedge lb_{\varepsilon
}(c)=s\wedge path_{\Re }(c))$$ for all $s\in \Sigma ^{\ast },$
where $path_{\Re }$ is defined in the same way as in an $\ell
-$valued automaton without $\varepsilon -$moves. The defining
equation of $rec_\Re$ may be rewritten in terms of truth valued as
follows: $$\lceil rec_\Re (s)\rceil = \vee \{I(b(c))\wedge
T(e(c))\lceil path_\Re (c)\rceil : c\in T_\varepsilon (Q,\Sigma)\
{\rm and}\ lb_\varepsilon (c)=s\},$$ where $$\lceil path_\Re
(c)\rceil = \wedge_{i=0}^{k-1}\delta(q_i,\sigma_{i+1},q_{i+1})$$
if $c=q_0 \sigma_1 q_1 ... q_{k-1} \sigma_k q_k.$

For any $\ell -$valued automaton $\Re =<Q,I,T,\delta >$ with
$\varepsilon -$ moves, its $\varepsilon-$reduction is defined to
be the $\ell -$valued automaton $\Re ^{-\varepsilon
}=<Q,I,T^{\prime },\delta ^{\prime }>$ (without $\varepsilon -
$moves) in which

(i) for any $q\in Q,$
$$q\in T^{\prime }\stackrel{def}{=}(\exists q\in Q,m\geq 0)(q\in
T\wedge \delta (q_{0},\varepsilon ^{m},q));$$ that is,
$$T^{\prime }(q)=\vee _{q\in Q,m\geq 0}(T(q)\wedge \delta
(q_{0},\varepsilon ^{m},q));$$

(ii) for any $p,q\in Q$ and $\sigma \in \Sigma ,$ $$\delta
^{\prime }(p,\sigma ,q)\stackrel{def}{=}(\exists m,n\geq 0)\delta
(p,\varepsilon ^{m}\sigma \varepsilon ^{n},q);$$ that is,
$$\delta ^{\prime }(p,\sigma ,q)=\vee _{m,n\geq 0}\delta
(p,\varepsilon ^{m}\sigma \varepsilon ^{n},q),$$ where
$$\delta (q_{0},\sigma _{1}...\sigma
_{k},q_{k})\stackrel{def}{=}(\exists q_{1},...,q_{k-1}\in
Q)(\delta (q_{0},\sigma _{1},q_{1})\wedge \delta (q_{1},\sigma
_{2},q_{2})\wedge \delta (q_{k-1},\sigma _{k},q_{k})).$$ In other
words,
$$\delta (q_{0},\sigma _{1}...\sigma _{k},q_{k})=\vee \{(\delta
(q_{0},\sigma _{1},q_{1})\wedge \delta (q_{1},\sigma
_{2},q_{2})\wedge \delta (q_{k-1},\sigma
_{k},q_{k}):q_{1},...,q_{k-1}\in Q\}$$ for all $k\geq 1,$
$q_{0},q_{k}\in Q$ and $\sigma _{1},...,\sigma _{k}\in \Sigma .$

The following theorem gives a clear relation between the language
accepted by an orthomodular lattice-valued automaton with
$\varepsilon-$moves and that accepted by its
$\varepsilon-$reduction. In general, the $\varepsilon-$reduction
of an automaton with $\varepsilon-$moves has a stronger power of
acceptance than itself. A certain commutativity between basic
actions of the automaton implies the equivalence between an
automaton with $\varepsilon-$moves and its
$\varepsilon-$reduction. However, an universal validity of such an
equivalence requires that the underlying logic degenerates to the
classical Boolean logic.

\smallskip\

\textbf{Theorem 5.1.} Let $\ell =<L,\leq ,\wedge ,\vee ,\bot
,0,1>$ be an orthomodular lattice, and let $\longrightarrow$ be an
implication operator satisfying the Birkhoff-von Neumann
requirement.

(1) For any $\ell -$valued automaton $\Re $ with $\varepsilon
-$moves over $ \Sigma ,$ and for any $s\in \Sigma ^{\ast },$
$$\stackrel{\ell }{\models }rec_{\Re }(s)\rightarrow rec_{\Re
^{-\varepsilon }}(s).$$

(2) For any $\ell -$valued automaton $\Re $ with $\varepsilon
-$moves over $ \Sigma ,$ and for any $s\in \Sigma ^{\ast },$
$$\stackrel{\ell }{\models }\gamma (atom(\Re ))\wedge rec_{\Re
^{-\varepsilon }}(s)\rightarrow rec_{\Re }(s),$$ and in particular
if $\rightarrow =\rightarrow _{3}$ then
$$\stackrel{\ell }{\models }\gamma (atom(\Re ))\rightarrow (rec_{\Re
}(s)\leftrightarrow rec_{\Re ^{-\varepsilon }}(s)).$$

(3) The following two statements are equivalent:

\ \ \ \ \ \ (i) $\ell $ is a Boolean algebra;

\ \ \ \ \ \ (ii) For all $\ell -$valued automaton $\Re $ with
$\varepsilon -$moves over $ \Sigma ,$ and for all $s\in \Sigma
^{\ast },$
$$\stackrel{\ell }{\models }rec_{\Re }(s)\leftrightarrow rec_{\Re
^{-\varepsilon }}(s).$$

\smallskip\

\textbf{Proof.} The proof of (1) is similar to that of (2), so we
omit it. We now prove (2). First, we use induction on the length
$|c|$ of $c$ to show that for any $c\in T(Q,\Sigma ),$
$$claim:\ \ \lceil \gamma (atom(\Re )\rceil \wedge \lceil path_{\Re
^{-\varepsilon }}(c)\rceil \leq \vee \{\lceil path_{\Re
}(c^{\prime })\rceil :c^{\prime }\in T_{\varepsilon }(Q,\Sigma
),b(c^{\prime })=b(c),\ \ \ \ \ \ \ \ \ \ $$
$$\ \ \ \ \ \ \ \ \ \ \ \ \ \ \ \ \ \ \ \ \ \ \ e(c^{\prime })=e(c)\ {\rm and}\ lb_{\varepsilon }(c^{\prime
})=lb(c)\}.$$ For the case of $|c|=1,$ it is immediate from the
definition of transition relation $\delta ^{\prime }$ in $\Re
^{-\varepsilon }.$ If $c=c^{\prime }\sigma q,$ then with the
induction hypothesis and Lemmas 2.5 and 2.6, we have
$$\lceil \gamma (atom(\Re ))\rceil \wedge \lceil path_{\Re
^{-\varepsilon }}(c)\rceil =\lceil \gamma (atom(\Re ))\rceil
\wedge \lceil path_{\Re ^{-\varepsilon }}(c^{\prime })\rceil
\wedge \delta ^{\prime }(e(c^{\prime }),\sigma ,q)$$
$$=\lceil \gamma (atom(\Re ))\rceil \wedge \lceil \gamma (atom(\Re
))\rceil \wedge \lceil path_{\Re ^{-\varepsilon }}(c^{\prime
})\rceil \wedge \vee _{m,n\geq 0}\delta (e(c^{\prime
}),\varepsilon ^{m}\sigma \varepsilon ^{n},q) $$
$$\leq \vee _{m,n\geq 0}(\lceil \gamma (atom(\Re ))\rceil \wedge
\lceil path_{\Re ^{-\varepsilon }}(c^{\prime })\rceil \wedge
\delta (e(c^{\prime }),\varepsilon ^{m}\sigma \varepsilon
^{n},q))$$
$$\leq \vee _{m,n\geq 0}[\lceil \gamma (atom(\Re ))\rceil \wedge
\vee \{\lceil path_{\Re }(d^{\prime })\rceil :d^{\prime }\in
T_{\varepsilon }(Q,\Sigma ),b(d^{\prime })=b(c^{\prime
}),e(d^{\prime })=e(c^{\prime })$$ $${\rm and}\ lb_{\varepsilon
}(d^{\prime })=lb(c^{\prime })\}\wedge \delta (e(c^{\prime
}),\varepsilon ^{m}\sigma \varepsilon ^{n},q)]$$
$$\leq \vee \{\lceil \gamma (atom(\Re ))\rceil \wedge \lceil
path_{\Re }(d^{\prime })\rceil \wedge \delta (e(c^{\prime
}),\varepsilon ^{m}\sigma \varepsilon ^{n},q):m,n\geq 0,d^{\prime
}\in T_{\varepsilon }(Q,\Sigma ),$$
$$b(d^{\prime })=b(c^{\prime }),e(d^{\prime })=e(c^{\prime
}),lb_{\varepsilon }(d^{\prime })=lb(c^{\prime })\}.$$
Furthermore, we know that
$$\delta (e(c^{\prime }),\varepsilon ^{m}\sigma \varepsilon
^{n},q)=\vee \{\delta (e(c^{\prime }),\varepsilon ,p_{1})\wedge
\delta (p_{1},\varepsilon ,p_{2})\wedge ...\wedge \delta
(p_{m-1},\varepsilon ,p_{m})\wedge \delta (p_{m},\sigma ,q_{n})$$
$$\wedge \delta (q_{n},\varepsilon ,q_{n-1})\wedge ...\wedge \delta
(q_{2},\varepsilon ,q_{1})\wedge \delta (q_{1},\varepsilon
,q):p_{1},...,p_{m},q_{1},...,q_{n}\in Q\}.$$ Again, we use Lemmas
2.5 and 2.6 and obtain
$$\lceil \gamma (atom(\Re ))\rceil \wedge \lceil path_{\Re
^{-\varepsilon }}(c)\rceil \leq \vee \{\lceil path_{\Re
}(d^{\prime })\rceil \wedge \delta (e(c^{\prime }),\varepsilon
,p_{1})\wedge \delta (p_{1},\varepsilon ,p_{2})\wedge ...$$
$$\wedge \delta (p_{m-1},\varepsilon ,p_{m})\wedge \delta
(p_{m},\sigma ,q_{n})\wedge \delta (q_{n},\varepsilon
,q_{n-1})\wedge ...\wedge \delta (q_{2},\varepsilon ,q_{1})\wedge
\delta (q_{1},\varepsilon ,q):$$ $$m,n\geq 0,d^{\prime }\in
T_{\varepsilon }(Q,\Sigma )\ {\rm with}\ b(d^{\prime
})=b(c^{\prime }),e(d^{\prime })=e(c^{\prime })$$ $${\rm and}\
lb_{\varepsilon }(d^{\prime })=lb(c^{\prime
}),p_{1},...,p_{m},q_{1},...,q_{n}\in Q\}.$$ We put $d=d^{\prime
}\varepsilon p_{1}\varepsilon p_{2}...p_{m-1}\varepsilon
p_{m}\sigma q_{n}\varepsilon q_{n-1}...q_{2}\varepsilon
q_{1}\varepsilon q.$ Then $b(d)=b(d^{\prime })=b(c^{\prime }),$
$e(d)=q=e(c),$ $lb_{\varepsilon }(d)=lb_{\varepsilon }(d^{\prime
})\sigma =lb(c^{\prime })\sigma =lb(c),$ and
$$\lceil path_{\Re }(d)\rceil =\lceil path_{\Re }(d^{\prime
})\rceil \wedge \delta (e(c^{\prime }),\varepsilon ,p_{1})\wedge
\delta (p_{1},\varepsilon ,p_{2})\wedge ...$$
$$\wedge \delta (p_{m-1},\varepsilon ,p_{m})\wedge \delta
(p_{m},\sigma ,q_{n})\wedge \delta (q_{n},\varepsilon
,q_{n-1})\wedge ...\wedge \delta (q_{2},\varepsilon ,q_{1})\wedge
\delta (q_{1},\varepsilon ,q).$$ Therefore,
$$\lceil \gamma (atom(\Re ))\rceil \wedge \lceil path_{\Re
^{-\varepsilon }}(c)\rceil \leq \vee \{\lceil path_{\Re }(d)\rceil
:d\in T_{\varepsilon }(Q,\Sigma ),b(d)=b(c),$$ $$e(d)=e(c)\ {\rm
and}\ lb_{\varepsilon }(d)=lb(c)\}$$ and the claim holds for the
case of $|c|=|c^{\prime }|+1.$

Now it follows from the above claim and Lemmas 2.5 and 2.6 that
$$\lceil \gamma (atom(\Re ))\rceil \wedge \lceil rec_{\Re
^{-\varepsilon }}(s)\rceil =\lceil \gamma (atom(\Re ))\rceil
\wedge \lceil \gamma (atom(\Re ))\rceil \wedge \vee
\{I(b(c))\wedge T^{\prime }(e(c))$$
$$\wedge \lceil path_{\Re ^{-\varepsilon }}(c)\rceil :c\in
T(Q,\Sigma ),lb(c)=s\}$$
$$\leq \vee \{\lceil \gamma (atom(\Re ))\rceil \wedge I(b(c))\wedge
T^{\prime }(e(c))\wedge \lceil path_{\Re ^{-\varepsilon
}}(c)\rceil :c\in T(Q,\Sigma ),lb(c)=s\}$$
$$\leq \vee \{\lceil \gamma (atom(\Re ))\rceil \wedge I(b(c))\wedge
T^{\prime }(e(c))\wedge \vee \{\lceil path_{\Re }(c^{\prime
})\rceil :c^{\prime }\in T_{\varepsilon }(Q,\Sigma ),b(c^{\prime
})=b(c),$$
$$e(c^{\prime })=e(c)\ {\rm and}\ lb_{\varepsilon }(c^{\prime
})=lb(c)\}:c\in T(Q,\Sigma ),lb(c)=s\}$$
$$\leq \vee \{\lceil \gamma (atom(\Re ))\rceil \wedge I(b(c))\wedge
T^{\prime }(e(c))\wedge \lceil path_{\Re }(c^{\prime })\rceil
:c\in T(Q,\Sigma ),$$
$$c^{\prime }\in T_{\varepsilon }(Q,\Sigma ),b(c^{\prime
})=b(c),e(c^{\prime })=e(c)\ {\rm and}\ lb_{\varepsilon
}(c^{\prime })=lb(c)=s\}$$
$$=\vee \{\lceil \gamma (atom(\Re ))\rceil \wedge I(b(c))\wedge
\vee _{m\geq 0}(T(q)\wedge \delta (e(c),\varepsilon ^{m},q))\wedge
\lceil path_{\Re }(c^{\prime })\rceil :c\in T(Q,\Sigma ),$$
$$c^{\prime }\in T_{\varepsilon }(Q,\Sigma ),b(c^{\prime
})=b(c),e(c^{\prime })=e(c)\ {\rm and}\ lb_{\varepsilon
}(c^{\prime })=lb(c)=s\}$$ $$\leq \vee \{\lceil \gamma (atom(\Re
))\rceil \wedge I(b(c))\wedge T(q)\wedge \lceil path_{\Re
}(c^{\prime })\rceil \wedge \delta (e(c),\varepsilon ^{m},q):m\geq
0,c\in T(Q,\Sigma ),$$
$$c^{\prime }\in T_{\varepsilon }(Q,\Sigma ),b(c^{\prime
})=b(c),e(c^{\prime })=e(c)\ {\rm and}\ lb_{\varepsilon
}(c^{\prime })=lb(c)=s\}$$
$$\leq \vee \{\lceil \gamma (atom(\Re ))\rceil \wedge I(b(c))\wedge
T(q)\wedge \lceil path_{\Re }(c^{\prime })\rceil \wedge \delta
(e(c),\varepsilon ,q_{1})\wedge \delta (q_{1},\varepsilon
,q_{2})\wedge ...$$
$$\wedge \delta (q_{m-2},\varepsilon ,q_{m-1})\wedge \delta
(q_{m-1},\varepsilon q):m\geq 0,c\in T(Q,\Sigma ),c^{\prime }\in
T_{\varepsilon }(Q,\Sigma ),b(c^{\prime })=b(c),$$
$$e(c^{\prime })=e(c),lb_{\varepsilon }(c^{\prime
})=lb(c)=s,q_{1},...,q_{m-1}\in Q\}.$$ If we write $d=c^{\prime
}\varepsilon q_{1}\varepsilon q_{2}...q_{m-2}\varepsilon
q_{m-1}\varepsilon q,$ then $b(d)=b(c^{\prime })=b(c),$ $e(d)=q,$
$lb_{\varepsilon }(d)=lb(c^{\prime })=s$ and
$$\lceil path_{\Re }(d)\rceil =\lceil path_{\Re }(c^{\prime
})\rceil \wedge \delta (e(c),\varepsilon ,q_{1})\wedge \delta
(q_{1},\varepsilon ,q_{2})\wedge ...\wedge \delta
(q_{m-2},\varepsilon ,q_{m-1})\wedge \delta (q_{m-1},\varepsilon
,q).$$ Thus,
$$\lceil \gamma (atom(\Re ))\rceil \wedge \lceil rec_{\Re
^{-\varepsilon }}(s)\rceil \leq \vee \{I(b(d))\wedge T(e(d))\wedge
\lceil path_{\Re }(d)\rceil :d\in T_{\varepsilon }(Q,\Sigma
),lb_{\varepsilon }(d)=s\}$$
$$=\lceil rec_{\Re }(s)\rceil .$$

For (3), the part from (i) to (ii) is immediate from (2) by noting
that $ \lceil \gamma (atom(\Re ))\rceil =1$ always holds in a
Boolean algebra $\ell .$ Conversely, we demonstrate that (ii)
implies (i). For any $a,b,c\in L,$ consider $\ell -$valued
automaton $\Re
=<\{q_{0},q_{1},...,q_{5}\},\{q_{0}\},\{q_{5}\},\delta >$ with
$\varepsilon - $moves in which $\sigma \in \Sigma ,$ $\delta
(q_{0},\sigma ,q_{1})=a,$ $ \delta (q_{1},\varepsilon
,q_{2})=\delta (q_{1},\varepsilon ,q_{3})=\delta (q_{4},\sigma
,q_{5})=1,$ $\delta (q_{2},\varepsilon ,q_{4})=b,$ $\delta
(q_{3},\varepsilon ,q_{4})=c,$ and $\delta $ takes values $0$ for
other arguments (see Figure 4).

\begin{figure}\centering
\includegraphics{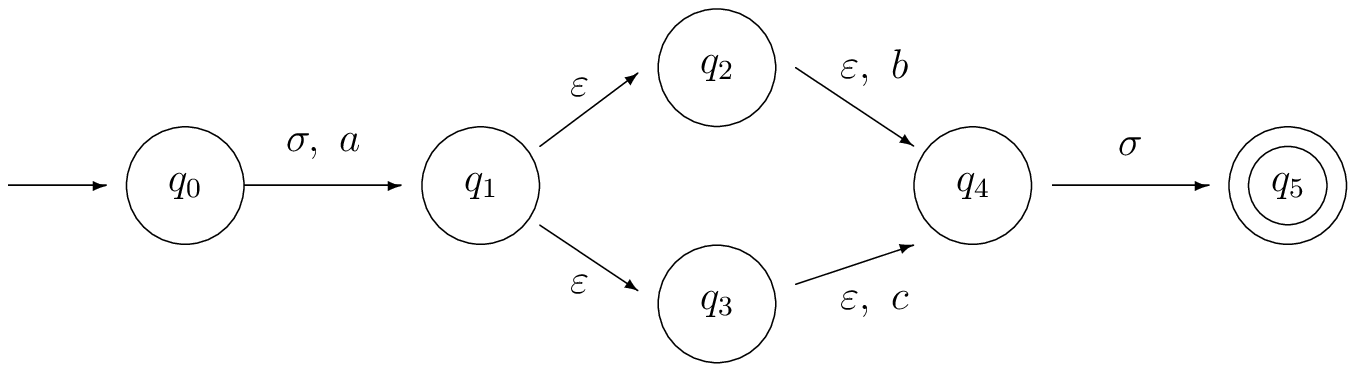}\caption{Automaton b} \label{fig 4}
\end{figure}

By a routine calculation we know that its $\varepsilon -$reduction
is $\Re ^{-\varepsilon
}=<\{q_{0},q_{1},...,q_{5}\},\{q_{0}\},\{q_{5}\},\delta ^{\prime
}>$ where $\delta ^{\prime }(q_{0},\sigma ,q_{1})=\delta ^{\prime
}(q_{0},\sigma ,q_{2})=\delta ^{\prime }(q_{0},\sigma ,q_{3})=a,$
$\delta ^{\prime }(q_{0},\sigma ,q_{4})=(a\wedge b)\vee (a\wedge
c),$ $\delta ^{\prime }(q_{1},\sigma ,q_{5})=b\vee c,$ $\delta
^{\prime }(q_{2},\sigma ,q_{5})=b,$ $\delta ^{\prime
}(q_{3},\sigma ,q_{5})=c,$ $\delta ^{\prime }(q_{4},\sigma
,q_{5})=1,$ and $\delta $ takes value $0$ for other arguments (see
Figure 5). Then it follows from (ii) that
$$a\wedge (b\vee c)=[a\wedge (b\vee c)]\vee (a\wedge b)\vee
(a\wedge c)\vee \lbrack (a\wedge b)\vee (a\wedge c)]$$
$$=\lceil rec_{\Re ^{-\varepsilon }}(\sigma \sigma )\rceil $$
$$=\lceil rec_{\Re }(\sigma \sigma )\rceil $$
$$=(a\wedge b)\vee (a\wedge c).$$
This shows that $\ell $ enjoys the distributivity of meet over
union, and it is a Boolean algebra.$\heartsuit$

\begin{figure}\centering
\includegraphics[width=1\textwidth]{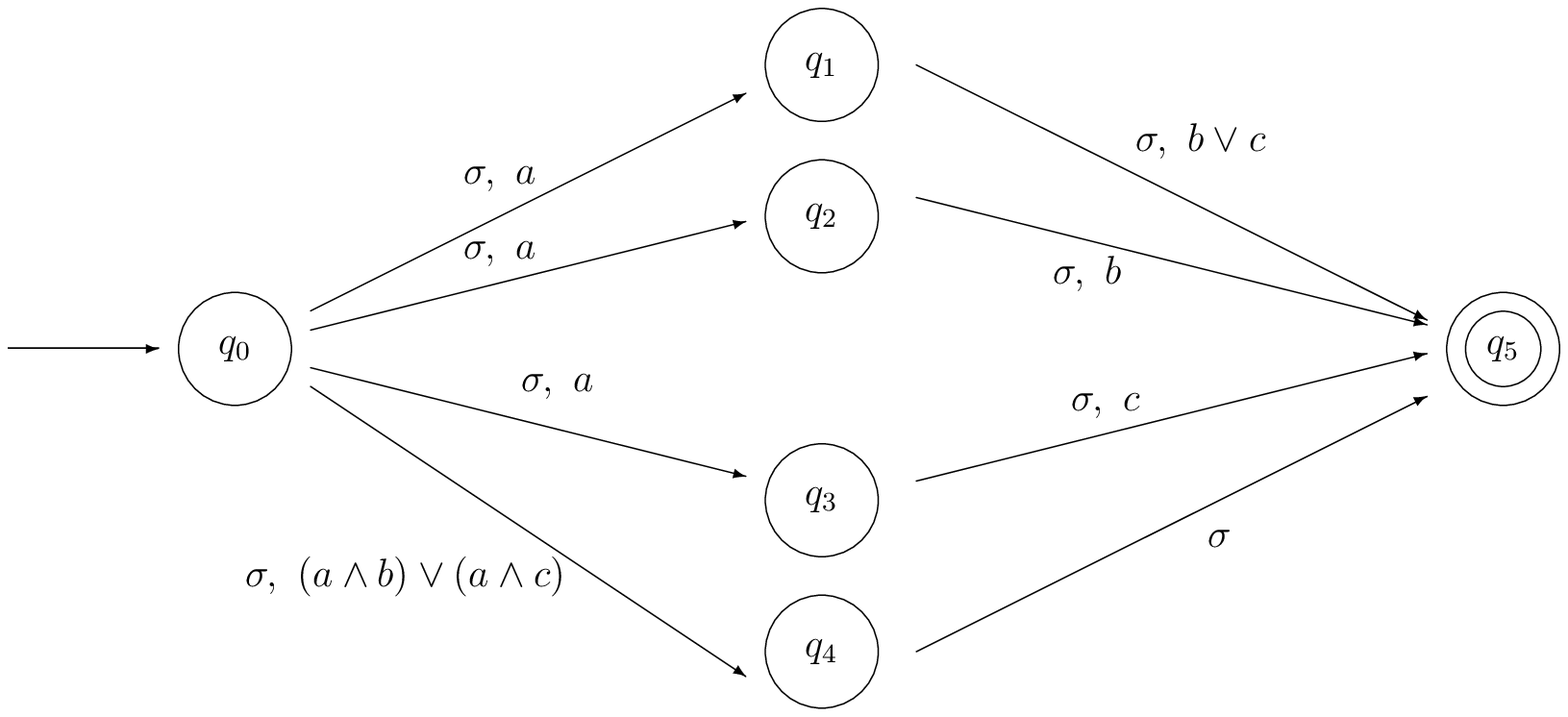}\caption{Automaton c} \label{fig 5}
\end{figure}

\bigskip\

\textbf{6. Closure Properties of Orthomodular Lattice-Valued
Regularity}

\smallskip\

It was shown in the classical automata theory that the class of
regular languages is closed under various operations such as
union, intersection, complement, concatenation, the Kleene
closure, substitution and homomorphism. In this section, we are
going to examine the closure properties of orthomodular
lattice-valued languages under these operations. We first consider
the inverse of an $\ell-$valued language. Let $A\in L^{\Sigma
^{*}}.$ Then the inverse $A^{-1}\in L^{\Sigma ^{*}}$ of $ A$ is
defined as follows:
$$A^{-1}(\sigma _1...\sigma _m)=A(\sigma _m...\sigma _1)$$ for any
$m\in \omega $ and for any $\sigma _1,...,\sigma _m\in \Sigma .$

The following proposition shows that both regularity and
commutative regularity are preserved by the inverse operation.

\smallskip\

\textbf{Proposition 6.1.} Let $\ell=<L,\leq ,\wedge ,\vee ,\bot
,0,1>$ be a complete orthomodular lattice, and let
$\longrightarrow $ fulfil the property that $a\leftrightarrow a=1$
for any $a\in L$. Then for any $A\in L^{\Sigma ^{*}},$
$$\stackrel{\ell }{ \models }Reg_\Sigma (A)\leftrightarrow
Reg_\Sigma (A^{-1}),$$ and
$$\stackrel{\ell }{ \models }CReg_\Sigma (A)\leftrightarrow
CReg_\Sigma (A^{-1}).$$

\smallskip\

\textbf{Proof}. Noting that $A=(A^{-1})^{-1},$ it suffices to show
that $$ \lceil Reg_\Sigma (A)\rceil \leq \lceil Reg_\Sigma
(A^{-1})\rceil .$$ For any $\ell -$valued automaton $\Re
=(Q,I,T,\delta ),$ we define the inverse of $ \Re $ to be the
$\ell -$valued automaton $\Re ^{-1}=(Q,T,I,\delta ^{-1}),$ where
$\delta ^{-1}(p,\sigma ,q)=\delta (q,\sigma ,p)$ for any $p,q\in
Q$ and $\sigma \in \Sigma .$ Then it is easy to see that $rec_{\Re
^{-1}}=(rec_\Re )^{-1},$ and furthermore we have
$$\lceil
Reg_\Sigma (A)\rceil =\vee \{\lceil A\equiv rec_\Re \rceil :\Re
\in \mathbf{A}(\Sigma ,\ell )\}$$
$$=\vee \{\lceil A^{-1}\equiv
(rec_\Re )^{-1}\rceil :\Re \in \mathbf{A} (\Sigma ,\ell )\}$$
$$=\vee \{\lceil A^{-1}\equiv rec_{\Re ^{-1}}\rceil :\Re \in
\mathbf{A} (\Sigma ,\ell )\}$$
$$\leq \vee \{\lceil A^{-1}\equiv
rec_\wp \rceil :\wp \in \mathbf{A}(\Sigma ,\ell )\}=\lceil
Rec_\Sigma (A^{-1})\rceil .$$. The proof for commutative
regularity is similar. $\heartsuit$

\smallskip\

The commutative regularity is preserved by the complement
operation, but it is not the case for the regularity predicate.

\smallskip\

\textbf{Proposition 6.2.} If $\ell =<L,\leq ,\wedge ,\vee ,\perp
,0,1>$ is a finite orthomodular lattice, and $\rightarrow
=\rightarrow _{3},$ then for any $A\in L^{\Sigma ^{\ast }},$
$$\stackrel{\ell }{\models }CReg_{\Sigma }(A)\rightarrow
CReg_{\Sigma }(A^{c}).$$

\smallskip\

\textbf{Proof.} For any $\Re =<Q,I,T,\delta >\in \mathbf{A}(\Sigma
,\ell ),$ we observe that $\ell ^{\Re
}=<L^{Q},I_{1},\overline{T},\overline{\delta }>$ is an $\ell
-$valued deterministic automaton and only $\overline{T}$ carries
$\ell -$valued information. Then we set $(\ell ^{\Re
})^{c}=<L^{Q},I_{1}, \overline{T}^{c},\overline{\delta }>,$ where
for any $X\in L^{Q},$ $
\overline{T}^{c}(X)=(\overline{T}(X))^{c}.$ It is easy to see that
for all $ s\in \Sigma ^{\ast },$ $rec_{(\ell ^{\Re
})^{c}}(s)=(rec_{(\ell ^{\Re })}(s))^{\perp }.$

Now by using Theorem 4.1 and Lemmas 2.5 and 2.6 we obtain
$$\lceil \gamma (atom(\Re )\cup r(A)\rceil \wedge \lceil A\equiv
rec_{\Re }\rceil \leq \lceil \gamma (atom(\Re )\cup r(A)\rceil
\wedge \lceil A\equiv rec_{\Re }\rceil \wedge \lceil rec_{\Re
}\equiv rec_{(\ell ^{\Re })}\rceil $$
$$=\wedge _{s\in \Sigma ^{\ast }}(\lceil \gamma (atom(\Re )\cup
r(A)\rceil \wedge (A(s)\rightarrow rec_{\Re }(s))\wedge (rec_{\Re
}(s)\rightarrow rec_{(\ell ^{\Re })}(s)))\wedge $$
$$\wedge _{s\in \Sigma ^{\ast }}(\lceil \gamma (atom(\Re )\cup
r(A)\rceil \wedge (rec_{(\ell ^{\Re })}(s)\rightarrow rec_{\Re
}(s))\wedge (rec_{\Re }(s)\rightarrow A(s)))$$
$$\leq \wedge _{s\in \Sigma ^{\ast }}(\lceil \gamma (atom(\Re )\cup
r(A)\rceil \wedge (A(s)\rightarrow rec_{(\ell ^{\Re })}(s)))\wedge
$$
$$\wedge _{s\in \Sigma ^{\ast }}(\lceil \gamma (atom(\Re )\cup
r(A)\rceil \wedge (rec_{(\ell ^{\Re })}(s)\rightarrow A(s))).$$
Then Lemma 2.11(2) yields
$$\lceil \gamma (atom(\Re )\cup r(A)\rceil \wedge \lceil A\equiv
rec_{\Re }\rceil \leq \wedge _{s\in \Sigma ^{\ast }}(rec_{(\ell
^{\Re })}^{\perp }(s)\rightarrow A^{\perp }(s))\wedge \wedge
_{s\in \Sigma ^{\ast }}(A^{\perp }(s)\rightarrow rec_{(\ell ^{\Re
})}^{\perp }(s))$$
$$=\wedge _{s\in \Sigma ^{\ast }}(A^{\perp }(s)\leftrightarrow
rec_{(\ell ^{\Re })}^{\perp }(s))$$
$$=\wedge _{s\in \Sigma ^{\ast }}(A^{\perp }(s)\leftrightarrow
rec_{(\ell ^{\Re })^{c}}(s))$$
$$=\lceil A^{c}\equiv rec_{(\ell ^{\Re })^{c}}\rceil .$$

In addition, we have $$\lceil \gamma (atom(\Re )\cup r(A)\rceil
\leq \lceil \gamma (atom(\Re )\cup r(A^{c})\rceil $$ from Lemma
2.5. Therefore,
$$\lceil \gamma (atom(\Re )\cup r(A)\rceil \wedge \lceil A\equiv
rec_{\Re }\rceil \leq \lceil \gamma (atom(\Re )\cup r(A^{c})\rceil
\wedge \lceil A^{c}\equiv rec_{(\ell ^{\Re })^{c}}\rceil $$
$$\leq \lceil CReg_{\Sigma }(A^{c})\rceil .$$

Finally, since $\Re $ is allowed to be arbitrary, it follows that
$$\lceil CReg_{\Sigma }(A)\rceil =\vee _{\Re \in
\mathbf{A}(\Sigma ,\ell )}\lceil \gamma (atom(\Re )\cup r(A)\rceil
\wedge \lceil A\equiv rec_{\Re }\rceil $$
$$\leq \lceil CReg_{\Sigma }(A^{c})\rceil .\heartsuit$$

\smallskip\

We now turn to deal with the union of two $\ell-$valued language.
Let $\Re =<Q_{A},I_{A},T_{A},\delta _{A})$ and $\wp
=<Q_{B},I_{B},T_{B},\delta _{B}>\in \mathbf{A}(\Sigma ,\ell )$ be
two $\ell - $valued automata over $\Sigma .$ We assume that
$Q_{A}\cap Q_{B}=\phi .$ Then the (disjoint) union $\Re \cup \wp $
of $\Re $ and $\wp $ is defined to be $\Im
=(Q_{C},I_{C},T_{C},\delta _{C}),$ where:

(i) $Q_{C}=Q_{A}\cup Q_{B};$

(ii) $I_{C}=I_{A}\cup I_{B};$

(iii) $T_{C}=T_{A}\cup T_{B};$ and

(iv) $\delta _{C}:Q_{C}\times \Sigma \times Q_{C}\longrightarrow
L$ is given as follows: for any $p,q\in Q_{C}$ and $\sigma \in
\Sigma ,$
$$\delta _{C}(p,\sigma ,q)=\left\{
\begin{array}{c}
\delta _{A}(p,\sigma ,q)\ {\rm if }\ p,q\in Q_{A}, \\
\delta _{B}(p,\sigma ,q)\ {\rm if }\ p,q\in Q_{B}, \\
0\ {\rm otherwise.}
\end{array}
\right. $$

The following proposition describes the recognizability of the
union of two $\ell-$valued automata. As in the classical theory, a
word $s$ in $\Sigma^{\ast}$ is recognized by the union of two
$\ell-$valued automata if and only if $s$ is recognized by one of
them.

\smallskip\

\textbf{Proposition 6.3.} Let $\ell=<L,\leq ,\wedge ,\vee ,\bot
,0,1>$ be a complete orthomodular lattice. If the implication
operator $\longrightarrow $ satisfies that $a\leftrightarrow a=1$
for any $a\in L,$ then for any $\Re ,\wp \in \mathbf{A(\Sigma
,\ell )}$ and for any $s\in \Sigma ^{\ast },$
$$\stackrel{\ell }{ \models } rec_{\Re \cup \wp
}(s)\longleftrightarrow rec_{\Re }(s)\vee rec_{\wp }(s).$$

\smallskip\

\textbf{Proof}. Let $s=\sigma _{1}...\sigma _{k}.$ Then
$$\lceil rec_{\Re \cup \wp }(s)\rceil =\vee \{(I_{A}\cup
I_{B})(q_{0})\wedge (T_{A}\cup T_{B})(q_{k})\wedge \wedge
_{i=0}^{k-1}\delta _{A\cup B}(q_{i},\sigma
_{i+1},q_{i+1}):q_{0},q_{1},...,q_{k}\in Q_{A}\cup Q_{B}\}$$
$$=[\vee \{(I_{A}\cup I_{B})(q_{0})\wedge (T_{A}\cup
T_{B})(q_{k})\wedge \wedge _{i=0}^{k-1}\delta _{A\cup
B}(q_{i},\sigma _{i+1},q_{i+1}):q_{0},q_{1},...,q_{k}\in
Q_{A}\}]$$
$$\ \ \ \ \vee \lbrack \vee \{(I_{A}\cup I_{B})(q_{0})\wedge (T_{A}\cup
T_{B})(q_{k})\wedge \wedge _{i=0}^{k-1}\delta _{A\cup
B}(q_{i},\sigma _{i+1},q_{i+1}):q_{0},q_{1},...,q_{k}\in
Q_{B}\}]$$
$$\ \ \ \ \ \ \vee \lbrack \vee \{(I_{A}\cup I_{B})(q_{0})\wedge (T_{A}\cup
T_{B})(q_{k})\wedge \wedge _{i=0}^{k-1}\delta _{A\cup
B}(q_{i},\sigma _{i+1},q_{i+1}):q_{0},q_{1},...,q_{k}\in Q_{A}\cup
Q_{B},$$ $${\rm and\ there\ are}\ i,j\ {\rm such\ that}\ 0\leq
i,j\leq k\ {\rm and}\ q_{i}\in Q_{A}\ {\rm and}\ q_{j}\in
Q_{B}\}].$$ From the definition of $\Re \cup \wp ,$ we know that
for any $ q_{0},q_{1},...,q_{k}\in Q_{A},$ $$(I_{A}\cup
I_{B})(q_{0})=I_{A}(q_{0}),$$ $$(T_{A}\cup
T_{B})(q_{k})=T_{A}(q_{k}),$$
$$\wedge _{i=0}^{k-1}\delta _{A\cup B}(q_{i},\sigma
_{i+1},q_{i+1})=\wedge _{i=0}^{k-1}\delta _{A}(q_{i},\sigma
_{i+1},q_{i+1}),$$ and for any $q_{0},q_{1},...,q_{k}\in Q_{B},$
$$(I_{A}\cup I_{B})(q_{0})=I_{B}(q_{0}),$$
$$(T_{A}\cup T_{B})(q_{k})=T_{B}(q_{k}),$$
$$\wedge _{i=0}^{k-1}\delta _{A\cup B}(q_{i},\sigma
_{i+1},q_{i+1})=\wedge _{i=0}^{k-1}\delta _{B}(q_{i},\sigma
_{i+1},q_{i+1}).$$ If $q_{0},q_{1},...,q_{k}\in Q_{A}\cup Q_{B},$
and there are $i,j$ such that $0\leq i,j\leq k$ and $q_{i}\in
Q_{A}$ and $q_{j}\in Q_{B},$ then we can find some $m\in
\{0,1,...,k-1\}$ such that $q_{m}\in Q_{A}$ and $q_{m+1}\in
Q_{B},$ or $q_{m}\in Q_{B}$ and $q_{m+1}\in Q_{A}.$ Then $\delta
_{A\cup B}(q_{m},\sigma _{m+1},q_{m+1})=0,$ and $$\wedge
_{i=0}^{k-1}\delta _{A\cup B}(q_{i},\sigma _{i+1},q_{i+1})=0.$$
Therefore, it follows that
$$\lceil rec_{\Re \cup \wp }(s)\rceil =[\vee \{I_{A}(q_{0})\wedge
T_{A}(q_{k})\wedge \wedge _{i=0}^{k-1}\delta _{A}(q_{i},\sigma
_{i+1},q_{i+1}):q_{0},q_{1},...,q_{k}\in Q_{A}\}]$$
$$\vee \lbrack \vee \{I_{B}(q_{0})\wedge T_{B}(q_{k})\wedge \wedge
_{i=0}^{k-1}\delta _{B}(q_{i},\sigma
_{i+1},q_{i+1}):q_{0},q_{1},...,q_{k}\in Q_{B}\}]$$
$$=\lceil rec_{\Re }(s)\vee rec_{\wp }(s)\rceil .\heartsuit $$

\smallskip\

The following corollary slightly generalizes Example 3.1.

\smallskip\

\textbf{Corollary 6.4}. If $Range(A)=\{A(s):s\in \Sigma^{\ast}\}$
is finite, and $A_\lambda =\{s\in\Sigma^{\ast}:A(s)\geq \lambda\}$
is a regular language (in classical automata theory) for every
$\lambda\in Range(A)$, then
$$\stackrel{\ell }{ \models }Reg_\Sigma (A).$$

\smallskip\

\textbf{Proof.} Suppose that
$Range(A)=\{\lambda_1,...,\lambda_n\}.$ Then it is easy to see
that
$$A=\cup_{i=1}^{n} \lambda_i A_{\lambda_i}.$$ From Example 3.3 we
know that there exists an $\ell-$valued automaton $\Re_i$ such
that $rec_{\Re_i}=\lambda_i A_{\lambda_i}$ for each $i\leq n$.
Thus, by proposition 6.3 we obtain $$rec_{\cup_{i=1}^{n} \Re_i}
=\cup_{i=1}^{n} \lambda_i A_{\lambda_i}=A$$ and complete the
proof. $\heartsuit$

\smallskip\

We now consider the product of two $\ell-$valued automata. Let
$\Re =<Q_{A},I_{A},T_{A},\delta _{A})$ and $\wp
=<Q_{B},I_{B},T_{B},\delta _{B}>\in \mathbf{A}(\Sigma ,\ell )$ be
two $\ell -$valued automata over $ \Sigma .$ Then their product
$\Re \times \wp $ is defined to be $\Im =(Q_{C},I_{C},T_{C},\delta
_{C}),$ where:

\smallskip\

(i) $Q_{C}=Q_{A}\times Q_{B};$

(ii) $I_{C}=I_{A}\times I_{B};$

(iii) $T_{C}=T_{A}\times T_{B};$ and

(iv) $\delta _{C}:Q_{C}\times \Sigma \times Q_{C}\longrightarrow
L$ and for any $p_{a},q_{a}\in Q_{A},$ $p_{b},q_{b}\in Q_{B}$ and
$\sigma \in \Sigma ,$
$$\delta _{C}((p_{a},p_{b}),\sigma ,(q_{a},q_{b}))=\delta
_{A}(p_{a},\sigma ,q_{a})\wedge \delta _{B}(p_{b},\sigma
,q_{b}).$$

\smallskip\

It is well-known in the classical automata theory that the
language accepted by the union of two automata is the union of the
languages accepted by these two automata, and the language
accepted by the product of two automata is the intersection of the
languages accepted by the factor automata. Proposition 6.3 shows
that the conclusion about the union of two automata can be
generalized into the theory of automata based on quantum logic
without appealing any additional condition. One may naturally
expect that the conclusion for product of automata can also be
easily generalized into the framework of quantum logic. However,
the case for the product of two automata is much more complicated,
and the following proposition tells us that in order to make the
above conclusion about product of automata still valid in the
automata theory based on quantum logic, a certain commutativity is
necessary to be added on the basic actions of the factor automata.

\smallskip\

\textbf{Proposition 6.5.} Let $\ell =<L,\leq ,\wedge ,\vee ,\perp
,0,1>$ be a complete orthomodular lattice.

(1) For any $\Re ,\wp \in \mathbf{A}(\Sigma ,\ell ),$ and for any
$s\in \Sigma ^{\ast },$
$$\stackrel{\ell }{\models }rec_{\Re \times \wp }(s)\longrightarrow
rec_{\Re }(s)\wedge rec_{\wp }(s).$$

(2) For any $\Re ,\wp \in \mathbf{A}(\Sigma ,\ell ),$ and for any
$s\in \Sigma ^{\ast },$
$$\stackrel{\ell }{\models }\gamma (atom(\Re )\cup atom(\wp ))\wedge
rec_{\Re }(s)\wedge rec_{\wp }(s)\longrightarrow rec_{\Re \times
\wp }(s),$$ and in particular if $\longrightarrow =\longrightarrow
_{3},$ then
$$\stackrel{\ell }{\models}\gamma (atom(\Re )\cup atom(\wp
))\longrightarrow (rec_{\Re }(s)\wedge rec_{\wp
}(s)\longleftrightarrow rec_{\Re \times \wp }(s)).$$

(3) The following two statements are equivalent:

\ \ \ \ \ \ (i) $\ell $ is a Boolean algebra.

\ \ \ \ \ \ (ii) for all $\Re ,\wp \in \mathbf{A}(\Sigma ,\ell ),$
and for all $s\in \Sigma ^{\ast },$
$$\stackrel{\ell }{\models }rec_{\Re }(s)\wedge rec_{\wp
}(s)\longleftrightarrow rec_{\Re \times \wp }(s).$$

\smallskip\

\textbf{Proof}. We have directly
$$\lceil rec_{\Re \times \wp }(s)\rceil =\vee \{(I_{A}\times
I_{B})(q_{a0},q_{b0})\wedge (T_{A}\times
T_{B})(q_{ak},q_{bk})\wedge \wedge _{i=0}^{k-1}\delta _{A\times
B}((q_{ai},q_{bi}),$$ $$\ \ \ \ \ \ \ \ \ \ \ \ \ \ \ \ \ \ \ \ \
\ \sigma _{i+1},(q_{a(i+1)},q_{b(i+1)})):
q_{a0},q_{a1},...,q_{ak}\in Q_{A}\ {\rm and}\
q_{b0},q_{b1},...,q_{bk}\in Q_{B}\}$$
$$=\vee \{I_{A}(q_{a0})\wedge I_{B}(q_{b0})\wedge
T_{A}(q_{ak})\wedge T_{B}(q_{bk})\wedge \wedge _{i=0}^{k-1}\delta
_{A}(q_{ai},\sigma _{i+1},q_{a(i+1)})\wedge$$
$$\ \ \ \ \ \ \ \ \ \ \ \ \ \ \ \ \wedge
_{i=0}^{k-1}\delta _{B}(q_{bi},\sigma _{i+1},q_{b(i+1)}):
q_{a0},q_{a1},...,q_{ak}\in Q_{A}\ {\rm and}\
q_{b0},q_{b1},...,q_{bk}\in Q_{B}\}, $$ and
$$\lceil rec_{\Re }(s)\wedge rec_{\wp }(s)\rceil =[\vee
\{I_{A}(q_{0})\wedge T_{A}(q_{k})\wedge \wedge _{i=0}^{k-1}\delta
_{A}(q_{i},\sigma _{i+1},q_{i+1}):q_{0},q_{1},...,q_{k}\in
Q_{A}\}]$$
$$\ \ \ \ \ \ \ \ \ \ \ \ \ \ \ \ \ \ \ \ \ \ \ \wedge \lbrack \vee \{I_{B}(q_{0})\wedge
T_{B}(q_{k})\wedge \wedge _{i=0}^{k-1}\delta _{B}(q_{i},\sigma
_{i+1},q_{i+1}):q_{0},q_{1},...,q_{k}\in Q_{B}\}]$$ from the
definitions of product and recognizability of $\ell -$valued
automata. It is clear that $$\lceil rec_{\Re \times \wp }(s)\rceil
\leq \lceil rec_{\Re }(s)\wedge rec_{\wp }(s)\rceil .$$ This
indicates that (1) is true. By using Lemmas 2.4(2), 2.5 and 2.6
twice, we can deduce that $$\lceil \gamma (atom(\Re )\cup atom(\wp
))\wedge rec_{\Re }(s)\wedge rec_{\wp }(s)\rceil \leq \lceil
rec_{\Re \times \wp }(s)\rceil .$$ Thus, (2) is proved. The first
part of (3) that (i) implies (ii) is immediately derivable from
(2) because we have $\lceil \gamma (atom(\Re )\cup atom(\wp
))\rceil =1$ provided $\ell $ is a Boolean algebra. Conversely, we
show that (ii) implies (i) by constructing two $\ell -$valued
automata and examining the behavior of their product. For any
$a,b,c\in L,$ we choose some $\sigma _{0}\in \Sigma $ and set
$$\Re =(\{p\},\{p\},\{p\},\delta _{A}),$$ where $\delta
_{A}(p,\sigma ,p)=a$ if $\sigma =\sigma _{0}$ and $0$ otherwise,
and $$\wp =(\{q,r,s\},\{q\},\{r,s\},\delta _{B}),$$ where $\delta
_{B}(x,\sigma ,y)=b$ if $x=q,$ $y=r,$ and $\sigma =\sigma _{0};$
$c$ if $x=q,$ $y=s,$ and $\sigma =\sigma _{0},$ $0$ otherwise.
Then $\Re ,\wp \in \mathbf{A}(\Sigma ,\ell ),$ and it is easy to
see that
$$\Re \times \wp
=(\{(p,q),(p,r),(p,s)\},\{(p,q)\},\{(p,r),(p,s)\},\delta _{A\times
B}),$$ where $\delta _{A\times B}((p,x),\sigma ,(p,y))=a\wedge b$
if $x=q,$ $y=r$ and $\sigma =\sigma _{0};$ $a\wedge c$ if $x=q,$
$y=s$ and $\sigma =\sigma _{0};$ and $0$ otherwise (see Figure 6).
Furthermore, by a routine calculation we have
$$\lceil rec_{\Re }(\sigma _{0})\rceil =a,$$
$$\lceil rec_{\wp }(\sigma _{0})\rceil =b\vee c,\ {\rm and}$$
$$\lceil rec_{\Re \times \wp }(\sigma _{0})\rceil =(a\wedge b)\vee
(a\wedge c).$$ Therefore, with (ii) we finally obtain
$$a\wedge b\vee c=\lceil rec_{\Re }(\sigma _{0})\rceil \wedge
\lceil rec_{\wp }(\sigma _{0})\rceil $$ $$=\lceil rec_{\Re \times
\wp }(\sigma _{0})\rceil =(a\wedge b)\vee (a\wedge c).\heartsuit$$

\begin{figure}\centering
\includegraphics[width=0.5\textwidth]{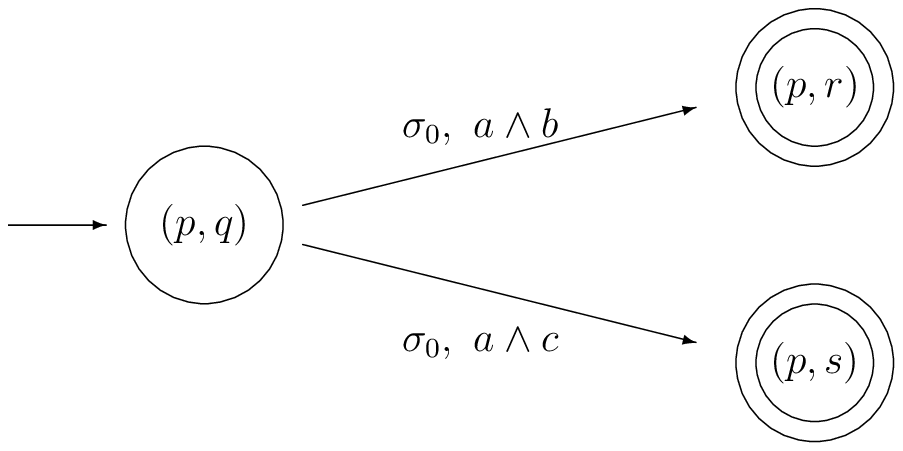}\caption{Automaton d} \label{fig 6}
\end{figure}

\smallskip\

To prove the closure property of orthomodular lattice-valued
regularity under the concatenation operation of languages, we
propose the concept of concatenation of two orthomodular
lattice-valued automata. Suppose that $\Re
_{1}=<Q_{1},I_{1},T_{1},\delta _{1}>,$ $\Re
_{2}=<Q_{2},I_{2},T_{2},\delta _{2}>\in \mathbf{A}(\Sigma ,\ell )$
be two $ \ell -$valued automata, and $Q_{1}\cap Q_{2}=\phi .$ We
define the concatenation of $\Re _{1}$ and $\Re _{2}$ to be $\ell
-$valued automaton $ \Re _{1}\Re _{2}=<Q_{1}\cup
Q_{2},I_{1},T_{2},\delta _{2}>$ with $ \varepsilon -$moves, where
$\delta :Q\times (\Sigma \cup \{\varepsilon \})\times Q\rightarrow
L$ is given by
$$\delta (p,\sigma ,q)=\left\{
\begin{array}{c}
\delta _{1}(p,\sigma ,q)\ {\rm if }\ p,q\in Q_{1}\ {\rm and }\
\sigma \neq
\varepsilon  \\
\delta _{2}(p,\sigma ,q)\ {\rm if }\ p,q\in Q_{2}\ {\rm and }\
\sigma \neq
\varepsilon  \\
T_{1}(p)\wedge I_{2}(q)\ {\rm if }\ p\in Q_{1},q\in Q_{2}\ {\rm
and }\ \sigma
=\varepsilon  \\
0\ {\rm otherwise.}
\end{array}
\right. $$

The following proposition clarifies the relation between the
language recognized by the concatenation of two orthomodular
lattice-valued automata and the concatenation of the languages
recognized by the two automata.

\smallskip\

\textbf{Proposition 6.6}. Let $\ell =<L,\leq ,\wedge ,\vee ,\perp
,0,1>$ be an orthomodular lattice and $\rightarrow $ fulfil the
Birkhoff-von Neumann requirement.

(1) For all $\Re _{1},\Re _{2}\in \mathbf{A}(\Sigma ,\ell ),$ and
for each $ s\in \Sigma ^{\ast },$
$$\stackrel{\ell }{\models }rec_{\Re _{1}\Re _{2}}(s)\rightarrow
(\exists s_{1},s_{2}\in \Sigma ^{\ast })(s_1 s_2=s\wedge rec_{\Re
_{1}}(s_{1})\wedge rec_{\Re _{2}}(s_{2})).$$

(2) For all $\Re _{1},\Re _{2}\in \mathbf{A}(\Sigma ,\ell ),$ and
for each $ s\in \Sigma ^{\ast },$
$$\stackrel{\ell }{\models }\gamma (atom(\Re _{1})\cup atom(\Re
_{2}))\wedge (\exists s_{1},s_{2}\in \Sigma ^{\ast })(s_1
s_2=s\wedge $$ $$rec_{\Re _{1}}(s_{1})\wedge rec_{\Re
_{2}}(s_{2}))\rightarrow rec_{\Re _{1}\Re _{2}}(s),$$ and if
$\rightarrow =\rightarrow _{3}$ then
$$\stackrel{\ell }{\models }\gamma (atom(\Re _{1})\cup atom(\Re
_{2}))\rightarrow (rec_{\Re _{1}\Re _{2}}(s)\leftrightarrow
(\exists s_{1},s_{2}\in \Sigma ^{\ast })$$ $$(s_1 s_2=s\wedge
rec_{\Re _{1}}(s_{1})\wedge rec_{\Re _{2}}(s_{2}))).$$

(3) The following two statements are equivalent:

\ \ \ \ \ \ (i) $\ell $ is a Boolean algebra;

\ \ \ \ \ \ (ii) for all $\Re _{1},\Re _{2}\in \mathbf{A}(\Sigma
,\ell ),$ and for each $ s\in \Sigma ^{\ast },$
$$\stackrel{\ell }{\models }rec_{\Re _{1}\Re _{2}}(s)\leftrightarrow
(\exists s_{1},s_{2}\in \Sigma ^{\ast })(s_1 s_2=s\wedge rec_{\Re
_{1}}(s_{1})\wedge rec_{\Re _{2}}(s_{2})).$$

\smallskip\

\textbf{Proof}. (1) For any $q_{0},q_{1},...,q_{m}\in Q_{1}\cup
Q_{2},$ $ \sigma _{1},...,\sigma _{m}\in \Sigma \cup \{\varepsilon
\}$ with $\sigma _{1}...\sigma _{m}=s$ (note that it is possible
that $|s|<m$ since $\sigma _{1},...,\sigma _{m}$ may contain
$\varepsilon$'s), if
$$I_{1}(q_{0})\wedge T_{2}(q_{m})\wedge \wedge _{i=1}^{m}\delta
(q_{i-1},\sigma _{i},q_{i})>0,$$ then there exists $j\leq m$ such
that $\sigma _{j}=\varepsilon ,$ $\sigma _{i}\neq \varepsilon $
$(i\neq j),$ $q_{0},...,q_{j-1}\in Q_{1},$ $ q_{j},...,q_{m}\in
Q_{2}.$ Thus, $s=\sigma _{1}...\sigma _{j-1}\sigma _{j+1}...\sigma
_{n},$ and
$$I_{1}(q_{0})\wedge T_{2}(q_{m})\wedge \wedge _{i=1}^{m}\delta
(q_{i-1},\sigma _{i},q_{i})=I_{1}(q_{0})\wedge T_{2}(q_{m})\wedge
\wedge _{i=1}^{j-1}\delta _{1}(q_{i-1},\sigma _{i},q_{i})$$
$$\wedge T_{1}(q_{j-1})\wedge I_{2}(q_{j})\wedge \wedge
_{i=j+1}^{m}\delta _{2}(q_{i-1},\sigma _{i},q_{i})$$
$$=[I_{1}(q_{0})\wedge T_{1}(q_{j-1})\wedge \wedge
_{i=1}^{j-1}\delta _{1}(q_{i-1},\sigma _{i},q_{i})]\wedge \lbrack
I_{2}(q_{j})\wedge T_{2}(q_{m})\wedge \wedge _{i=j+1}^{m}\delta
_{2}(q_{i-1},\sigma _{i},q_{i})] $$
$$\leq rec_{\Re _{1}}(\sigma _{1}...\sigma _{j-1})\wedge rec_{\Re
_{2}}(\sigma _{j+1}...\sigma _{m})$$
$$\leq \vee \{rec_{\Re _{1}}(s_{1})\wedge rec_{\Re
_{2}}(s_{2}):s_{1}s_{2}=s\}.$$

(2) First, we can use Lemmas 2.5 and 2.6 to derive that
$$\lceil \gamma (atom(\Re _{1})\cup atom(\Re _{2}))\wedge (\exists
s_{1},s_{2}\in \Sigma ^{\ast })(s_{1}s_{2}=s\wedge rec_{\Re
_{1}}(s_{1})\wedge rec_{\Re _{2}}(s_{2}))\rceil $$
$$=\lceil \gamma (atom(\Re _{1})\cup atom(\Re _{2}))\rceil \wedge
\vee _{s_{1}s_{2}=s}(rec_{\Re _{1}}(s_{1})\wedge rec_{\Re
_{2}}(s_{2}))$$
$$\leq \vee _{s_{1}s_{2}=s}(\lceil \gamma (atom(\Re _{1})\cup
atom(\Re _{2}))\rceil \wedge rec_{\Re _{1}}(s_{1})\wedge rec_{\Re
_{2}}(s_{2})).$$

For any $s_{1},s_{2}\in \Sigma ^{\ast }$ with $s_{1}s_{2}=s,$ we
use Lemmas 2.5 and 2.6 again, and this yields
$$\lceil \gamma (atom(\Re _{1})\cup atom(\Re _{2}))\rceil \wedge
rec_{\Re _{1}}(s_{1})\wedge rec_{\Re _{2}}(s_{2})=\lceil \gamma
(atom(\Re _{1})\cup atom(\Re _{2}))\rceil \wedge $$
$$\vee _{lb(c_{1})=s_{1}}(I_{1}(b(c_{1}))\wedge
T_{1}(e(c_{1}))\wedge \lceil path_{\Re _{1}}(s_{1})\rceil )\wedge
\vee _{lb(c_{2})=s_{2}}(I_{2}(b(c_{2}))\wedge
T_{2}(e(c_{2}))\wedge \lceil path_{\Re _{2}}(s_{2})\rceil )$$
$$\leq \vee
_{lb(c_{1})=s_{1},lb(c_{2})=s_{2}}(I_{1}(b(c_{1}))\wedge
T_{1}(e(c_{1}))\wedge \lceil path_{\Re _{1}}(s_{1})\rceil \wedge
I_{2}(b(c_{2}))\wedge T_{2}(e(c_{2}))\wedge \lceil path_{\Re
_{2}}(s_{2})\rceil ).$$

Furthermore, for any $c_{1}=p_{0}\sigma _{1}p_{1}...p_{m-1}\sigma
_{m}p_{m}$ and $c_{2}=q_{0}\tau _{1}q_{1}...q_{n-1}\tau _{n}q_{n}$
with $s_{1}=\sigma _{1}...\sigma _{m}$ and $s_{2}=\tau _{1}...\tau
_{n},$
$$I_{1}(b(c_{1}))\wedge T_{1}(e(c_{1}))\wedge \lceil path_{\Re
_{1}}(s_{1})\rceil \wedge I_{2}(b(c_{2}))\wedge
T_{2}(e(c_{2}))\wedge \lceil path_{\Re _{2}}(s_{2})\rceil =$$
$$I_{1}(p_{0})\wedge T_{2}(q_{m})\wedge \wedge _{i=1}^{m}\delta
_{1}(p_{i-1},\sigma _{i},p_{i})\wedge T_{1}(p_{m})\wedge
I_{2}(q_{0})\wedge \wedge _{j=1}^{n}\delta _{2}(q_{j-1},\tau
_{j},q_{j})$$
$$=I_{1}(p_{0})\wedge T_{2}(q_{m})\wedge \lceil path_{\Re _{1}\Re
_{2}}(p_{0}\sigma _{1}p_{1}...p_{m-1}\sigma _{m}p_{m}\varepsilon
q_{0}\tau _{1}q_{1}...q_{n-1}\tau _{n}q_{n})\rceil $$
$$\leq rec_{\Re _{1}\Re _{2}}(s).$$

(3) The part that (i) implies (ii) is a simple corollary of (2).
Conversely, it suffices to show that $\ell $ enjoys
distributivity; that is, for any $ a,b,c\in L,$ $a\wedge (b\vee
c)=(a\wedge b)\vee (a\wedge c).$ Let $\Re
_{1}=<\{p_{0},p_{1}\},\{p_{0}\},\{p_{1}\},\delta _{1}>$ and $\Re
_{2}=<\{q_{0},q_{1},q_{2}\},\{q_{0}\},\{q_{1},q_{2}\},\delta
_{2}>,$ where $ \delta _{1}(p_{0},\sigma ,p_{1})=a,$ $\delta
_{2}(q_{0},\sigma ,q_{1})=b,$ $ \delta _{2}(q_{0},\sigma
,q_{2})=c,$ and $\delta _{1},$ $\delta _{2}$ take value $0$ for
other arguments (see Figure 7). Then it follows that
$$a\wedge (b\vee c)=\lceil (\exists s_{1},s_{2}\in \Sigma ^{\ast
})(s_{1}s_{2}=\sigma \sigma \wedge rec_{\Re _{1}}(s_{1})\wedge
rec_{\Re _{2}}(s_{2}))\rceil $$
$$=\lceil rec_{\Re _{1}\Re _{2}}(\sigma \sigma )\rceil $$
$$=(a\wedge b)\vee (a\wedge c).\heartsuit$$

\begin{figure}\centering
\includegraphics[width=1\textwidth]{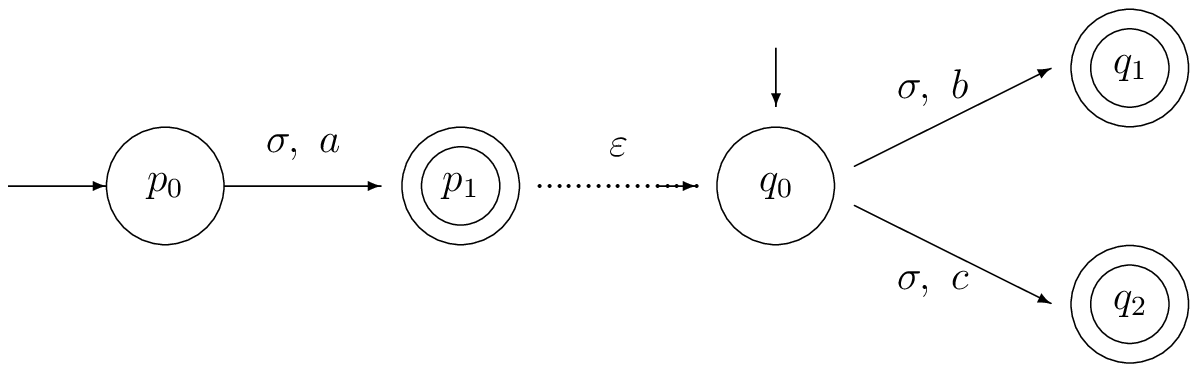}\caption{Automaton e} \label{fig 7}
\end{figure}

\smallskip\

We now turn to consider the Kleene closure of an orthomodular
lattice-valued language. For this purpose, we need to introduce
the fold construction of an orthomodular lattice-valued automaton.
Let $\Re =<Q,I,T,\delta >\in \mathbf{A}(\Sigma ,\ell )$ be an
$\ell -$valued automaton, and let $q_{0}\notin Q$ be a new state.
We define the fold of $ \Re $ to be $\ell -$valued automaton $\Re
^{\ast }=<Q\cup \{q_{0}\},\{q_{0}\},T\cup \{q_{0}\},\delta ^{\ast
}>$ with $\varepsilon -$ moves, where
$$\delta ^{\ast }:(Q\cup \{q_{0}\})\times (\Sigma \cup
\{\varepsilon \})\times (Q\cup \{q_{0}\})\rightarrow L$$ is given
by
$$\delta ^{\ast }(p,\sigma ,q)=\left\{
\begin{array}{c}
I(q)\ {\rm if }\ p=q_{0}\ {\rm and }\ \sigma =\varepsilon , \\
\delta (p,\sigma ,q)\ {\rm if }\ p,q\in Q\ {\rm and }\ \sigma \neq
\varepsilon ,
\\
T(p)\wedge I(q)\ {\rm if }\ p,q\in Q\ {\rm and }\ \sigma =\varepsilon , \\
0\ {\rm otherwise.}
\end{array}
\right. $$

The language accepted by the fold of an orthomodular
lattice-valued automaton is then clearly presented by the
following proposition.

\smallskip\

\textbf{Proposition 6.7}. Let $\ell =<L,\leq ,\wedge ,\vee ,\perp
,0,1>$ be an orthomodular lattice, and let $\rightarrow $ enjoy
the Birkhoff-von Neumann requirement.

(1) For any $\Re \in \mathbf{A}(\Sigma ,\ell )$ and for all $s\in
\Sigma ^{\ast },$
$$\stackrel{\ell }{\models }rec_{\Re ^{\ast }}(s)\rightarrow (\exists
m\geq 0,s_{1},...,s_{m}\in \Sigma ^{\ast })(s_{1}...s_{m}=s\wedge
\wedge _{i=1}^{m}rec_{\Re }(s_{i})).$$

(2) For any $\Re \in \mathbf{A}(\Sigma ,\ell )$ and for each $s\in
\Sigma ^{\ast },$
$$\stackrel{\ell }{\models }\gamma (atom(\Re ))\wedge (\exists m\geq
0,s_{1},...,s_{m}\in \Sigma ^{\ast })(s_{1}...s_{m}=s\wedge \wedge
_{i=1}^{m}rec_{\Re }(s_{i}))\rightarrow rec_{\Re ^{\ast }}(s),$$
and in particular if $\rightarrow =\rightarrow _{3},$ then
$$\stackrel{\ell }{\models }\gamma (atom(\Re ))\rightarrow (rec_{\Re
^{\ast }}(s)\leftrightarrow (\exists m\geq 0,s_{1},...,s_{m}\in
\Sigma ^{\ast })(s_{1}...s_{m}=s\wedge \wedge _{i=1}^{m}rec_{\Re
}(s_{i}))).$$

(3) The following two statements are equivalent:

\ \ \ \ \ \ (i) $\ell $ is a Boolean algebra;

\ \ \ \ \ \ (ii) for all $\Re \in \mathbf{A}(\Sigma ,\ell )$ and
$s\in \Sigma ^{\ast },$
$$\stackrel{\ell }{\models }rec_{\Re ^{\ast }}(s)\leftrightarrow
(\exists m\geq 0,s_{1},...,s_{m}\in \Sigma ^{\ast
})(s_{1}...s_{m}=s\wedge \wedge _{i=1}^{m}rec_{\Re }(s_{i})).$$

\smallskip\

\textbf{Proof.} For (1), (2) and the part from (i) to (ii) of (3),
it is similar to the proof of Proposition 6.6, and here we omit
the details. To show that (ii) implies (i), we assume that
$a,b,c\in L$ and want to construct an $\ell -$valued automaton for
which the validity of (ii) leads to $a\wedge (b\vee c)=(a\wedge
b)\vee (a\wedge c).$ Let $\Re
=<\{q_{1},q_{2},...,q_{6}\},\{q_{1},q_{2},q_{3}\},\{q_{6}\},\delta
>$ in which $\delta (q_{1},\sigma ,q_{4})=\delta (q_{3},\sigma
,q_{5})=1,$ $\delta (q_{2},\sigma ,q_{6})=a,$ $\delta
(q_{4},\sigma ,q_{6})=b,$ $\delta (q_{5},\sigma ,q_{6})=c,$ and
$\delta $ takes value $0$ for the other arguments. Then $\Re
^{\ast }$ is visualized as Figure 8.

\begin{figure}\centering
\includegraphics[width=1\textwidth]{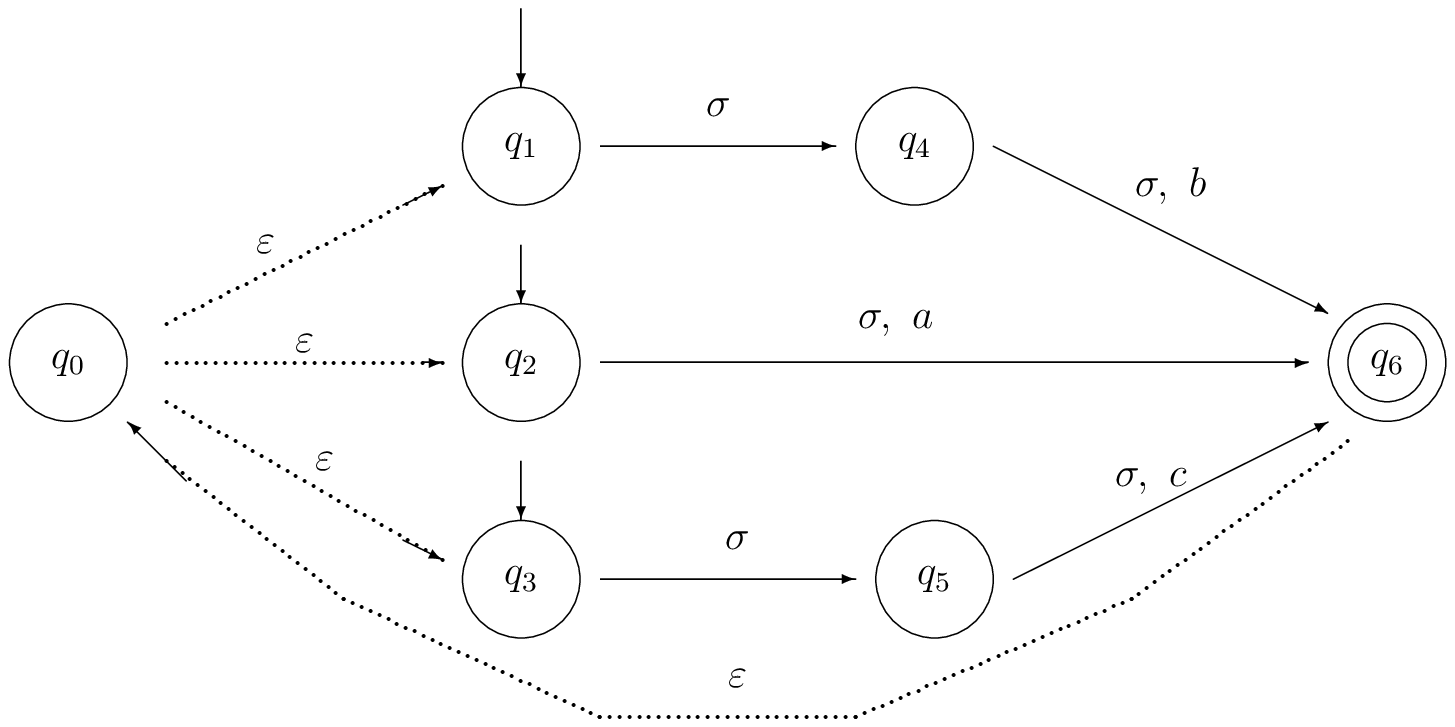}\caption{Automaton f} \label{fig 8}
\end{figure}

We now have
$$a\wedge (b\vee c)=\lceil (\exists m\geq 0,s_{1},...,s_{m}\in
\Sigma ^{\ast })(s_{1}...s_{m}=\sigma ^{3}\wedge \wedge
_{i=1}^{m}rec_{\Re }(s_{i}))\rceil $$
$$=rec_{\Re ^{\ast }}(\sigma ^{3})$$
$$=(a\wedge b)\vee (a\wedge c).\heartsuit$$

\smallskip\

From the above proposition, we are able to demonstrate that the
predicate $CReg_\Sigma$ is preserved by the Kleene closure. The
corresponding result for the predicate $Reg_\Sigma$ is not true in
general.

\smallskip\

\textbf{Corollary 6.8.} Let $\ell =<L,\leq ,\wedge ,\vee ,\perp
,0,1>$ be an orthomodular lattice, and let $\rightarrow
=\rightarrow _{3}.$ Then for any $ A\in L^{\Sigma ^{\ast }},$
$$\stackrel{\ell }{\models }CReg_{\Sigma }(A)\longrightarrow
CReg_{\Sigma }(A^{\ast }).$$

\smallskip\

\textbf{Proof}. It is similar to the proof of Proposition 6.2. The
point here is to show the following inequality:
$$\lceil \gamma (atom(\Re )\cup r(A))\rceil \wedge \lceil A\equiv
rec_{\Re }\rceil \leq \lceil A^{\ast }\equiv rec_{\Re ^{\ast
}}\rceil $$ for any $\Re\in \mathbf{A}(\Sigma, \ell).$ In fact, by
using Lemma 2.11(1) we have
$$\lceil \gamma (atom(\Re )\cup r(A))\rceil \wedge \lceil A\equiv
rec_{\Re }\rceil =\lceil \gamma (atom(\Re )\cup r(A))\rceil \wedge
\wedge _{s\in \Sigma ^{\ast }}(A(s)\leftrightarrow rec_{\Re
}(s))$$
$$\leq \lceil \gamma (atom(\Re )\cup r(A))\rceil \wedge \wedge
_{s\in \Sigma ^{\ast }}(\vee _{s_{1}...s_{m}=s}\wedge
_{i=1}^{m}A(s_{i})\leftrightarrow \vee _{s_{1}...s_{m}=s}\wedge
_{i=1}^{m}rec_{\Re }(s_{i}))$$
$$=\lceil \gamma (atom(\Re )\cup r(A))\rceil \wedge \lceil A^{\ast
}\equiv (rec_{\Re })^{\ast }\rceil .$$

On the other hand, it follows from Proposition 6.7 that $$\lceil
\gamma (atom(\Re ))\rceil \leq \lceil (rec_{\Re })^{\ast }\equiv
rec_{\Re ^{\ast }}\rceil .$$ Then with Lemma 2.11(3) we obtain
$$\lceil \gamma (atom(\Re )\cup r(A))\rceil \wedge \lceil A\equiv
rec_{\Re }\rceil \leq \lceil \gamma (atom(\Re )\cup r(A))\rceil
\wedge \lceil A^{\ast }\equiv (rec_{\Re })^{\ast }\rceil$$ $$
\wedge \lceil (rec_{\Re })^{\ast }\equiv rec_{\Re ^{\ast }}\rceil
$$
$$\leq \lceil A^{\ast }\equiv rec_{\Re ^{\ast }}\rceil .\heartsuit$$

\smallskip\

To conclude this section, we show that both the predicate
$Reg_\Sigma$ and $CReg_\Sigma$ are preserved by the pre-image of a
homomorphism between two languages. But the closure property of an
orthomodular lattice-valued language under homomorphism is left to
be examined in the next section, after the notion of orthomodular
lattice-valued regular expression is proposed. Let $\Sigma $ and
$\Gamma $ be two alphabets of input symbols. Then each mapping
$h:\Sigma \rightarrow \Gamma ^{\ast }$ can be uniquely extended to
a homomorphism $h:\Sigma ^{\ast }\rightarrow \Gamma ^{\ast }$ such
that $ h(\varepsilon )=\varepsilon $ and
$$h(xy)=h(x)h(y)$$ for all $x,y\in \Sigma ^{\ast }.$ Furthermore, we
may define images of $\ell -$valued subsets of $ \Sigma ^{\ast }$
under $h$ and pre-images of $\ell -$valued subsets of $ \Gamma
^{\ast }$ under $h.$ Recall that for any $A\in L^{\Sigma ^{\ast
}}$ and $B\in L^{\Gamma ^{\ast }},$ $h(A)\in L^{\Gamma ^{\ast }}$
and $h^{-1}(B)\in L^{\Sigma ^{\ast }}$ are given as follows:
$$h(A)(t)=\vee \{A(s):s\in \Sigma ^{\ast }\ {\rm and}\ h(s)=t\}$$
for each $t\in \Gamma ^{\ast },$ and
$$h^{-1}(B)(s)=B(h(s))$$
for each $s\in \Sigma ^{\ast }.$

Let $\Re =<Q,I,T,\delta >\in \mathbf{A}(\Gamma ,\ell )$ be an
$\ell -$valued automaton over $\Gamma .$ Then the pre-image of
$\Re $ under $h$ is defined to be an $\ell -$valued automaton
$$h^{-1}(\Re )=<Q,I,T,h^{-1}(\delta )>\in \mathbf{A}(\Sigma ,\ell
)$$ over $\Sigma ,$ where for any $p,q\in Q$ and $ \sigma \in
\Sigma ,$
$$h^{-1}(\delta )(p,\sigma ,q)=\delta (p,h(\sigma ),q).$$

The pre-image of a homomorphism has a very nice compatibility with
the predicates $reg_\Sigma$ and $CReg_\Sigma$, and no
commutativity is needed here.

\smallskip\

\textbf{Proposition 6.9.} Let $\ell =<L,\leq ,\wedge ,\vee ,\perp
,0,1>$ be an orthomodular lattice, let $\rightarrow $ enjoy the
Birkhoff-von Neumann requirement, and let $h:\Sigma \rightarrow
\Gamma ^{\ast }$ be a mapping. Then for any $\Re \in
\mathbf{A}(\Gamma ,\ell )$ and for any $s\in \Sigma ^{\ast },$
$$\stackrel{\ell }{\models} rec_{h^{-1}(\Re )}(s)\longleftrightarrow rec_{\Re
}(h(s)).$$

\smallskip\

\textbf{Proof.} Suppose that $s=\sigma _{1}\sigma _{2}...\sigma
_{n}.$ Then
$$\lceil rec_{h^{-1}(\Re )}(s)\rceil =\vee \{I(q_{0})\wedge
T(q_{n})\wedge \wedge _{i=0}^{n-1}h^{-1}(\delta )(q_{i},\sigma
_{i+1},q_{i+1}):q_{0},q_{1},...,q_{n}\in Q\}$$
$$=\vee \{I(q_{0})\wedge T(q_{n})\wedge \wedge _{i=0}^{n-1}\delta
(q_{i},h(\sigma _{i+1}),q_{i+1}):q_{0},q_{1},...,q_{n}\in Q\}$$
$$=\lceil rec_{\Re }(h(\sigma _{1})h(\sigma _{2})...h(\sigma
_{n}))\rceil $$
$$=\lceil rec_{\Re }(h(s))\rceil .\heartsuit$$

\smallskip\

\textbf{Corollary 6.10.} Let $\ell =<L,\leq ,\wedge ,\vee ,\perp
,0,1>$ be an orthomodular lattice, let $\rightarrow $ enjoy the
Birkhoff-von Neumann requirement, and let $h:\Sigma \rightarrow
\Gamma ^{\ast }$ be a mapping. Then for any $B\in L^{\Gamma ^{\ast
}},$
$$\stackrel{\ell }{\models} Reg_{\Gamma }(B)\rightarrow Reg_{\Sigma
}(h^{-1}(B)),$$ and
$$\stackrel{\ell }{\models} CReg_{\Gamma }(B)\rightarrow CReg_{\Sigma
}(h^{-1}(B)),$$

\smallskip\

\textbf{Proof.} From the above proposition we have
$$h^{-1}(rec_{\Re })(s)=rec_{\Re }(h(s))=rec_{h^{-1}(\Re )}(s)$$
for all $s\in \Sigma ^{\ast }.$ Then with Lemma 2.12 we obtain
$$\lceil Rec_{\Gamma }(B)\rceil =\vee \{\lceil B\equiv
rec_{\Re }\rceil :\Re \in \mathbf{A}(\Gamma ,\ell )\}$$
$$\leq \vee \{\lceil h^{-1}(B)\equiv h^{-1}(rec_{\Re })\rceil :\Re
\in \mathbf{A}(\Gamma ,\ell )\}$$
$$=\vee \{\lceil h^{-1}(B)\equiv rec_{h^{-1}(\Re )}\rceil :\Re \in \mathbf{A}
(\Gamma ,\ell )\}$$
$$\leq \vee \{\lceil h^{-1}(B)\equiv rec_{\wp }\rceil :\wp \in \mathbf{A}
(\Sigma ,\ell )\}$$
$$=\lceil Reg_{\Sigma }(h^{-1}(B))\rceil .$$

It is similar for the case of commutative regularity.$\heartsuit$

\bigskip\

\textbf{7. Orthomodular Lattice-Valued Regular Expressions}

\smallskip\

One of the most interesting results in classical automata theory
is the Kleene theorem which shows the equivalence between finite
automata and regular expressions. The main aim of this section is
to present an orthomodular lattice-valued generalization of the
Kleene theorem. Let $\ell =<L,\leq ,\wedge ,\vee ,\bot ,0,1>$ be
an orthomodular lattice, and let $\Sigma$ be an nonempty set of
input symbols. Then the language of $\ell-$valued regular
expressions over $\Sigma$ has the alphabet
$(\Sigma\cup\{\varepsilon,\phi\})\cup(L\cup\{+,\cdot ,*\})$. The
symbols in $\Sigma\cup\{\varepsilon,\phi\}$ will be used to stand
for atomic expressions, and the symbols in $L\cup\{+,\cdot ,*\}$
will be used to denote operators for building up compound
expressions: $*$ and all $\lambda\in L$ are unary operators, and
$+,\cdot$ are binary ones. We use $\alpha,\beta,...$ to act as
meta-symbols for regular expressions and $L(\alpha)$ for the
language denoted by expression $\alpha$. More explicitly,
$L(\alpha)$ will be used to denote an $\ell-$valued subset of
$\Sigma^{*}$; that is, $L(\alpha)\in L^{\Sigma^{*}}$. Orthomodular
lattice-valued regular expressions and the orthomodular
lattice-valued languages denoted by them are formally defined as
follows:

(i) For each $a\in\Sigma$, $a$ is a regular expression, and
$L(a)={a}$; $\varepsilon$ and $\phi$ are regular expressions, and
$L(\varepsilon)={\varepsilon}$, $L(\phi)=\phi$.

(ii) If both $\alpha$ and $\beta$ are regular expressions, then
for each $\lambda\in L$, $\lambda\alpha$ is a regular expression,
and $$L(\lambda\alpha)=\lambda L(\alpha);$$ and $\alpha + \beta$,
$\alpha \cdot \beta$ and $\alpha ^{*}$ are all regular
expressions, and
$$L(\alpha +\beta)=L(\alpha)\cup L(\beta),$$ $$L(\alpha \cdot
\beta)=L(\alpha)\cot L(\beta),$$ $$L(\alpha ^{*})=L(\alpha)
^{*}.$$ It is easy to see that the only difference between
orthomodular lattice-valued regular expressions and the classical
ones is that the additional unary (scalar) operators $\lambda\in
L$ are permitted to occur in the former.

The central part of the Kleene theorem is a mechanism to transform
a finite automaton into a regular expression. This mechanism has a
straightforward extension in the framework of orthomodular
lattice-valued automata. Let $\Re =<Q,I,T,\delta >\in
\mathbf{A}(\Sigma ,\ell )$ be an $\ell -$valued automaton over
$\Sigma .$ For any $u,v\in Q$ and $X\subseteq Q,$ $\alpha
_{uv}^{X}$ is defined by induction on the cardinality $|X| $ of
$X:$

(1) $$\alpha _{uv}^{\phi }=\left\{
\begin{array}{c}
\Sigma_{\sigma \in \Sigma }\delta (u,\sigma ,v)\sigma \ {\rm if}\
u\neq
v, \\
\varepsilon +\Sigma_{\sigma \in \Sigma }\delta (u,\sigma ,v)\sigma
\ {\rm if}\ u=v.
\end{array}
\right. $$

(2) if $X\neq \phi ,$ then we choose any $q\in X$ and define
$$\alpha _{uv}^{X}=\alpha _{uv}^{X-\{q\}}+\alpha
_{uq}^{X-\{q\}}\cdot (\alpha _{qq}^{X-\{q\}})^{\ast }\cdot \alpha
_{qv}^{X-\{q\}}.$$ Then the $\ell -$valued regular expression
$$k(\Re )=\Sigma_{u,v\in Q}(I(u)\wedge T(v))\alpha
_{uv}^{Q}$$ is called a Kleene representation of $\Re .$

The following theorem describes properly the relationship between
the language recognized by an orthomodular lattice-valued
automaton and the language expressed by its Kleene representation.

\smallskip\

\textbf{Theorem 7.1. } Let $\ell =<L,\leq ,\wedge ,\vee ,\bot
,0,1>$ be an orthomodular lattice, and let $\rightarrow$ satisfy
the Birkhoff-von Neumann requirement.

(1) For any $\Re \in \mathbf{A}(\Sigma ,\ell )$ and $s\in \Sigma
^{\ast },$ if $k(\Re)$ is a Kleene representation of $\Re$, then
$$\stackrel{\ell }{\models} rec_{\Re }(s)\longrightarrow s\in L(k(\Re
)).$$

(2) For any $\Re \in \mathbf{A}(\Sigma ,\ell )$ and $s\in \Sigma
^{\ast },$ and for any Kleene representation $k(\Re)$ of $\Re$,
$$\stackrel{\ell }{\models} \gamma(atom(\Re))\wedge s\in L(k(\Re))\longrightarrow rec_\Re
(s),$$ and especially if $\longrightarrow =\longrightarrow_3$,
then $$\stackrel{\ell }{\models} \gamma(atom(\Re))\longrightarrow
(rec_\Re (s)\longleftrightarrow s\in L(k(\Re))).$$

(3) The following two statements are equivalent:

\ \ \ \ \ \ (i) $\ell $ is a Boolean algebra.

\ \ \ \ \ \ (ii) For any $\Re \in \mathbf{A}(\Sigma ,\ell )$ and
$s\in \Sigma ^{\ast },$ and for any Kleene representation $k(\Re)$
of $\Re$,
$$\stackrel{\ell }{\models} rec_{\Re }(s)\longleftrightarrow s\in L(k(\Re
)).$$

\smallskip\

\textbf{Proof.} We prove (1) and (2) together. To this end, we
have to demonstrate that for any $u,v\in Q,$ $X\subseteq Q$ and
$s\in \Sigma ^{\ast },$ $$(a)\ \vee \{\lceil path_{\Re
}(c)\rceil:c\in T(Q,\Sigma ),b(c)=u,e(c)=v,M(c)\subseteq
X,lb(c)=s\}\leq L(\alpha _{uv}^{X})(s),$$
$$(b)\ \lceil\gamma(atom(\Re))\rceil\wedge L(\alpha _{uv}^{X})(s)\leq\ \ \ \ \ \
\ \ \ \ \ \ \ \  \ \ \ \ \ \ \ \ \ \ \ \ \ \ \ \ \ \ \ \  \ \ \ \
\ \ \ \ \ \ \ \ \ \ \ \ \ \ \ \  \ \ \ \ \ \ \ \ \ \ \ \ \ \ \ \ \
$$
$$\ \ \ \ \ \ \ \ \vee \{\lceil path_{\Re }(c)\rceil:c\in T(Q,\Sigma ),b(c)=u,e(c)=v,M(c)\subseteq
X,lb(c)=s\},$$ where $M(c)$ stands for the set of states along $c$
except $u$ and $v;$ more exactly, $M(c)=\{q_{1},...,q_{k-1}\}$ if
$c=u\sigma _{1}q_{1}...q_{k-1}\sigma _{k}v.$ This claim may be
proved by induction on $ |X|.$ For the case of $X=\phi ,$ it is
easy. We now suppose that $q\in X\neq \phi $ and $$\alpha
_{uv}^{X}=\alpha _{uv}^{X-\{q\}}+[\alpha _{uq}^{X-\{q\}}(\alpha
_{qq}^{X-\{q\}})^{\ast }]\alpha _{qv}^{X-\{q\}}.$$ We first show
that (a) is valid in this case. From the induction hypothesis we
have $$(c)\ \vee \{\lceil path_{\Re }(c)\rceil :c\in T(Q,\Sigma
),b(c)=e(c)=q,\ \ \ \ \ \ \ \ \ \ \ \ \  \ \ \ \ \ \ \ \ \ \ \ \ \
\ \ \ \ \ \ \ \ \ \ \ \ \ \ \ \ \ \ \ \ \ \ $$
$$\ \ \ \ \ \ \ \ \ \ \ \ \ \ \ \ \ \ \ \ \ \ \ \ \ \ \ \ \ \ \ \ \ \ \ \ \ \
M(c)\subseteq X-\{q\}, lb(c)=s\}\leq L(\alpha
_{qq}^{X-\{q\}})(s)$$ for each $s\in \Sigma ^{\ast }.$ Then we
assert that for all $s\in \Sigma ^{\ast},$
$$(d)\ \vee \{\lceil path_{\Re }(c)\rceil :c\in
T(Q,\Sigma ),b(c)=e(c)=q,M(c)\subseteq X,lb(c)=s\}
 \ \ \ \ \ \ \ \ \ \ \ \ \ \ \ \ \ \ \ \ \ \ \ \ \ \ \ \ $$
$$\leq L((\alpha
_{qq}^{X-\{q\}})^{\ast })(s).$$ In fact, for any $c\in T(Q,\Sigma
)$ if $b(c)=e(c)=q,M(c)\subseteq X$ and $ lb(c)=s,$ we write
$c_{i}$ for the substring of $c$ beginning with the $i$th $q $ and
ending at the $(i+1)$th $q.$ If the number of occurrences of $q$
in $c$ is $m+1,$ then
$$\lceil path_{\Re }(c)\rceil =\wedge _{i=1}^{m}\lceil path_{\Re }(c_{i})\rceil .$$
Furthermore, by using (c) and noting that
$s=lb(c_{1})...lb(c_{m})$ we obtain
$$\lceil path_{\Re }(c)\rceil =\wedge _{i=1}^{m}L(\alpha
_{qq}^{X-\{q\}})(lb(c_{i}))\ \ \ \ \ \ \ \ \ \ \ \ \ \ \ \ \ \ \ \
\ \ \ \ \ \ \ \ \ \ \ \ \ \ \ \ \ \ \ \ \ \ \ \ \ \ \ \ \ \ \ \ \
\ \ \ \ \ \ \ $$
$$\leq\vee\{\wedge_{i=1}^{n} L(\alpha_{qq}^{X-\{q\}})(s_i):n\geq
0, s_1,...,s_n\in \Sigma^{*},s=s_1...s_n\}$$
$$=(L(\alpha _{qq}^{X-\{q\}}))^{\ast }(s)$$
$$= L((\alpha _{qq}^{X-\{q\}})^{\ast })(s).$$
Let $c$ range over $\{c\in T(Q,\Sigma ):b(c)=e(c)=q,M(c)\subseteq
X,lb(c)=s\}.$ Then (d) is proved.

Furthermore, from the induction hypothesis and (d) we have
$$([L(\alpha _{uq}^{X-\{q\}})L((\alpha _{qq}^{X-\{q\}})^{\ast
})]L(\alpha _{qv}^{X-\{q\}}))(s)=\ \ \ \ \ \ \ \ \ \ \ \ \ \ \ \ \
\ \ \ \ \ \ \ \ \ \ \ \ \ \ \ \ \ \ \ \ \ \ \ \ \ \ \ \ \ \ \ \ \
\ \ $$
$$\vee \{[L(\alpha_{uq}^{X-\{q\}})L((\alpha_{qq}^{X-\{q\}})^{\ast})](x)
\wedge L(\alpha_{qv}^{X-\{q\}})(y):s=xy\} $$
$$=\vee \{\vee\{L(\alpha_{uq}^{X-\{q\}})(x_1)\wedge
L((\alpha_{qq}^{X-\{q\}})^{\ast})(x_2):x=x_1 x_2\} \wedge
L(\alpha_{qv}^{X-\{q\}})(y):s=xy\}\ \ $$
$$\geq \vee \{L(\alpha_{uq}^{X-\{q\}})(x_1)\wedge
L((\alpha_{qq}^{X-\{q\}})^{\ast})(x_2) \wedge
L(\alpha_{qv}^{X-\{q\}})(y):s=x_1x_2y\}\ \ \ \ \ \ \ \ \ \ \ \ \ \
\ \ \ $$
$$\geq \vee \{\lceil path_{\Re }(c_{1})\rceil \wedge \lceil path_{\Re }(c_{2})\rceil \wedge \lceil path_{\Re
}(c_{3})\rceil : c_{1},c_{2},c_{3}\in T(Q,\Sigma),\ \ \ \ \ \ \ \
\ \ \ \ \ \ \ \ \ \ \ \ \ \ \ \ \ \ \ \ $$
$$\ \ \ \ b(c_{1})=u, e(c_{1})=b(c_{2})=e(c_{2})=b(c_{3})=q, e(c_{3})=v,s=lb(c_{1})lb(c_{2})lb(c_{3})\}$$
$$=\vee \{\lceil path_{\Re }(c)\rceil :c\in T(Q,\Sigma ),b(c)=u,e(c)=v,q\in
M(c)\}.\ \ \ \ \ \ \ \ \ \ \ \ \ \ \ \ \ \ \ \ \ \ \ \ \ \ \ \ \ \
\ \ $$ This yields further
$$L(\alpha _{uv}^{X})(s)=L(\alpha _{uv}^{X-\{q\}})(s)\vee
([L(\alpha _{uq}^{X-\{q\}})L((\alpha _{qq}^{X-\{q\}})^{\ast
}]L(\alpha _{qv}^{X-\{q\}}))(s)$$
$$\geq \ {\rm the\ left-hand\ side\ of\ (a)}.$$

We now turn to consider (b). The induction hypothesis gives
$$(e)\ \lceil \gamma(atom(\Re))\rceil \wedge L(\alpha_{uv}^{X-\{q\}})(s)\leq \{\lceil path_\Re (c)\rceil :c\in
T(Q,\Sigma),\ \ \ \ \ \ \ \ \ \ \ \ \ \ \ \ \ \ \ \ \ \ \ \ \ \ \
\ \ \ \ \ \ \ \ \ \ \ \ \ $$
$$b(c)=u, e(c)=v, M(c)\subseteq X-\{q\}, lb(c)=s \}.$$ For any
$n\geq 0$ and $s_{1},...,s_{n}\in \Sigma ^{\ast }$ with $
s=s_{1}...s_{n},$ from (e) we can apply Lemmas 2.5 and 2.6 to
obtain
$$\lceil \gamma(atom(\Re))\rceil \wedge \wedge
_{i=1}^{n}L(\alpha _{qq}^{X-\{q\}})(s_{i})=\lceil
\gamma(atom(\Re))\rceil \wedge \wedge_{i=1}^{n}[\lceil
\gamma(atom(\Re))\rceil \wedge L(\alpha_{qq}^{X-\{q\}})(s_i)]$$
$$\leq \lceil \gamma(atom(\Re))\rceil \wedge \wedge _{i=1}^{n}\vee \{\lceil path_{\Re
}(c_{i})\rceil :c_{i}\in T(Q,\Sigma ),\ \ \ \ \ \ \ \ \ \ \ \ \ \
\ \ \ \ \ \ \ \ \ \ \ \ \ \ \ \ \ \ \ \ \ \ \ \ \ \ \ \ \ \ \ \ \
\
$$
$$b(c_{i})=e(c_{i})=q,M(c_{i})\subseteq X-\{q\},lb(c_{i})=s_{i}\}$$
$$\leq \vee \{\wedge _{i=1}^{n}\lceil path_{\Re }(c_{i})\rceil :c_{i}\in T(Q,\Sigma
),b(c_{i})=e(c_{i})=q, M(c_{i})\subseteq X-\{q\},\ \ \ \ \ \ \ \ \
\ \ \ \ \ \ \ \ \ \ $$
$$lb(c_{i})=s_{i}\ {\rm for\ each}\ i=1,2,...,n\}$$
$$\leq \vee \{\lceil path_{\Re }(\overline{c_{1}...c_{n}})\rceil :c_{i}\in T(Q,\Sigma
),b(c_{i})=e(c_{i})=q, M(c_{i})\subseteq X-\{q\},\ \ \ \ \ \ \ \ \
\ \ \ \ \ \ \ \ \ \ $$
$$lb(c_{i})=s_{i}\ {\rm for\ each}\ i=1,2,...,n\},$$ where
$\overline{c_{1}...c_{n}}=c_{1}c_{2}^{\prime }...c_{n}^{\prime },$
$c_{i}^{\prime }$ is the resulting string after removing the first
$q$ in $c_{i}$ for each $i\geq 2.$ Note that
$lb(\overline{c_{1}...c_{n}} )=s_{1}...s_{n}=s$ whenever
$lb(c_{i})=s_{i}$ $(i=1,2,...,n).$ We write
$$\lambda=\vee \{\lceil path_\Re (c)\rceil :c\in T(Q,\Sigma), b(c)=e(c)=q, M(c)\subseteq X,
lb(c)=s\}.$$ Then it holds that
$$\lceil \gamma(atom(\Re))\rceil \wedge \wedge _{i=1}^{n}L(\alpha _{qq}^{X-\{q\}})(s_{i})\leq\lambda.$$
Moreover, note that $\lceil \gamma(atom(\Re))\rceil ,
L(\alpha_{qq}^{X-\{q\}})(s_i)\in [atom(\Re)].$ It follows that
$$\lceil \gamma(atom(\Re))\rceil \wedge L((\alpha_{qq}^{X-\{q\}})^{\ast})(s)=\lceil \gamma(atom(\Re))\rceil \wedge
\lceil \gamma(atom(\Re))\rceil \wedge\ \ \ \ \ \ \ \ \ \ \ \ \ \ \
\ \ \ \ \ \ \ \ \ \ \ \ $$
$$\vee\{\wedge_{i=1}^{n}L(\alpha_{qq}^{X-\{q\}})(s_i):n\geq 0,
s=s_1...s_n \}$$
$$\leq \vee\{\lceil \gamma(atom(\Re))\rceil \wedge \wedge_{i=1}^{n}L(\alpha_{qq}^{X-\{q\}})(s_i):n\geq 0,
s=s_1...s_n \}\leq \lambda.\ \ \ \ \ \ \ \ \ \ \ \ \ \ \ \ \ \ \
$$
This enables us to obtain
$$\lceil \gamma (atom(\Re ))\rceil \wedge \lbrack L(\alpha
_{uq}^{X-\{q\}})L((\alpha _{qq}^{X-\{q\}})^{\ast })](x)\ \ \ \ \ \
\ \ \ \ \ \ \ \ \ \ \ \ \ \ \ \ \ \ \ \ \ \ \ \ \ \ \ \ \ \ \ \ \
\ \ \ \ \ \ \ \ \ \ \ \ \ \ $$
$$=\lceil \gamma
(atom(\Re ))\rceil \wedge \lceil \gamma (atom(\Re ))\rceil \wedge
\vee \{L(\alpha _{uq}^{X-\{q\}})(x_{1})\wedge L((\alpha
_{qq}^{X-\{q\}})^{\ast })(x_{2}):x=x_{1}x_{2}\}\ \ \ $$
$$\leq \vee \{\lceil \gamma (atom(\Re ))\rceil \wedge
L(\alpha _{uq}^{X-\{q\}})(x_{1})\wedge L((\alpha
_{qq}^{X-\{q\}})^{\ast })(x_{2}):x=x_{1}x_{2}\}\ \ \ \ \ \ \ \ \ \
\ \ \ \ \ \ \ \ \ \ \ \ \ \ \ \ \ \ \ \ \ $$
$$=\vee \{\lceil \gamma
(atom(\Re ))\rceil \wedge [\lceil \gamma (atom(\Re ))\rceil \wedge
L(\alpha _{uq}^{X-\{q\}})(x_{1})]\wedge\ \ \ \ \ \ \ \ \ \ \ \ \ \
\ \ \ \ \ \ \ \ \ \ \ \ \ \ \ \ \ \ \ \ \ \ \ \ \ \ \ \ \ \ \ \ \
\ \ $$
$$[\lceil \gamma (atom(\Re ))\rceil \wedge
L((\alpha _{qq}^{X-\{q\}})^{\ast })(x_{2})]:x=x_{1}x_{2}\}$$
$$\leq
\vee \{\lceil \gamma (atom(\Re ))\rceil \wedge [ \vee \{\lceil
path_{\Re }(c_{1})\rceil :c_{1}\in T(Q,\Sigma
),b(c_{1})=u,e(c_{1})=q,\ \ \ \ \ \ \ \ \ \ \ \ \ \ \ \ \ \ \ \ \
\ \ \ \ \ \ $$
$$M(c_{1})\subseteq X-\{q\},lb(c_{1})=x_{1}\}]
\wedge \lbrack \vee \{\lceil path_{\Re }(c_{2})\rceil :c_{2}\in
T(Q,\Sigma ),b(c_{2})=$$ $$e(c_{2})=q,M(c_{2})\subseteq
X,lb(c_{2})=x_{2}\}]:x=x_{1}x_{2}\}$$
$$\leq \vee \{\lceil path_{\Re
}(c_{1})\rceil \wedge \lceil path_{\Re }(c_{2})\rceil
:c_{1},c_{2}\in T(Q,\Sigma
),b(c_{1})=u,e(c_{1})=b(c_{2})=e(c_{2})=q,$$
$$M(c_{1})\subseteq X-\{q\}, M(c_{2})\subseteq
X,x=lb(c_{1})lb(c_{2})\}.$$ Furthermore, we can derive in a
similar way that
$$\lceil \gamma (atom(\Re
))\rceil \wedge ([L(\alpha _{uq}^{X-\{q\}})L((\alpha
_{qq}^{X-\{q\}})^{\ast })]L(\alpha _{qv}^{X-\{q\}}))(s)\ \ \ \ \ \
\ \ \ \ \ \ \ \ \ \ \ \ \ \ \ \ \ \ \ \ \ \ \ \ \ \ \ \ \ \ \ \ $$
$$\leq \vee \{\lceil path_{\Re }(c_{1})\rceil \wedge \lceil path_{\Re }(c_{2})\rceil \wedge \lceil path_{\Re
}(c_{3})\rceil :c_{1},c_{2},c_{3}\in T(Q,\Sigma ),b(c_{1})=u,\ \ \
\ \ \ \ \ \ \ \ \ \ \ \ \ $$
$$e(c_{1})=b(c_{2})=e(c_{2})=b(c_{3})=q,e(c_{3})=v,s=lb(c_{1})lb(c_{2})lb(c_{3})\}
$$
$$=\vee \{\lceil path_{\Re }(c)\rceil :c\in T(Q,\Sigma ),b(c)=u,e(c)=v,q\in
M(c),s=lb(c)\}.\ \ \ \ \ \ \ \ \ \ \ \ \ \ \ \ \ \ \ \ $$
Consequently, it holds that
$$\lceil \gamma (atom(\Re ))\rceil \wedge L(\alpha _{uv}^{X})(s)=\lceil \gamma (atom(\Re
))\rceil \wedge \{L(\alpha _{uv}^{X-\{q\}})(s)\vee\ \ \ \ \ \ \ \
\ \ \ \ \ \ \ \ \ \ \ \ \ \ \ \ \ \ \ \ \ \ \ \ \ \ \ \ \ \ $$
$$\ \ \ \ \ \ \ \ \ \ \ \ \ \ \ \ \ \ ([L(\alpha _{uq}^{X-\{q\}})L((\alpha _{qq}^{X-\{q\}})^{\ast
})]L(\alpha _{qv}^{X-\{q\}}))(s)\}$$
$$\leq [\lceil \gamma (atom(\Re ))\rceil \wedge L(\alpha
_{uv}^{X-\{q\}})(s)]\vee \{\lceil \gamma (atom(\Re ))\rceil \wedge
([L(\alpha _{uq}^{X-\{q\}})L((\alpha _{qq}^{X-\{q\}})^{\ast
})]L(\alpha _{qv}^{X-\{q\}}))(s)\}$$
$$\leq\ {\rm the\ right-hand\ side\ of\ (b)}.\ \ \ \ \ \
\ \ \ \ \ \ \ \ \ \ \ \ \ \ \ \ \ \ \ \ \ \ \ \ \ \ \ \ \ \ \ \ \
\ \ \ \ \ \ \ \ \ \ \ \ \ \ \ \ \ \ \ \ \ \ \ \ \ \ \ \ \ \ \ \ \
$$

After proving (a), we can assert that
$$\lceil s\in L(k(\Re ))\rceil =\vee
_{u,v\in Q}[I(u)\wedge T(v)\wedge L(\alpha _{uv}^{Q})(s)]\ \ \ \ \
\ \ \ \ \ \ \ \ \ \ \ \ \ \ \ \ \ \ \ \ \ \ \ \ \ \ \ \ \ \ \ \ \
\ \ \ \ \ $$
$$\geq \vee _{u,v\in Q}[(I(u)\wedge T(v))\wedge \vee \{\lceil path_{\Re
}(c)\rceil :c\in T(Q,\Sigma ),b(c)=u,e(c)=v,lb(c)=s\}]$$
$$\geq \vee _{u,v\in Q}\vee \{I(u)\wedge T(v)\wedge \lceil path_{\Re }(c)\rceil :c\in
T(Q,\Sigma ),b(c)=u,e(c)=v,lb(c)=s\}\ \ $$
$$=\lceil rec_{\Re }(s)\rceil .\ \ \ \ \ \ \ \ \ \ \ \ \ \ \ \ \ \ \ \ \ \ \ \ \ \ \ \ \ \ \
\ \ \ \ \ \ \ \ \ \ \ \ \ \ \ \ \ \ \ \ \ \ \ \ \ \ \ \ \ \ \ \ \
\ \ \ \ \ \ \ \ \ \ \ \ \ \ \ \ \ \ \ \ \ \ \ \ $$

By using (b) and Lemmas 2.5 and 2.6, we have
$$\lceil \gamma (atom(\Re))\rceil \wedge \lceil s\in L(k(\Re ))\rceil =\lceil \gamma (atom(\Re))\rceil \wedge \vee
_{u,v\in Q}[I(u)\wedge T(v)\wedge L(\alpha _{uv}^{Q})(s)]$$
$$\leq \vee
_{u,v\in Q}[I(u)\wedge T(v)\wedge \lceil \gamma (atom(\Re))\rceil
\wedge L(\alpha _{uv}^{Q})(s)]\ \ \ \ \ \ \ \ \ \ \ \ \ \ \ \ \ \
\ \ \ \ \ \ \ \ \ \ \ \ \ \ \ \ \ \ \ \ $$
$$\leq \vee _{u,v\in Q}[(I(u)\wedge T(v))\wedge \lceil \gamma (atom(\Re))\rceil\wedge \vee \{\lceil path_{\Re
}(c)\rceil :c\in T(Q,\Sigma ),\ \ \ \ \ \ \ \ \ \ \ \ \ \ $$
$$b(c)=u,e(c)=v,lb(c)=s\}]$$
$$\leq \vee _{u,v\in Q}\vee \{I(u)\wedge T(v)\wedge \lceil path_{\Re }(c)\rceil :c\in
T(Q,\Sigma ),b(c)=u,e(c)=v,lb(c)=s\}$$
$$=\lceil rec_{\Re }(s)\rceil .\ \ \ \ \ \ \
\ \ \ \ \ \ \ \ \ \ \ \ \ \ \ \ \ \ \ \ \ \ \ \ \ \ \ \ \ \ \ \ \
\ \ \ \ \ \ \ \ \ \ \ \ \ \ \ \ \ \ \ \ \ \ \ \ \ \ \ \ \ \ \ \ \
\ \ \ \ \ \ \ \ \ \ \ \ \ $$

Thus, (1) and (2) are proved, and the part that (i) implies (ii)
of (3) is a simple corollary of (2). We now turn to prove that
(ii) implies (i). For any $a,b,c\in L,$ we consider the $\ell
-$valued automaton
$$\Re =<\{u,v\},\delta ,u_{a},\{u,v\}>,$$ where $\delta (u,\sigma
,u)=b,$ $\delta (u,\sigma ,v)=c,$ and $\delta $ takes value $0$
for other cases (see Figure 9). Then
$$\lceil rec_{\Re }(\sigma )\rceil =\vee \{I(q_{0})\wedge
T(q_{1})\wedge \delta (q_{0},\sigma ,q_{1}):q_{0},q_{1}\in Q\}$$
$$=(a\wedge b)\vee (a\wedge c).$$

\begin{figure}\centering
\includegraphics{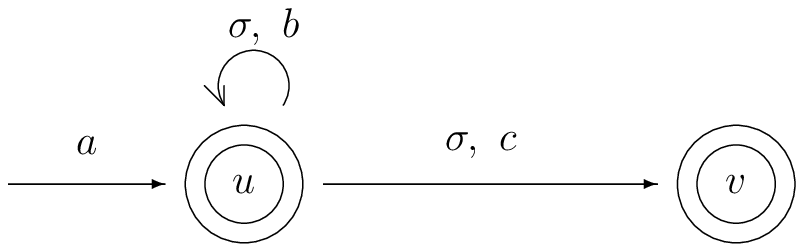}\caption{Automaton g} \label{fig 9}
\end{figure}

On the other hand, we have
$$\left\{
\begin{array}{c}
\alpha _{uu}^{\phi }=\varepsilon +b\sigma , \\
\alpha _{uv}^{\phi }=c\sigma , \\
\alpha _{vv}^{\phi }=\varepsilon , \\
\alpha _{vu}^{\phi }=\phi .
\end{array}
\right. $$ Therefore,
$$\alpha _{uu}^{\{v\}}=\alpha
_{uu}^{\phi }+[\alpha _{uv}^{\phi }(\alpha _{vv}^{\phi })^{\ast
}]\alpha _{vu}^{\phi }$$
$$=(\varepsilon +b\sigma )+[c\sigma
(\varepsilon )^{\ast }]\phi $$
$$=\varepsilon +b\sigma ,$$
$$\alpha _{uv}^{\{v\}}=\alpha _{uv}^{\phi }+[\alpha _{uv}^{\phi
}(\alpha _{vv}^{\phi })^{\ast }]\alpha _{vv}^{\phi }$$
$$=c\sigma +[c\sigma (\varepsilon )^{\ast }]\varepsilon $$
$$=c\sigma ,$$
and
$$\alpha _{uv}^{\{u,v\}}=\alpha _{uu}^{\{v\}}+[\alpha _{uu
}^{\{v\}}(\alpha _{u}^{\{v\}})^{\ast }]\alpha _{uv}^{\{v\}}$$
$$=\varepsilon +b\sigma +[(\varepsilon +b\sigma )(\varepsilon
+b\sigma )^{\ast }](c\sigma ).$$

From the assumption (ii) we know that
$$(a\wedge b)\vee (a\wedge
c)=\lceil rec_{\Re }(\sigma )\rceil $$
$$=L(k(\Re ))(\sigma )$$
$$=[L(a\alpha _{u}^{\{u,v\}})\cup L(a\alpha
_{uv}^{\{u,v\}})](\sigma )$$
$$\geq L(a\alpha _{u}^{\{u,v\}})(\sigma )$$
$$=a\wedge L(\alpha _{u}^{\{u,v\}})(\sigma )$$
$$=a\wedge L(\varepsilon +b\sigma +[(\varepsilon +b\sigma
)(\varepsilon +b\sigma )^{\ast }](c\sigma ))(\sigma )$$
$$\geq
a\wedge (b\vee c).$$ This completes the proof.$\heartsuit$

\smallskip\

\textbf{Corollary 7.2.} Let $\ell =<L,\leq ,\wedge ,\vee ,\bot
,0,1>$ be an orthomodular lattice, and let
$\rightarrow=\rightarrow_3$. Then for any $A\in L^{\Sigma^{*}}$,
$$\stackrel{\ell}{\models} CReg_{\Sigma}(A)\rightarrow (\exists\ {\rm regular\ expression}\ \alpha)(A\equiv L(\alpha)).$$

\smallskip\

\textbf{Proof.} It can be derived from Theorem 7.1 in a way
similar to the proof of Corollary 4.2.$\heartsuit$

\smallskip\

We now turn to consider homomorphisms of $\ell-$valued regular
expressions. Let $\Sigma$ and $\Gamma$ be two alphabet, and let
$h:\Sigma \rightarrow \Gamma ^{\ast }$ be a mapping. Then it can
be uniquely extended to a mapping, denoted still by $h,$ from
$\ell -$valued regular expressions over $\Sigma $ into $\ell
-$valued regular expressions over $\Gamma .$ For any $\ell
-$valued regular expression $\alpha $ over $ \Sigma ,$ $h(\alpha
)$ is defined to be the $\ell -$valued regular expression over
$\Gamma $ obtained by replacing each letter $\sigma \in \Sigma $
appearing in $\alpha $ with the string $h(\sigma )\in \Gamma
^{\ast }.$ Formally, $h(\alpha )$ is defined by induction on the
length of $\alpha : $
$$h(\varepsilon )=\varepsilon ,$$
$$h(\phi )=\phi ,$$
$$h(\sigma )\ {\rm is\ already\ given\ for\ each}\ \sigma \in \Sigma
,$$
$$h(\lambda \alpha )=\lambda h(\alpha ),$$
$$h(\alpha _{1}+\alpha _{2})=h(\alpha _{1})+h(\alpha _{2}),$$
$$h(\alpha _{1}\cdot\alpha _{2})=h(\alpha _{1})\cdot h(\alpha _{2}),$$
$$h(\alpha ^{\ast })=(h(\alpha ))^{\ast }.$$

\smallskip\

For each $\ell-$valued regular expression $\alpha$ over $\Sigma$,
we write $\Lambda (\alpha )$ for the set of scalar values $\lambda
\in L$ occurring in $\alpha .$ Indeed, $\Lambda (\alpha )$ may be
formally defined by induction on the length of $\alpha $ as
follows:
$$\Lambda (\varepsilon )=\Lambda (\phi )=\Lambda (\sigma )=\phi\ {\rm for\ every}\
\sigma \in \Sigma ,$$
$$\Lambda (\lambda \alpha )=\{\lambda \}\cup
\Lambda (\alpha ),$$
$$\Lambda (\alpha _{1}+\alpha _{2})=\Lambda (\alpha _{1}\cdot \alpha
_{2})=\Lambda (\alpha _{1})\cup \Lambda (\alpha _{2}),$$
$$\Lambda
(\alpha ^{\ast })=\Lambda (\alpha ).$$ It is easy to see that
$\Lambda (\alpha )$ is a finite subset of $L.$ Moreover, we write
$$\Delta(\alpha)=\{\mathbf{a}:a\in \Lambda (\alpha)\}$$ for the
set of (constant) propositions in our logical language
corresponding to the elements in $\Lambda (\alpha)$.

The following two lemmas are very useful when we are dealing with
orthomodular lattice-valued expressions, they evaluate the range
of language generated by an orthomodular lattice-valued regular
expression. In particular, it will be shown in Lemma 7.4 that this
range is a finite set whenever the lattice $\ell$ of truth values
is a Boolean algebra.

\smallskip\

\textbf{Lemma 7.3.} Let $\ell =<L,\leq ,\wedge ,\vee ,\bot ,0,1>$
be an orthomodular lattice. Then for any $\ell -$valued regular
expression $\alpha ,$ $ \{L(\alpha )(s):s\in \Sigma ^{\ast
}\}\subseteq [\Lambda (\alpha )],$ where $[A]$ denotes the
subalgebra of $\ell$ generated by $A$ for any $A\subseteq L.$

\smallskip\

\textbf{Proof.} We use an induction argument on the length of
$\alpha .$ For simplicity, we only consider the following two
cases, and the other cases are easy or similar.

(1) From the induction hypothesis we know that
$$L(\lambda .\alpha
)(s)=\lambda \wedge L(\alpha )(s)\in \overline{\{\lambda \}\cup
\Lambda (\alpha )}=\overline{\Lambda (\lambda .\alpha )}$$ for
each $s\in \Sigma ^{\ast }.$

(2) Let $s\in \Sigma ^{\ast }.$ For any $s_{1},...,s_{n}\in \Sigma
^{\ast }$ with $s_{1}...s_{n}=s,$ we suppose that
$s_{i_{1}},...,s_{i_{m}}\neq \varepsilon $ and $s_{i}=\varepsilon
$ for every $i\in \{1,...,n\}-\{i_{1},...,i_{m}\}.$ Then
$s_{i_{1}}...s_{i_{m}}=s$ and
$$L(\alpha )(s_{1})\wedge ...\wedge
L(\alpha )(s_{n})=\left\{
\begin{array}{c}
L(\alpha )(s_{i_{1}})\wedge ...\wedge L(\alpha )(s_{i_{m}})\ {\rm
if }\ m=n,
\\
L(\alpha )(s_{i_{1}})\wedge ...\wedge L(\alpha )(s_{i_{m}})\wedge
L(\alpha )(\varepsilon )\ {\rm if }\ m<n.
\end{array}
\right. $$ Furthermore, we note that $$\{(s_{1},...,s_{n}):n\geq
0,s_{1},...,s_{n}\in \Sigma ^{\ast }-\{\varepsilon \}\ {\rm and}\
s_{1}...s_{n}=s\}$$ is finite. Therefore, $$\{L(\alpha
)(s_{1})\wedge ...\wedge L(\alpha )(s_{n}):s_{1}...s_{n}=s\}$$ is
also finite, and with the induction hypothesis we have
$$L(\alpha ^{\ast })(s)=\vee \{L(\alpha )(s_{1})\wedge ...\wedge
L(\alpha )(s_{n}):s_{1}...s_{n}=s\}\in \overline{\Lambda (\alpha
)}.\heartsuit$$

\smallskip\

\textbf{Lemma 7.4.} If $\ell =<L,\leq ,\wedge ,\vee ,\bot ,0,1>$
is a Boolean algebra, then for any $\ell -$valued regular
expression $\alpha ,$ $ \{L(\alpha )(s):s\in \Sigma ^{\ast }\}$ is
a finite set.

\smallskip\

\textbf{Proof. }From Lemma 7.3 and the distributivity of $\wedge $
over $\vee $ we know that for any $s\in \Sigma ^{\ast },$ there
are $\lambda _{ij_{i}}\in \Lambda (\alpha )$
$(i=1,...,m;j_{i}=1,...,n_{i})$ such that
$$L(\alpha )(s)=\vee
_{i=1}^{m}(\wedge _{j_{i}=1}^{n_{i}}\lambda _{ij_{i}}).$$ Since
$\Lambda (\alpha )$ is finite, both $$\Lambda (\alpha )^{(\wedge
)}=\{\lambda _{1}\wedge ...\wedge \lambda _{n}:n\geq 0,\lambda
_{1},...,\lambda \in \Lambda (\alpha )\}$$ and $$\Lambda (\alpha
)^{(\wedge )(\vee )}=\{\vee M:M\subseteq \Lambda (\alpha
)^{(\wedge )}\}$$ are also finite. Therefore, $$\Lambda (\alpha
)^{(\wedge )(\vee )}\supseteq \{L(\alpha )(s):s\in \Sigma ^{\ast
}\}$$ is a finite set.$\heartsuit$

\smallskip\

The following proposition shows that a homomorphism preserves the
language generated by an orthomodular lattice-valued regular
expression under the condition that all elements in the range of
the expression under consideration are commutative.

\smallskip\

\textbf{Proposition 7.5.} Let $\ell =<L,\leq ,\wedge ,\vee ,\bot
,0,1>$ be an orthomodular lattice and $\rightarrow$ fulfil the
Birkhoff-von Neumann requirement, and let $\Sigma$ and $\Gamma$ be
two alphabets.

(1) For any mapping $h:\Sigma\rightarrow \Gamma ^{\ast }$, and for
any $\ell -$valued regular expression $\alpha $ over $\Sigma ,$
$$\stackrel{\ell}{\models} h(L(\alpha ))\subseteq L(h(\alpha )).$$

(2) For any mapping $h:\Sigma\rightarrow \Gamma ^{\ast }$, for any
$\ell -$valued regular expression $\alpha $ over $\Sigma ,$ and
for any $t\in \Gamma^{*}$,
$$\stackrel{\ell}{\models}\gamma(\Delta(\alpha))\wedge t\in L(h(\alpha))\rightarrow
 t\in h(L(\alpha)),$$ and if $\rightarrow=\rightarrow_3$ then
$$\stackrel{\ell}{\models}\gamma(\Delta(\alpha))\rightarrow
L(h(\alpha))\equiv h(L(\alpha)).$$

(3) The following two statements are equivalent:

\ \ \ \ \ \ (i) $\ell $ is a Boolean algebra.

\ \ \ \ \ \ (ii) for any mapping $h:\Sigma\rightarrow \Gamma
^{\ast }$, and for any $\ell-$valued regular expression $ \alpha $
over $\Sigma ,$
$$\stackrel{\ell}{\models} h(L(\alpha ))\equiv L(h(\alpha )).$$

\smallskip\

\textbf{Proof.} We only prove (2) and (3), and (1) can be observed
from the proof of (2). The part that (i) implies (ii) of (3) may
be derived from (2); and it can also be proved directly by using
Lemma 7.4.

Our first aim is to prove that $$\lceil
\gamma(\Delta(\alpha))\rceil\wedge L(h(\alpha ))(t)=h(L(\alpha
))(t)$$ for any $t\in \Gamma ^{\ast }$ and for any $\ell -$valued
regular expression $\alpha $ over $ \Sigma .$ We proceed by
induction on the length of $\alpha .$

(a) It is obvious for the case of $\alpha =\varepsilon $ or $\phi
,$ or $ \alpha \in \Sigma .$

(b) With the definitions of $h(\cdot )$ and $L(\cdot )$ and the
induction hypothesis we derive that
$$L(h(\lambda .\alpha ))(t)=L(\lambda .h(\alpha ))(t)$$
$$=\lambda \wedge L(h(\alpha ))(t)$$
$$=\lambda \wedge h(L(\alpha ))(t)$$
$$=\lambda \wedge \vee \{L(\alpha )(s):s\in \Sigma ^{\ast }\ {\rm
and}\ h(s)=t\}.$$ Then from Lemmas 2.5, 2.6 and 7.3, it follows
that
$$\lceil \gamma(\Delta(\alpha))\rceil\wedge L(h(\lambda
.\alpha ))(t)\leq \vee \{\lambda \wedge L(\alpha )(s):s\in \Sigma
^{\ast }\ {\rm and}\ h(s)=t\}$$
$$=\vee \{L(\lambda .\alpha )(s):s\in
\Sigma ^{\ast }\ {\rm and}\ h(s)=t\}$$
$$=h(L(\lambda .\alpha
))(t).$$

(c) It is easy to observe that $h(A\cup B)=h(A)\cup h(B)$ for all
$A,B\in L^{\Sigma ^{\ast }}.$ This together with the induction
hypothesis as well as Lemmas 2.5 and 2.6 yields
$$\lceil \gamma(\Delta(\alpha_1 + \alpha_2))\rceil\wedge L(h(\alpha _{1}+\alpha
_{2}))(t) =\lceil \gamma(\Delta(\alpha_1 + \alpha_2))\rceil\wedge
L(h(\alpha _{1})+h(\alpha _{2}))(t)$$
$$=\lceil \gamma(\Delta(\alpha_1 + \alpha_2))\rceil\wedge\lceil \gamma(\Delta(\alpha_1 + \alpha_2))\rceil\wedge
[L(h(\alpha _{1}))(t)\vee L(h(\alpha _{2}))(t)]$$ $$\leq [\lceil
\gamma(\Delta(\alpha_1 + \alpha_2))\rceil\wedge
L(h(\alpha_1))(t)]\wedge [\lceil \gamma(\Delta(\alpha_1 +
\alpha_2))\rceil\wedge L(h(\alpha_2))(t)]$$ $$\leq [\lceil
\gamma(\Delta(\alpha_1))\rceil\wedge L(h(\alpha_1))(t)]\wedge
[\lceil \gamma(\Delta(\alpha_2))\rceil\wedge L(h(\alpha_2))(t)]$$
$$h(L(\alpha _{1}))(t)\vee h(L(\alpha _{2}))(t)$$
$$=h(L(\alpha _{1})\cup L(\alpha _{2}))(t)$$
$$=h(L(\alpha _{1}+\alpha _{2}))(t).$$

(d) For any $t\in \Gamma ^{\ast },$ Lemmas 2.5, 2.6 and 7.3 enable
us to assert that $$\lceil \gamma(\Delta(\alpha_1\cdot
\alpha_2))\rceil\wedge L(h(\alpha_1\cdot \alpha_2))(t)=\lceil
\gamma(\Delta(\alpha_1\cdot \alpha_2))\rceil\wedge
L(h(\alpha_1)\cdot h(\alpha_2))(t)$$ $$=\lceil
\gamma(\Delta(\alpha_1\cdot \alpha_2))\rceil\wedge
L(h(\alpha_1))L(h(\alpha_2))(t)$$
$$=\lceil \gamma(\Delta(\alpha_1\cdot
\alpha_2))\rceil\wedge \vee\{L(h(\alpha_1))(t_1)\wedge
L(h(\alpha_2))(t_2):t_1t_2=t\}$$
$$\leq \vee\{\lceil \gamma(\Delta(\alpha_1\cdot
\alpha_2))\rceil\wedge L(h(\alpha_1))(t_1)\wedge
L(h(\alpha_2))(t_2):t_1t_2=t\}$$
$$=\vee\{\lceil \gamma(\Delta(\alpha_1\cdot
\alpha_2))\rceil\wedge (\lceil
\gamma(\Delta(\alpha_1))\rceil\wedge L(h(\alpha_1))(t_1))\wedge
(\lceil \gamma(\Delta(\alpha_2))\rceil\wedge
L(h(\alpha_2))(t_2)):t_1t_2=t\}$$
$$\leq \vee\{\lceil \gamma(\Delta(\alpha_1\cdot
\alpha_2))\rceil\wedge h(L(\alpha_1))(t_1)\wedge
h(L(\alpha_2))(t_2):t_1t_2=t\}.$$ Furthermore, by using Lemmas
2.5, 2.6 and 7.3 again we obtain $$\lceil
\gamma(\Delta(\alpha_1\cdot \alpha_2))\rceil\wedge
h(L(\alpha_1))(t_1)\wedge h(L(\alpha_2))(t_2)=\lceil
\gamma(\Delta(\alpha_1\cdot \alpha_2))\rceil\wedge
(\vee\{L(\alpha_1)(s_1):h(s_1)=t_1\})$$
$$\wedge (\vee\{L(\alpha_2)(s_2):h(s_2)=t_2\})$$
$$\leq \vee\{L(\alpha_1)(s_1)\wedge L(\alpha_2)(s_2):h(s_1)=t_1\ {\rm and}\
h(s_2)=t_2\}.$$ Therefore, it follows that
$$\lceil \gamma(\Delta(\alpha_1\cdot
\alpha_2))\rceil\wedge L(h(\alpha_1\cdot \alpha_2))(t)\leq
\vee\{L(\alpha_1)(s_1)\wedge$$ $$L(\alpha_2)(s_2):h(s_1)=t_1,
h(s_2)=t_2\ {\rm and}\ t_1t_2=t\}$$
$$=\vee\{L(\alpha_1)(s_1)\wedge L(\alpha_2)(s_2):h(s_1s_2)=t\}$$
$$=h(L(\alpha_1)L(\alpha_2))(t)$$ $$=h(L(\alpha_1\alpha_2))(t).$$

(e) For every $t\in \Gamma ^{\ast },$ Lemmas 2.5, 2.6 and 7.3
guarantee that $$\lceil \gamma (\Delta (\alpha ^{\ast }))\rceil
\wedge L(h(\alpha ^{\ast }))(t)=\lceil \gamma (\Delta (\alpha
^{\ast }))\rceil \wedge L((h(\alpha ))^{\ast })(t)$$
$$=\lceil \gamma (\Delta (\alpha ^{\ast }))\rceil \wedge
(L(h(\alpha )))^{\ast }(t)$$
$$=\lceil \gamma (\Delta (\alpha ^{\ast }))\rceil \wedge \vee
\{\wedge _{i=1}^{n}L(h(\alpha ))(t_{i}):n\geq 0,t_{1},...,t_{n}\in
\Gamma ^{\ast },t_{1}...t_{n}=t\}$$
$$\leq \vee \{\lceil \gamma (\Delta (\alpha ^{\ast }))\rceil \wedge
\wedge _{i=1}^{n}L(h(\alpha ))(t_{i}):n\geq 0,t_{1},...,t_{n}\in
\Gamma ^{\ast },t_{1}...t_{n}=t\}$$
$$=\vee \{\lceil \gamma (\Delta (\alpha ))\rceil \wedge \wedge
_{i=1}^{n}(\lceil \gamma (\Delta (\alpha ))\rceil \wedge
L(h(\alpha ))(t_{i})):n\geq 0,t_{1},...,t_{n}\in \Gamma ^{\ast
},t_{1}...t_{n}=t\}$$
$$\leq \vee \{\lceil \gamma (\Delta (\alpha ))\rceil \wedge \wedge
_{i=1}^{n}h(L(\alpha ))(t_{i}):n\geq 0,t_{1},...,t_{n}\in \Gamma
^{\ast },t_{1}...t_{n}=t\}.$$ On the other hand, we have
$$\lceil \gamma (\Delta (\alpha ))\rceil \wedge \wedge
_{i=1}^{n}h(L(\alpha ))(t_{i})=\lceil \gamma (\Delta (\alpha
))\rceil \wedge \wedge _{i=1}^{n}(\vee \{L(\alpha
)(s_{i}):h(s_{i})=t_{i}\})$$
$$\leq \vee \{\wedge _{i=1}^{n}L(\alpha )(s_{i}):h(s_{i})=t_{i}\
(i=1,...,n)\}.$$ This further yields
$$\lceil \gamma (\Delta (\alpha ^{\ast }))\rceil \wedge L(h(\alpha
^{\ast }))(t)\leq \vee \{\wedge _{i=1}^{n}L(\alpha )(s_{i}):n\geq
0,h(s_{i})=t_{i}\ (i=1,...,n)\ {\rm and}\ t=t_{1}...t_{n}\}$$
$$=\vee \{\wedge _{i=1}^{n}L(\alpha )(s_{i}):n\geq
0,h(s_{1}...s_{n})=t\}$$
$$=\vee \{L(\alpha )^{\ast }(s):h(s)=t\}$$
$$=h((L(\alpha ))^{\ast })(t)$$
$$=h(L(\alpha ^{\ast }))(t).$$

What remains is to prove that (ii) implies (i) in (3). This needs
indeed to show that the distributivity of $\wedge$ over $\vee$ is
derivable from the statement (ii). Suppose that $a,b,c\in L.$ We
choose an symbol $\sigma \in \Sigma $ and an symbol $\gamma \in
\Gamma ,$ and define $h(\sigma )=\varepsilon $ and $h(\sigma
^{\prime })=\gamma $ for every $\sigma ^{\prime }\in \Sigma
-\{\sigma \}.$ We further set $\alpha _{1}=a.\sigma $ and $\alpha
_{2}=b.\varepsilon +c.\sigma .$ Then
$$L(\alpha _{1}.\alpha
_{2})(\sigma )=\left\{
\begin{array}{c}
a\wedge b\ {\rm if }\ n=1, \\
a\wedge c\ {\rm if }\ n=2, \\
0\ {\rm otherwise,}
\end{array}
\right. $$ and
$$h(L(\alpha _{1}.\alpha _{2}))(\varepsilon )=\vee
_{n=0}^{\infty }L(\alpha _{1}.\alpha _{2})(\sigma ^{n})$$
$$=(a\wedge b)\vee (a\wedge c).$$
On the other hand, we have
$$L(h(\alpha _{1}.\alpha
_{2}))(\varepsilon )=L((a.\varepsilon ).(b.\varepsilon
+c.\varepsilon ))(\varepsilon )$$
$$=L(a.\varepsilon )(\varepsilon
)\wedge L(b.\varepsilon +c.\varepsilon )(\varepsilon )$$
$$=a\wedge
(b\vee c).$$

From (ii) we know that $h(L(\alpha _{1}.\alpha _{2}))(\varepsilon
)=L(h(\alpha _{1}.\alpha _{2}))(\varepsilon ).$ This indicates
that $ (a\wedge b)\vee (a\wedge c)=a\wedge (b\vee c).\heartsuit$

\bigskip\

\textbf{8. Pumping Lemma for Orthomodular Lattice-Valued Regular
Languages}

\smallskip\

The pumping lemma in the classical automata theory is a powerful
tool to show that certain languages are not regular, and it
exposes some limitations of finite automata. The purpose of this
section is to establish a generalization of the pumping lemma for
orthomodular lattice-valued languages. It is worth noting that the
following orthomodular lattice-valued version of pumping lemma is
given for the commutative regularity $CReg_\Sigma$. In general,
the pumping lemma is not valid for the predicate $Reg_\Sigma$.

\smallskip\

\textbf{Theorem 8.1.} (The pumping lemma) Let $\ell =<L,\leq
,\wedge ,\vee ,\perp ,0,1>$ be an orthomodular lattice, and let
$\rightarrow =\rightarrow _{3}.$ Then for any $A\in L^{\Sigma
^{\ast }},$ if $Range(A)$ is finite, then
$$\stackrel{\ell }{\models}CReg_{\Sigma }(A)\rightarrow (\exists n\geq 0)(\forall s\in
\Sigma ^{\ast })[s\in A\wedge |s|\geq n\rightarrow $$
$$(\exists u,v,w\in \Sigma ^{\ast })(s=uvw\wedge |uv|\leq n\wedge
|v|\geq 1\wedge (\forall i\geq 0)(uv^{i}w\in A))],$$ where for any
word $t=\sigma _{1}...\sigma _{k}\in \Sigma ^{\ast },$ $|t|$
stands for the length $n$ of $t.$

\smallskip\

\textbf{Proof.} For simplicity, we use $X(s,n)$ to mean the
statement that $u,v,w\in \Sigma ^{\ast },\ s=uvw,\ |uv|\leq n,\
{\rm and}\ |v|\geq 1$ for each $s\in \Sigma^{\ast}$ and $n\geq 0$.
Then it suffices to show that
$$\lceil CReg_{\Sigma }(A)\rceil \leq \vee _{n\geq 0}\wedge _{s\in
\Sigma ^{\ast },|s|\geq n}(A(s)\rightarrow \vee _{X(s,n)}\wedge
_{i\geq 0}A(uv^{i}w)).$$ From Definition 3.3 we know that
$$\lceil CReg_{\Sigma }(A)\rceil =\vee _{\Re \in \mathbf{A}(\Sigma
,\ell )}(\lceil \gamma (atom(\Re )\cup r(A)\rceil \wedge \lceil
A\equiv rec_{\Re }\rceil ).$$ Thus, we only need to prove that for
any $\Re \in \mathbf{A}(\Sigma ,\ell ),$
$$\lceil \gamma (atom(\Re )\cup r(A)\rceil \wedge \lceil A\equiv
rec_{\Re }\rceil \leq \vee _{n\geq 0}\wedge _{s\in \Sigma ^{\ast
},|s|\geq n}(A(s)\rightarrow $$ $$\vee _{X(s,n)}\wedge _{i\geq
0}A(uv^{i}w)).$$

Let $Q$ be the set of states of $\Re .$ First, it holds that for
any $s\in \Sigma ^{\ast }$ with $|s|\geq |Q|,$ $$(1)\ \ \ \ \ \ \
\ \ \ \ \ \ \ \ \ \ \ \ \ \ \ \ \ \ \ \ \ \ \ \ \ rec_{\Re
}(s)\leq \vee _{X(s,n)}\wedge _{i\geq 0}rec_{\Re }(uv^{i}w).\ \ \
\ \ \ \ \ \ \ \ \ \ \ \ \ \ \ \ \ \ \ \ \ \ $$ In fact, suppose
that $s=\sigma _{1}...\sigma _{k}.$ Then
$$(2)\ \ \ \ \ \ \ \ \ \ \ \ \ \ \ \ rec_{\Re }(s)=\vee _{q_{0},q_{1},...,q_{k}}[I(q_{0})\wedge
T(q_{k})\wedge \wedge _{i=0}^{k-1}\delta (q_{i},\sigma
_{i+1},q_{i+1})].\ \ \ \ \ \ \ \ \ \ \ \ \ \ $$ Therefore, it
amounts to showing that for any $q_{0},q_{1},...,q_{k}\in Q,$
$$ (3)\ \ \ \ \ \ \ \ \ \  \ \ \ I(q_{0})\wedge T(q_{k})\wedge \wedge _{i=0}^{k-1}\delta
(q_{i},\sigma _{i+1},q_{i+1})\leq \vee _{X(s,n)}\wedge _{i\geq
0}rec_{\Re }(uv^{i}w).\ \ \ \ \ \ \ \ \ \ \ \ $$ Since $k=|s|\geq
|Q|,$ there are two identical states among $
q_{0},q_{1},...,q_{|Q|};$ in other words, there are $m\geq 0$ and
$n>0$ such that $m+n\leq |Q|$ and $q_{m}=q_{m+n}.$ We set
$u_{0}=\sigma _{1}...\sigma _{m},$ $v_{0}=\sigma _{m+1}...\sigma
_{m+n},$ and $w_{0}=\sigma _{m+n+1}...\sigma _{k}.$ Then
$s=u_{0}v_{0}w_{0},$ $|u_{0}v_{0}|=m+n\leq |Q|, $ $|v|=n\geq 1,$
and $$(4) \ \ \ \ \ \ \ \ \ \ \ \ \ \ \ \ \ \ \ \ \ \vee
_{X(s,n)}\wedge _{i\geq 0}rec_{\Re }(uv^{i}w)\geq \wedge _{i\geq
0}rec_{\Re }(u_{0}v_{0}^{i}w_{0}).\ \ \ \ \ \ \ \ \ \ \ \ \ \ \ \
\ \ \ \ \ \ \ \ \ \ \ \ $$ From the definition of $rec_{\Re },$ it
is easy to see that for all $i\geq 0, $ $$(5)\ \ \ \ \ \ \ \ \ \ \
\ \ rec_{\Re }(u_{0}v_{0}^{i}w_{0})\geq \lceil path_{\Re
}(q_{0}\sigma _{1}q_{1}...\sigma _{m}q_{m}\ \ \ \ \ \ \ \ \ \ \ \
\ \ \ \ \ \ \ \ \ \ \ \ \ \ \ \ \ \ \ \ \ \ \ \ \ \ \ \ \ \ \ \ \
\ \ \  \ \ \ \ \ \ $$
$$(\sigma _{m+1}q_{m+1}...\sigma _{m+n}q_{m+n})^{i}\sigma
_{m+n+1}q_{m+n+1}...\sigma _{k}q_{k})\rceil $$
$$=I(q_{0})\wedge T(q_{k})\wedge \wedge
_{j=0}^{m+n-1}\delta (q_{j},\sigma _{j+1},q_{j+1})\wedge \wedge
_{l=1}^{i-1}[\delta (q_{m+n},\sigma _{m+1},q_{m+1})\wedge $$
$$\wedge _{j=m+1}^{m+n-1}\delta (q_{j},\sigma
_{j+1},q_{j+1})]\wedge \wedge _{j=m+n}^{k-1}\delta (q_{j},\sigma
_{j+1},q_{j+1})$$
$$=I(q_{0})\wedge T(q_{k})\wedge \wedge _{j=0}^{k-1}\delta
(q_{j},\sigma _{j+1},q_{j+1})$$ because $q_{m+n}=q_{m}$ and
$\delta (q_{m+n},\sigma _{m+1},q_{m+1})=\delta (q_{m},\sigma
_{m+1},q_{m+1}).$ Thus, by combining (4) and (5), we obtain (3)
which, together with (2), yields (1).

Now we use Lemmas 2.11(1) and (3) and obtain
$$\vee _{X(s,|Q|)}\wedge _{i\geq 0}rec_{\Re }(uv^{i}w)\rightarrow
\vee _{X(s,|Q|)}\wedge _{i\geq 0}A(uv^{i}w)\geq \lceil \gamma
(atom(\Re )\cup r(A)\rceil \wedge $$ $$\wedge _{X(s,|Q|)}(\wedge
_{i\geq 0}rec_{\Re }(uv^{i}w)\rightarrow \wedge _{i\geq
0}A(uv^{i}w))$$ $$\geq \lceil \gamma (atom(\Re )\cup r(A)\rceil
\wedge \wedge _{X(s,|Q|)}\wedge _{i\geq 0}(rec_{\Re
}(uv^{i}w)\rightarrow A(uv^{i}w))$$
$$\geq \lceil \gamma (atom(\Re )\cup r(A)\rceil \wedge \wedge
_{t\in \Sigma ^{\ast }}(rec_{\Re }(t)\rightarrow A(t))$$
$$=\lceil \gamma (atom(\Re )\cup r(A)\rceil \wedge \lceil rec_{\Re
}\subseteq A\rceil .$$ Furthermore, from the above inequality we
have
$$\lceil \gamma (atom(\Re )\cup r(A)\rceil \wedge \lceil rec_{\Re
}\equiv A\rceil =\lceil \gamma (atom(\Re )\cup r(A)\rceil \wedge
\lceil A\subseteq rec_{\Re }\rceil \wedge \lceil rec_{\Re
}\subseteq A\rceil $$
$$=\lceil \gamma (atom(\Re )\cup r(A)\rceil \wedge \wedge _{s\in
\Sigma ^{\ast }}(A(s)\rightarrow rec_{\Re }(s))\wedge \lceil
rec_{\Re }\subseteq A\rceil $$
$$\leq \lceil \gamma (atom(\Re )\cup r(A)\rceil \wedge \wedge
_{s\in \Sigma ^{\ast },|s|\geq |Q|}(A(s)\rightarrow rec_{\Re
}(s))\wedge \lceil rec_{\Re }\subseteq A\rceil $$
$$=\wedge _{s\in \Sigma ^{\ast },|s|\geq |Q|}(\lceil \gamma
(atom(\Re )\cup r(A)\rceil \wedge (A(s)\rightarrow rec_{\Re
}(s))\wedge \lceil \gamma (atom(\Re )\cup r(A)\rceil \wedge \lceil
rec_{\Re }\subseteq A\rceil )$$
$$\leq \wedge _{s\in \Sigma ^{\ast },|s|\geq |Q|}(\lceil \gamma
(atom(\Re )\cup r(A)\rceil \wedge (A(s)\rightarrow rec_{\Re
}(s))$$
$$\wedge (\vee _{X(s,|Q|)}\wedge _{i\geq 0}rec_{\Re
}(uv^{i}w)\rightarrow \vee _{X(s,|Q|)}\wedge _{i\geq
0}A(uv^{i}w))).$$ Then from (1) it follows that
$$\lceil \gamma (atom(\Re )\cup r(A)\rceil \wedge \lceil rec_{\Re
}\equiv A\rceil \leq \wedge _{s\in \Sigma ^{\ast },|s|\geq
|Q|}(\lceil \gamma (atom(\Re )\cup r(A)\rceil \wedge$$
$$(A(s)\rightarrow \vee _{X(s,|Q|)}\wedge _{i\geq 0}rec_{\Re
}(uv^{i}w))\wedge$$ $$(\vee _{X(s,|Q|)}\wedge _{i\geq 0}rec_{\Re
}(uv^{i}w)\rightarrow \vee _{X(s,|Q|)}\wedge _{i\geq
0}A(uv^{i}w))).$$ By using Lemmas 2.11(1) and (3) we know that
$$\lceil \gamma (atom(\Re )\cup r(A)\rceil \wedge \lceil rec_{\Re
}\equiv A\rceil \leq \wedge _{s\in \Sigma ^{\ast },|s|\geq
|Q|}(A(s)\rightarrow \vee _{X(s,|Q|)}\wedge _{i\geq
0}A(uv^{i}w))$$
$$\leq \vee _{n\geq 0}\wedge _{s\in \Sigma ^{\ast },|s|\geq
n}(A(s)\rightarrow \vee _{X(s,n)}\wedge _{i\geq 0}A(uv^{i}w)),$$
and this completes the proof.$\heartsuit$

\bigskip\

\textbf{9. Conclusion}

\smallskip\

It is argued that a theory of computation based on quantum logic
has to be established as a logical foundation of quantum
computation. This paper is the first one of a series of papers
toward such a new theory. Quantum logic is treated as an
orthomodular lattice-valued logic in this paper, and the aim of
the paper is to develop elementally a theory of finite automata
based on such a logic by employing the semantical analysis
approach. The notions of orthomodular lattice-valued finite
automaton and regular language are introduced. Some modifications
of orthomodular lattice-valued automaton are presented, including
the orthomodular lattice-valued generalizations of deterministic
and nondeterministic automata and automata with
$\varepsilon-$moves, and their equivalence are thoroughly
analyzed. We also examine the closure properties of orthomodular
lattice-valued regular languages under various operations. The
concept of orthomodular lattice-valued regular expressions is
proposed, and the Kleene theorem concerning the equivalence
between finite automata and regular expressions is generalized
within the framework of quantum logic. Also, an orthomodular
lattice-valued version of the pumping lemma is found. Furthermore,
a theory of pushdown automata or Turing machines based on quantum
logic will be developed in the continuations of the present paper.

In the development of automata theory based on quantum logic, some
essential differences between the computation theory established
by using the classical Boolean logic as the underlying logical
tool and that whose meta-logic is quantum logic have been
discovered. First, it is found that the proofs of some even very
basic properties of automata appeal an essential application of
the distributivity for the lattice of truth values of the
underlying logic. This indicates that these properties holds only
in Boolean logic but not in quantum logic. We believe that there
are also many fundamental properties of pushdown automata and
Turing machines whose universal validity requires the
distributivity of meta-logic. In a sense, this observation
provides us with a set of negative results in the theory of
computation based on quantum logic. These negative results might
hints some limitations of quantum computers. More explicitly, some
methods based on certain properties of classical automata maybe
have been successfully used in the implementation of classical
computer systems, but they do not apply to quantum computers, or
at least they are only conditionally effective for quantum
computers. On the other hand, although these negative results are
found in the computation theory based on quantum logic, it seems
that some similar negative results exist in other mathematical
theories based on nonclassical logics. This stimulates us to
consider the problem of a logical revisit to mathematics. Various
classical mathematical results have been established based upon
classical logic, and sometimes, their universal validity can only
be established by exploiting the full power of classical logic.
Mathematicians usually use logic implicitly in their reasoning,
and they do not seriously care which logical laws they have
employed. But from a logician's point of view, it is very
interesting to determine how strong a logic we need to validate a
given mathematical theorem, and which logic guarantees this
theorem and which does not among the large population of
nonclassical logics. To be more explicit and also for a
comparison, let us present a short excerpt from A. Heyting [He63,
page 3]:
$$$$
\textit{"It may happen that for the proof of a theorem we do not
need all the axioms, but only some of them. Such a theorem is true
not only for models of the whole system, but also for those of the
smaller system which contains only the axioms used in the proof.
Thus it is important in an axiomatic theory to prove every theorem
from the least possible set of axioms."}
$$$$
We now are in a similar situation. The difference between our case
and A. Heyting's one is that we are concerned with the limitation
or redundance of power of the logic underlying an axiomatic
theory, whereas he considered that of axioms themselves. It seems
that the semantical analysis approach provides a nice framework
for solving this problem, much more suitable than a
proof-theoretical approach.

As stated above, some fundamental properties of automata are not
universally valid in quantum logic due to lack of distributivity.
However, a certain commutativity are able to regain a local
distributivity, and to give further a partial validity of these
properties in the theory of automata based on quantum logic. One
typical example of such properties is the equivalence of automata
and their various modifications. It is well-known that one
important witness for the Church-Turing thesis which asserts the
Turing machine is a general model of computation is that various
extensions of the Turing machine are all equivalent to itself. The
fact that the equivalence between automata and their modifications
depends upon the commutativity of their basic actions suggests us
to guess that the equivalence between the Turing machine and some
of its extensions may also need a support from a certain
commutativity. In the introduction, we already gave a physical
interpretation to the role of commutativity based on the
Heisenberg uncertainty principle, and pointed out that an
interesting connection may reside between the Heisenberg
uncertainty principle and the Church-Turing thesis. If this is
true, then it will give once again an evidence to the unity of the
whole science and to the fact that science is not only a simple
union of various subjects.

\bigskip\

\textbf{References}

\smallskip\

[Ba85] J. Barwise, in: J. Barwise and S. Feferman, eds.,
\textit{Model Theoretic Logics}, Springer-Verlag, Berlin, 1985,
pp. ?.

[Be80] P. A. Benioff, The computer as a physical system: a
microscopic quantum mechanical Hamiltonian model of computer s as
represented by Turing machines, \textit{Journal of Statistical
Physics} \textbf{22}(1980)563-591.

[Ben73] C. H. Bennet, Logical reversibility of computation,
\textit{IBM Journal of Research and Development}
\textbf{17}(1973)525-532.

[BN36] G. Birkhoff and J. von Neumann, The logic of quantum
mechanics, \textit{Annals of Mathematics},
\textbf{37}(1936)823-843.

[BH00] G. Bruns and J. Harding, Algebraic aspects of orthomodular
lattices, in: B. Coecke, D. Moore and A. Wilce (eds.),
\textit{Current Research in Operational Quantum Logic: Algebras,
Categories, Languages,} Kluwer, Dordrecht, 2000, pp. 37-65.

[CZ95] J. I. Cirac and P. Zoller, Quantum computation with cold
trapped ions, \textit{Physical Review Letters}
\textbf{74}(1995)4091-4094.

[CM00] J. P. Crutchfield and C. Moore, Quantum automata and
quantum grammar, \textit{Theoretical Computer Science}
\textbf{237}(2000)275-306.

[DC81] M. L. Dalla Chiara, Some meta-logical pathologies of
quantum logic, in: E. Beltrametti and B. C. van Fraassen (eds.),
\textit{Current Issues in Quantum Logics}, Plenum, New York, 1981,
pp. 147-159.

[DC86] M. L. Dalla Chiara, Quantum logic, in: D. Gabbay and E.
Guenthner (eds.), \textit{Handbook of Philosophical Logic, volume
III: Alternatives to Classical Logic}, D. Reidel Publishing
Company, Dordrecht, 1986, pp. 427-469. Also see arXiv:
quant-ph/0101028 for an extended and revised version.

[De85] D. Deutsch, Quantum theory, the Church-Turing principle and
the universal quantum computer, \textit{Proc. Roy. Soc. Lond}.
\textbf{A400}(1985)97-117.

[De89] D. Deutsch, Quantum computational networks, \textit{Proc.
Roy. Soc. Lond}. \textbf{A425}(1989)73-90.

[Di78] J. Dieudonne, The current trend of pure mathematics,
\textit{Advances in Mathematics} \textbf{27}(1978)235-255.

[E74] S. Eilenberg, \textit{Automata, Languages, and Machines,
volume A}, Academic Press, New York, 1974.

[Fe82] R. P. Feynman, Simulating physics with computers, \textit{
International Journal of Theoretical Physics},
\textbf{21}(1982)467-488.

[Fe86] R. P. Feynman, Quantum mechanical computers,
\textit{Foundations of Physics,} \textbf{16}(1986)507-531.

[Fi70] P. D. Finch, Quantum logic as an implication algebra,
\textit{Bulletin of Australian Mathematical Society}
\textbf{2}(1970)101-106.

[Gr96] L. K. Grover, A fast quantum mechanical algorithm for
database search, in: \textit{Proceedings of the 28th ACM STOC,}
1996, pp. 212-219.

[Gu00] S. Gudder, Basic properties of quantum automata,
\textit{Foundations of Physics}, \textbf{30}(2000)301-319.

[Ha82] W. Hatcher, \textit{The Logical Foundations of
Mathematics}, Pergamon, Oxford (1982).

[He63] A. Heyting, \textit{Axiomatic Projective Geometry,}
North-Holland, Amsterdam, 1963.

[Ho89] H. Hodes, Three-valued logics - an introduction, a
comparison of various logical lexica, and some philosophical
remarks, \textit{Annals of Pure and Applied Logic},
\textbf{43}(1989)99-145.

[Hu37] K. Husimi, Studies on the foundations of quantum mechanics
I, \textit{Proceedings of the Physicomath. Soc. of Japan}
\textbf{19}(1937)766-789.

[HMP75] L. Herman, E. Marsden and R. Piziak, Implication
connectives in orthomodular lattices, \textit{Notre Dame J. Formal
Logic} \textbf{16}(1975)305-328.

[HU79] J. E. Hopcroft and J. D. Ullman, \textit{Introduction to
Automata Theory, Languages, and Computation,} Addison-Wesley,
Reading, 1979.

[Ka74] G. Kalmbach, Orthomodular logic, \textit{Zeitschr. f. math.
Logik und Grundlagen d. Math} \textbf{20}(1974)395-406.

[Ka83] G. Kalmbach, \textit{Orthomodular Lattices}, Academic
Press, London, 1983.

[KW97] A. Kondacs and J. Watrous, On the power of quantum finite
state automata, in: \textit{Proc. of the 38th Annual Symposium on
Foundations of Computer Science}, pp.65-75, 1997.

[Ko97] D. C. Kozen, \textit{Automata and Computability,}
Springer-Verlag, New York, 1997.

[L93] S. Lloyd, A potentially realizable quantum computer,
\textit{Science} \textbf{261}(1993)1569-1571.

[Ma90] J. Malinowski, The deduction theorem for quantum logic -
some negative results, \textit{Journal of Symbolic Logic}
\textbf{55}(1990)615-625

[Mo65] A. Mostowski, \textit{Thirty Years of Foundational Studies:
Lectures on the Development of Mathematical Logic and the Study of
the Foundations of Mathematics in 1930-1964,} Acta Philosophica
Fennica 17, Helsinki, 1965.

[N62] J. von Neumann, Quantum logics (strict- and
probability-logics), summarized in: J. von Neumann,
\textit{Collected Works, vol. IV}, Macmillan, New York(1962).

[RS00] J. P. Rawling and S. A. Selesnick, Orthologic and quantum
logic: models and computational elements, \textit{Journal of the
ACM} \textbf{47}(2000)721-751.

[RR91] L. Rom\'{a}n and B. Rumbos, Quantum logic revisited,
\textit{ Foundations of Physics}, \textbf{21}(1991)727-734.

[RZ99] L. Rom\'{a}n and R. E. Zuazua, Quantum implication,
\textit{ International Journal of Theoretical Physics,}
\textbf{38}(1999)793-797.

[RT52] J. B. Rosser and A. R. Turquette, \textit{Many-Valued
Logics,} North-Holland, Amsterdam, 1952.

[Sc99] K. -G. Schlesinger, Toward quantum mathematics. I. From
quantum set theory to universal quantum mechanics, \textit{Journal
of Mathematics Physics,} \textbf{40}(1999)1344-1358.

[Sh94] P. W. Shor, Polynomial-time algorithm for prime
factorization and discrete logarithms on quantum computer, in:
\textit{Proc. 35th Annual Symp. on Foundations of Computer
Science, Santa Fe,} IEEE Computer Society Press, 1994.

[Sv98] K. Svozil, \textit{Quantum Logic,} Springer-Verlag, Berlin,
1998.

[T81] G. Takeuti, Quantum set theory, in: E. Beltrametti and B. C.
van Fraassen (eds.), \textit{Current Issues in Quantum Logics},
Plenum, New York, 1981, pp. 303-322.

[TD88] A. S. Troelstra and D. van Dalen, \textit{Constructivism in
Mathematics: An Introduction,}, volume I, II, North-Holland,
Amsterdam, 1988.

[VP] V. Vedral and M. B. Plenio, Basics of quantum computation,
\textit{ Prog. Quant. Electron}. \textbf{22}:1(1998).

[Ya93] A. C. Yao, Quantum circuit complexity, \textit{Proc. of the
34th Ann. IEEE Symp. on Foundations of Computer Science}, pp.
352-361, 1993.

[Yi91] M. S. Ying, Deduction theorem for many-valued inference,
\textit{ Zeitschr. f. math. Logik und Grundlagen d. Math}.
37:6(1991).

[Yi92a] M. S. Ying, The fundamental theorem of ultraproduct in
Pavelka's logic, \textit{Zeitschr. f. math. Logik und Grundlagen
d. Math.} 38:2(1992).

[Yi92b] M. S. Ying, Compactness, the Lowenheim-Skolem property and
the direct product of lattices of truth values, \textit{Zeitschr.
f. math. Logik und Grundlagen d. Math}. 38:4(1992).

[Yi91-93] M. S. Ying, A new approach for fuzzy topology (I), (II),
(III), \textit{Fuzzy Sets and Systems,} \textbf{39}(1991)303-321;
\textbf{47} (1992)221-232; \textbf{55}(1993)193-207.

[Yi93] M. S. Ying, Fuzzifying topology based on complete
residuated lattice-valued logic (I), \textit{Fuzzy Sets and
Systems,} \textbf{56}(1993)337-373.

[Yi94] M. S. Ying, A logic for approximate reasoning, \textit{J.
Symbolic Logic} \textbf{59}(1994).

[Yi00] M. S. Ying, Automata theory based on quantum logic (I),
(II), \textit{International Journal of Theoretical Physics}
\textbf{39} (2000), 985-995; 2545-2557.

\end{document}